\documentclass[10pt,journal,compsoc]{IEEEtran}

\usepackage[nocompress]{cite}
\usepackage[pdftex]{graphicx}   
\usepackage{hyperref}
\usepackage{url}

\usepackage{amsmath}
\usepackage{amssymb}

\makeindex

\newcommand{\citep}{\cite}

\begin{document}
	
	\title{An Ontological Analysis of Business Process Modeling and Execution}
	
	\author{
		Robert Singer\\
		FH JOANNEUM -- University of Applied Sciences\\
		Dep. of Applied Computer Sciences\\
		Alte Poststra\ss e 147\\
		8020 Graz, Austria\\
		robert.singer@fh-joanneum.at}
	
	\maketitle
	
	\begin{abstract}
This work presents a fully elaborated ontology, defined via the Ontology Web Language (OWL), of the Business Process Model and Notation (BPMN) standard to define business process models, and we demonstrate that any  BPMN model can be serialized as OWL file. Based on ontological analysis and a corresponding definition of a modeling notation as ontology we show that business process models can be transformed from one notation into another one as long as there are common underlying concepts; this is demonstrated with the case of an actor based, or subject-oriented, view on business processes. Furthermore, a reference architecture for Workflow Management Systems (WfMS) based on microservices is discussed which is capable of executing actor based business process models. As a transformation of BPMN models into the actor based view is generally possible, also BPMN models could be enacted. As a result, we can conclude that the actor system is a  promising way to stimulate new ways to design workflow management systems and to design business process modeling languages which are more comfortable to use by non-experts without losing necessary expressiveness. Another result is that an ontology is a productive way to define a modeling notation as it can be used as knowledge base, it is a formal conceptualization of the underlying notions, and can be semantically enriched for further use. 

	\end{abstract}
	

\section{Introduction}
\label{introduction}

Business process management (BPM) is a well-known concept known by management as well as experts of the information technology (IT) community. As already stated by Weske~\citep{weske2012}, members of these groups have different educational backgrounds and interests.

From a systems theoretical point of view, business process management constitutes a subsystem of the management system of a company, and it is a means to operationalize the business strategy. If the BPM subsystem works as intended, it is a management tool to steer the performance of the company. A detailed discussion of a systems view and the corresponding tools can be found, for example, in~\citep{malik2008a},~\citep{malik2011}, and~\citep{malik2013}, which is based on pioneer work of~\citep{beer1988}, for example.

In computer sciences, there are further interest groups, such as researchers with a background in formal methods to investigate structural properties of business processes and the software community which is interested in providing robust and scalable software systems to support the execution of business processes~\citep{weske2012}.

\section{Motivation and Definitions}
\label{motivationanddefinitions}

For a common understanding, it is worth to give definitions of the most important concepts used in this work. First, we define the notion of business process, and afterward the term business process management.

An elaborated definition of the notion of business process is as follows:

A business process is a network of connected activities (tasks), which are conducted by actors (humans or systems) in logical and temporal order with the help of tools (devices, software). The activities are executed on business objects (data or physical objects) to satisfy a customer requirement, and the business process has a defined beginning and input, and a defined end and a result (output). Furthermore, a business process is executed in a technical, organizational, and a social context.

Such a typical definition defines the architectural elements, or entities, which constitute a business process: activity, resource or actor, business object, customer, input, and output. That means a business process defines what is done (in logical order) and who is doing what.

A common definition~\citep{eabpm2014} of the term business process management is as follows:

\begin{quote}
Business process management (BPM) is a management discipline that integrates the strategy and goals of an organization with the expectations and needs of customers by focusing on end-to-end processes. BPM comprises strategies, goals, culture, organizational structures, roles, policies, methodologies, and IT tools to (a) analyze, design, implement, control, and continuously improve end-to-end processes, and (b) to establish process governance.
\end{quote}

Business process management is based on a closed life-cycle with typically four main phases. First, a business process has to be defined, then it has to be implemented, then it is executed, and finally, it has to be evaluated according to effectivity and efficiency. The life cycle is closed such that after evaluation a process is eventually redesigned to reflect the results from the evaluation phase. The BPM-System (the corresponding management system) ensures that each business process follows this life-cycle so that there is an implicit improvement circle.

Each phase has its challenges. In this work, we will focus on the modeling and the execution phase. Business processes can be executed with or without the support of software systems. We will discuss the relationship between these phases and the consequences of particular design and technology decisions. This discussion will lead to insights about business process modeling notations and about the design and architecture of supporting software systems, often named as workflow management systems (WfMS).

The analysis will be done based on an ontological analysis in several ways. First, we will do an ontological analysis to understand the core concepts of the notion of business process in general. Then we will discuss the use of ontologies to define business process models; for this purpose, an ontology of the Business Process Model and Notation (BPMN), the industry standard modeling notation, has been developed. The ontology is modeled in the Web Ontology Language (OWL), and a software application has been developed to store any BPMN-XML as OWL model.

Then we will discuss the capabilities of BPMN to define process choreographies (named as collaboration in the BPMN standard document) which are executable by a software system. This analysis will be based on a comparison with the well-known actor system concept. A business process modeling language based on the actor system is the so-called Subject-oriented Business Process Modelling (S-BPM) notation, which has been defined via an OWL ontology.

The nature of any model is that it has a purpose; therefore it is irrelevant which modeling language is used for the model as long as it fits for the intendet purpose. Nevertheless, the design of a modeling notation has consequences for the architecture of the supporting information technology as will be discussed in the case of WfMS. Furthermore, we will discuss some technical requirements for a modern WfMS to support business process networks, that means loosely coupled business processes. Interacting companies (buyer and seller, for erxample) realize loosely coupled business processes via communication between each process participant (named as choreography or collaboration in the domain of BPM).

We can summarize the research questions in the following way formulating some hypotheses:

\begin{itemize}
\item H1: it is possible to develop a BPMN ontology.

\item H2: it is possible to convert a BPMN model to OWL and back again without loss of information.

\item H3: it is possible to convert a BPMN model to an S-BPM model---based on an ontological fit of the two notations---and vice versa.

\item H4: an actor based modeling notation for business processes (as S-BPM) is a natural candidate to design a modern architecture of workflow engines.

\item H5: WfMS have a typical architecture, independent of the modeling notation so that it is possible to execute a transformed BPMN model on an S-BPM WfMS.

\end{itemize}

\section{The Notion of System Model}
\label{thenotionofsystemmodel}

\begin{figure}[htbp]
\centering
\includegraphics[keepaspectratio,width=200pt,height=0.75\textheight]{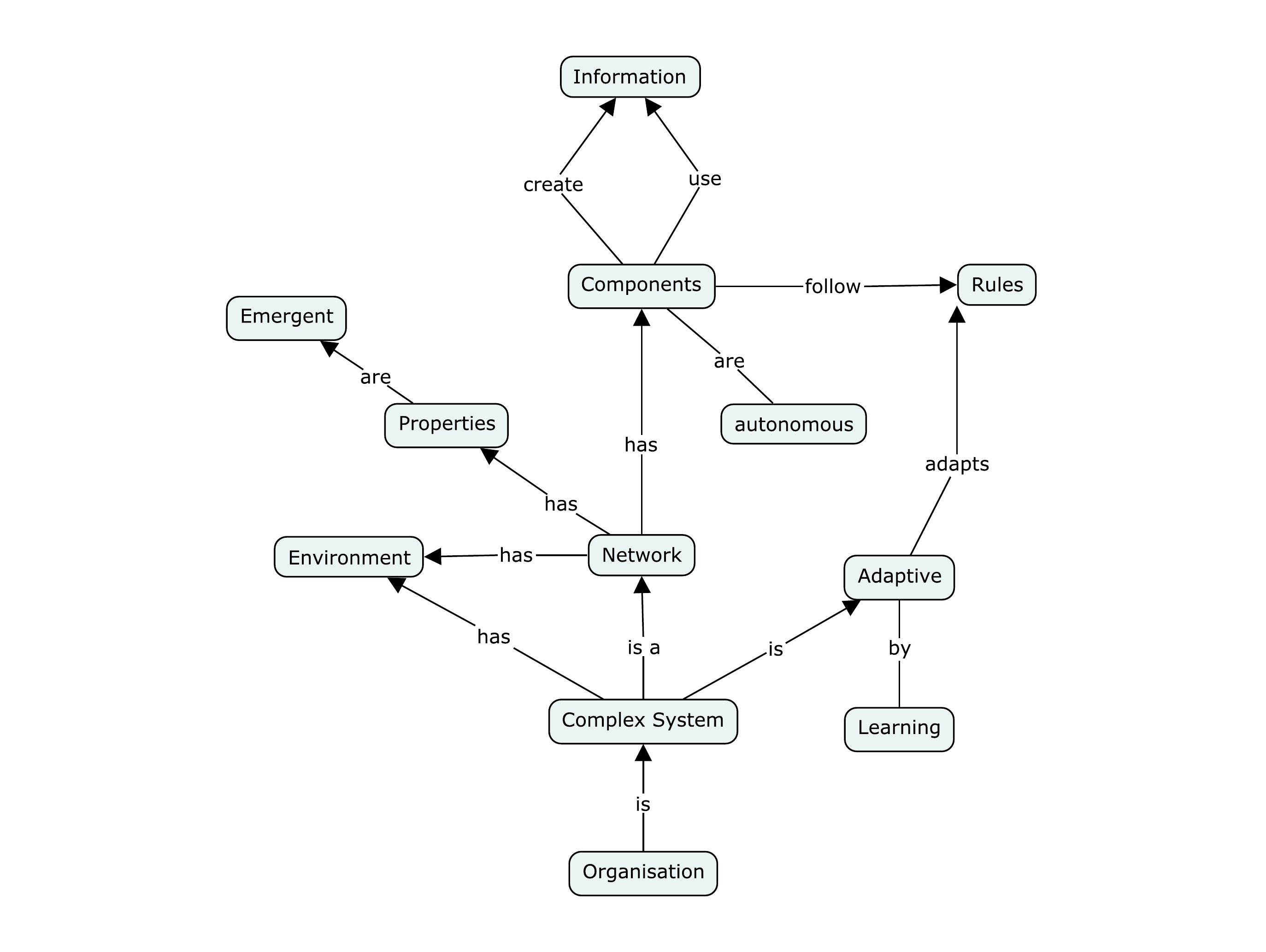}
\caption{Concept map: complex system.}
\label{complexsystem}
\end{figure}

For the following section, we provide some fundamental concepts about modeling in general and modeling in the domain of computer sciences or information technology (IT) in particular as a foundation for further discussions later in this work. Mainly we want to point out, that a model has a specific purpose (a selected view on reality) and involves some subjective aspects too. Furthermore, it is generally accepted that the organization of a company constitutes a complex system (see \autoref{complexsystem}). Modeling must be thought with this in mind.

\subsection{Meaning of a Model}
\label{meaningofamodel}

\begin{figure}[htbp]
\centering
\includegraphics[keepaspectratio,width=200pt,height=0.75\textheight]{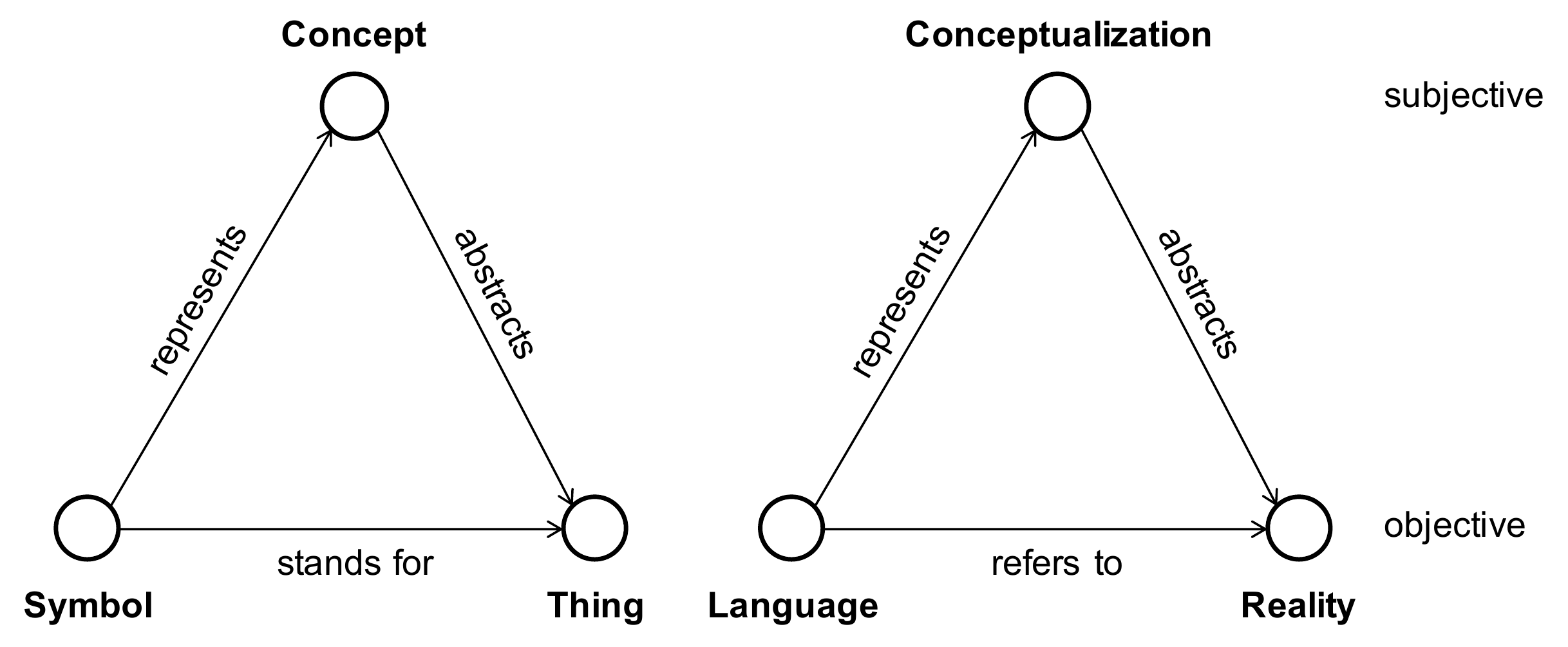}
\caption{Left: the meaning triangle for an individual; Right: the meaning triangle for a specific reality domain (adapted from~\citep{henderson-sellers2012a}~\citep{dietz2006})}
\label{meaningtriangle}
\end{figure}

If we want to model a world, we have to understand the so-called meaning triangle~\citep{ogden1923}~\citep{ullmann1962} from semiotics as shown in \autoref{meaningtriangle}.

A symbol or sign is an object that is used as a representation of something else, and it is used to communicate the concepts in our mind. A thing or object is an observable and identifiable individual thing; objects are concrete or abstract. A concept is a subjective individual thing (a thought in our mind). For a given domain, the set of all the individual concepts abstracted from that domain is called the conceptualization and to communicate the concepts we need a language.

Furthermore, we have to introduce the notion of a system as an organization can only be understood as a system. Therefore, any model is a model of a system. As discussed, models have a purpose and have a restricted view on reality; for example, an organizational chart and a business process model are static and dynamic views on the system of interest, respectively.

A precise formal definition of the construction of a system can be found, for example, in~\citep{bunge1977}, which can be described in the following way. Something is a system if it has the following properties: a composition (a set of elements of some category), an environment (a set of elements of the same category), whereby the composition and the environment are disjoint, and a structure (a set of influence bonds among the elements in the composition, and between them and the elements in the environment). Now, an organization or firm is a collection of socially linked human beings.

Dietz~\citep{dietz2006} furthermore adds the notion of production: the elements in the composition produce things (goods or services) that are delivered to the elements in the environment. That is important, as organizations are open systems.

\begin{figure}[htbp]
\centering
\includegraphics[keepaspectratio,width=250pt,height=0.75\textheight]{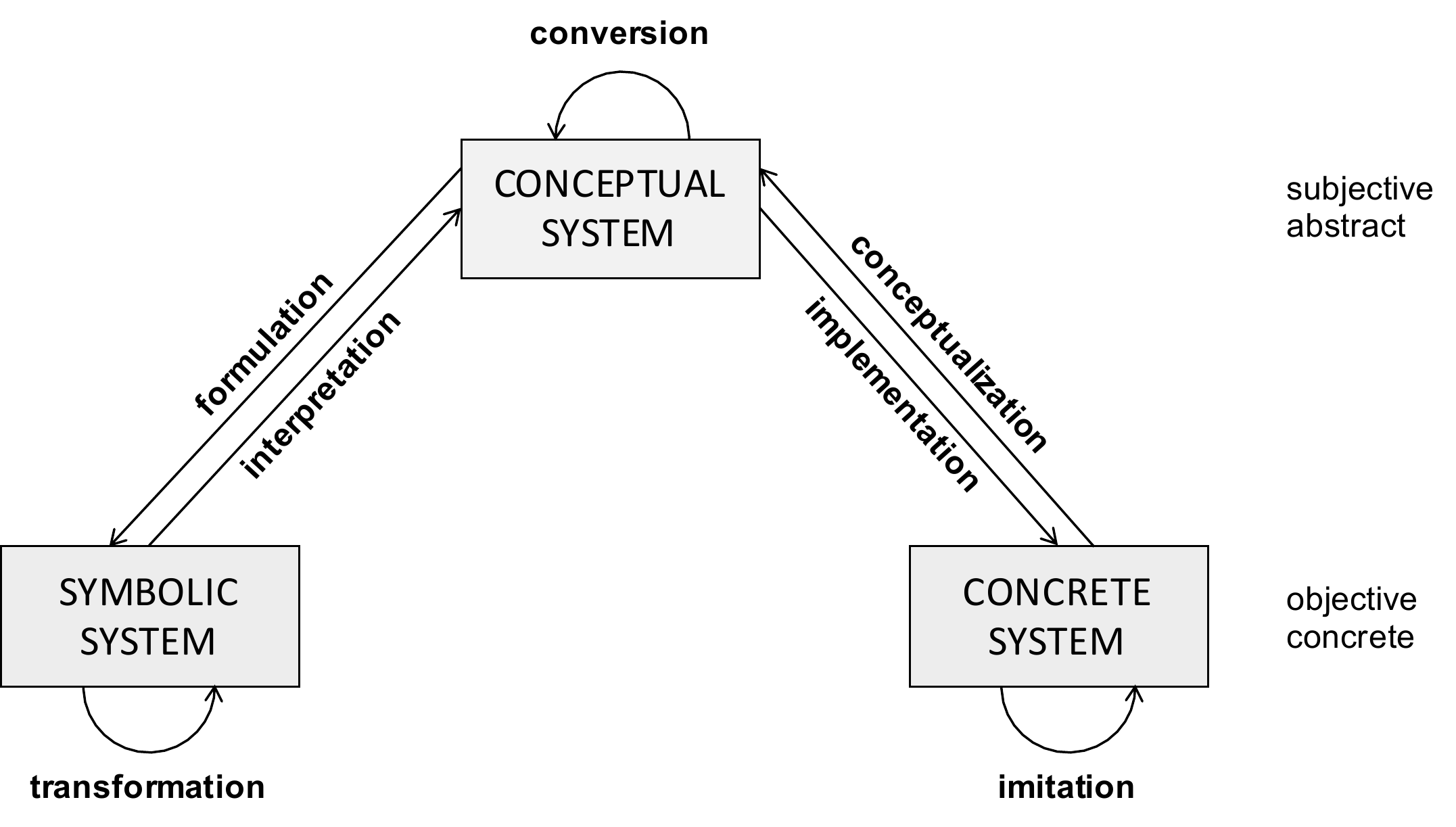}
\caption{The model triangle (adapted from ~\citep{dietz2006})}
\label{modeltriangle}
\end{figure}

Following the argumentation of Dietz~\citep{dietz2006}, three gross categories of systems can be distinguished: concrete systems, symbolic systems, and conceptual systems as depicted in \autoref{modeltriangle}.

The conceptual model of a concrete system is called a conceptualization; for example, a business process model is a conceptualization of the business processes of an organization or firm. A specific model of a conceptual system is called an implementation; for example, an enacted business process as an implementation of a business process model. A conceptual model of a conceptual system is called a conversion. A symbolic model of a conceptual system is called a formulation; a symbolic system is expressed in some formal language---the notation to represent the model. A conceptual model of a symbolic system is called an interpretation. A symbolic model of a symbolic system is called a transformation.

What is now the conclusion? Firstly, it is essential that the imitation of a concrete system never is the same. Secondly, concepts are only in our minds and therefore subjective. Thirdly, as a consequence, all stages include social interaction between human beings to construct a socially accepted view of the concrete system. That means the communication of a cognitive model needs a language: the cognitive model abstracts information from the system under study, and this is then represented in the communicated model by using symbols from some chosen (modeling) language.

\begin{figure}[htbp]
\centering
\includegraphics[keepaspectratio,width=250pt,height=0.75\textheight]{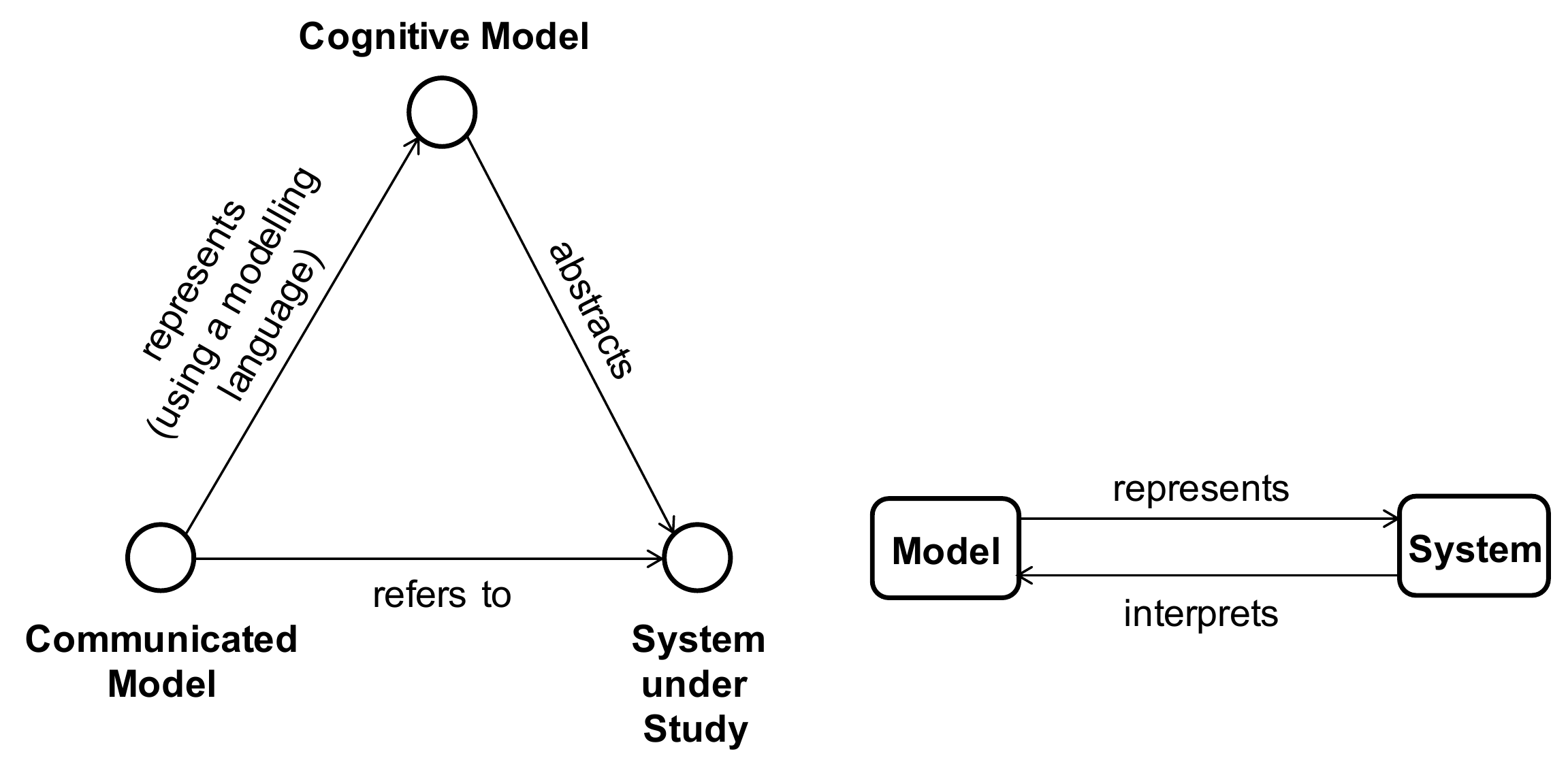}
\caption{In technical domains the meaning triangle (left) is often simplified to a binary relation (right)~\citep{henderson-sellers2012a}.}
\label{binarymeaning}
\end{figure}

The relation between subjective and objective is not reflected in much of the software engineering literature wherein the implicit ``refers-to'' link is generally named ``represents,'' and the ``represents'' and ``abstracts'' functions are typically confounded into a single abstraction notion between two representations as depicted in \autoref{binarymeaning}. That means, in software engineering (and other technical domains) these three relationships of the semiotic triangle typically are simplified as a binary relation between model and system.

\subsection{Requirements for Notations}
\label{requirementsfornotations}

Diagrams can convey information more precisely than conventional language~\citep{bertin1983}~\citep{larkin1987}. As discussed in~\citep{moody2009} the human mind has separate systems for processing pictorial and verbal material---according to \emph{dual channel theory}. Visual representations are processed in parallel by the visual system; textual representations are processed serially by the language system~\citep{bertin1983}. Only diagrammatic presentations are able to show (complex) relations at once.

The \emph{anatomy of a visual notation} is worked out very clearly by~\citep{moody2009}: a visual notation (or visual language, graphical notation, diagramming notation) consists of a set of graphical symbols (\emph{visual vocabulary}), a set of compositional rules (\emph{visual grammar}) and definitions of the meaning of each symbol (\emph{visual semantics}). The visual vocabulary and visual grammar together form the \emph{visual} (or \emph{concrete}) \emph{syntax}. Graphical symbols are used to \emph{symbolize} (perceptually represent) \emph{semantic constructs}, typically defined by a \emph{metamodel}. The meaning of graphical symbols is determined by mapping them to the constructs they represent. A valid expression in a visual notation is called a \emph{visual sentence} or \emph{diagram}. Diagrams are composed of \emph{symbol instances} (\emph{tokens}), arranged according to the rules of the visual grammar.

But, just presenting information in a graphical form does not guarantee that it will be worth \emph{a thousand of words}~\citep{cheng2001}. Most effort is spent on designing semantics, with visual syntax often an afterthought~\citep{moody2009}. For example, UML does not provide a design rationale for any of its graphical conventions~\citep{moody2009}. The same can be said for any process modeling language.

\begin{figure}[htbp]
\centering
\includegraphics[keepaspectratio,width=250pt,height=0.75\textheight]{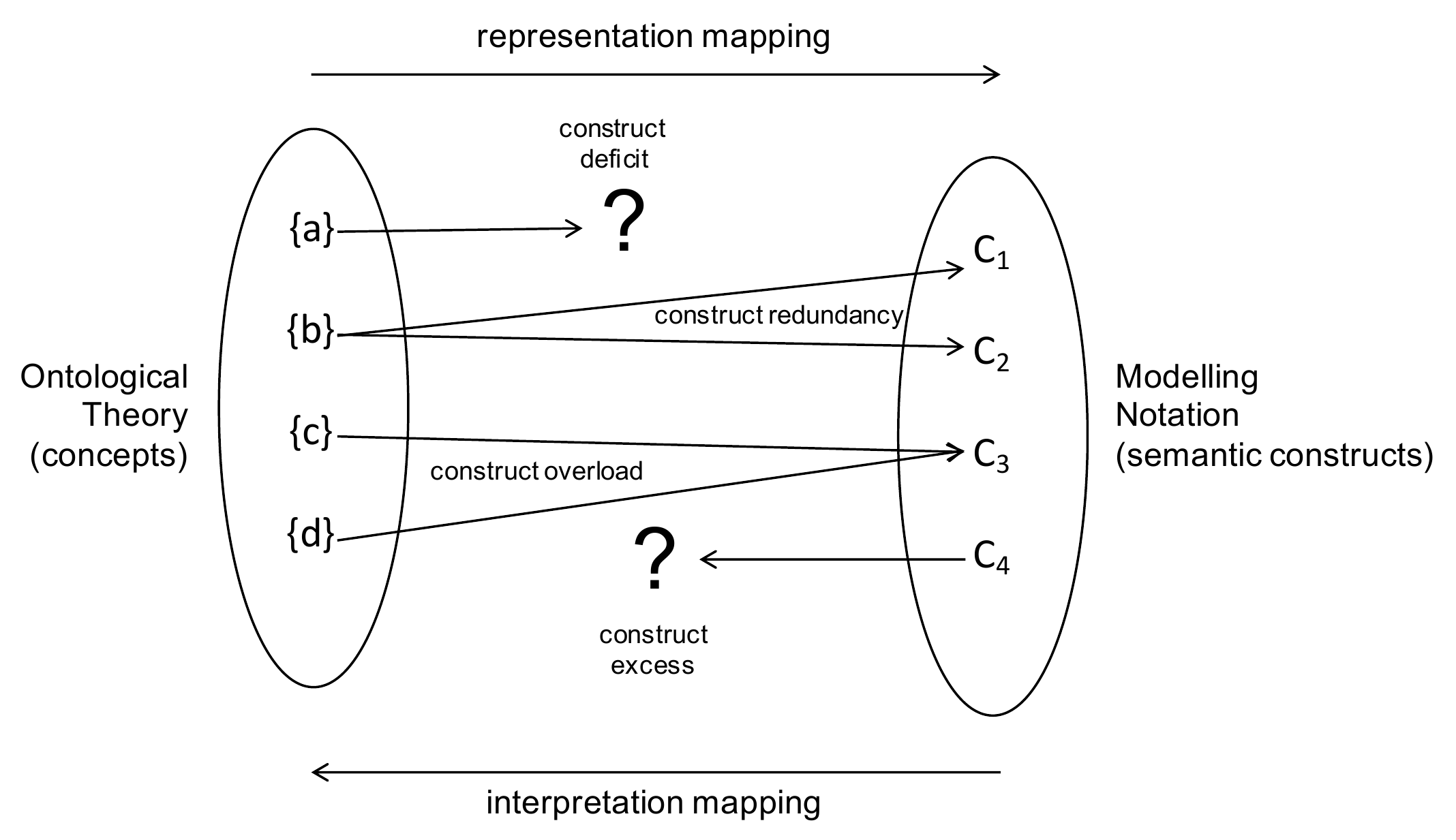}
\caption{Ontological analysis. There should be a 1:1 mapping between ontological concepts and notational constructs (adapted from ~\citep{moody2009}).}
\label{ontanalysis}
\end{figure}

A widely accepted way to evaluate notations is \emph{ontological analysis}; the most used ontology seems to be the \emph{Bunge-Wand-Weber} (BWW) ontology~\citep{wand1990}. The ontological analysis involves a two-way mapping between a modeling notation and an ontology. The \emph{interpretation mapping} describes the mapping from the notation to the ontology; the \emph{representation mapping} describes the inverse mapping~\citep{gehlert2007} as depicted in \autoref{ontanalysis}.

If construct deficits exist, the notation is ontologically incomplete; if any of the other three anomalies exist, it is ontologically unclear. The BWW ontology predicts that ontologically clear and complete notations will be more effective. As elaborated in~\citep{moody2009}, the ontological analysis focuses on content rather than form; if two notations have the same semantics but different syntax, ontological analysis cannot distinguish between them. Moody~\citep{moody2009} has developed a promising foundation to analyze the \emph{syntactic} aspects of notations in a similar stringent way based on scientific foundations:

\begin{quote}
{\ldots}, a diagram creator (sender) encodes information (message) in the form of a diagram (signal), and the diagram user (receiver) decodes this signal. The diagram is encoded using a visual notation (code), which defines a set of conventions that both sender and receiver understand. The medium (channel) is the physical form in which the diagram is presented (e.g., paper, whiteboard, and computer screen). Noise represents random variation in the signal which can interfere with communication. The effectiveness of communication is measured by the match between the intended message and the received message (information transmitted).
\end{quote}

\begin{figure}[htbp]
\centering
\includegraphics[keepaspectratio,width=250pt,height=0.75\textheight]{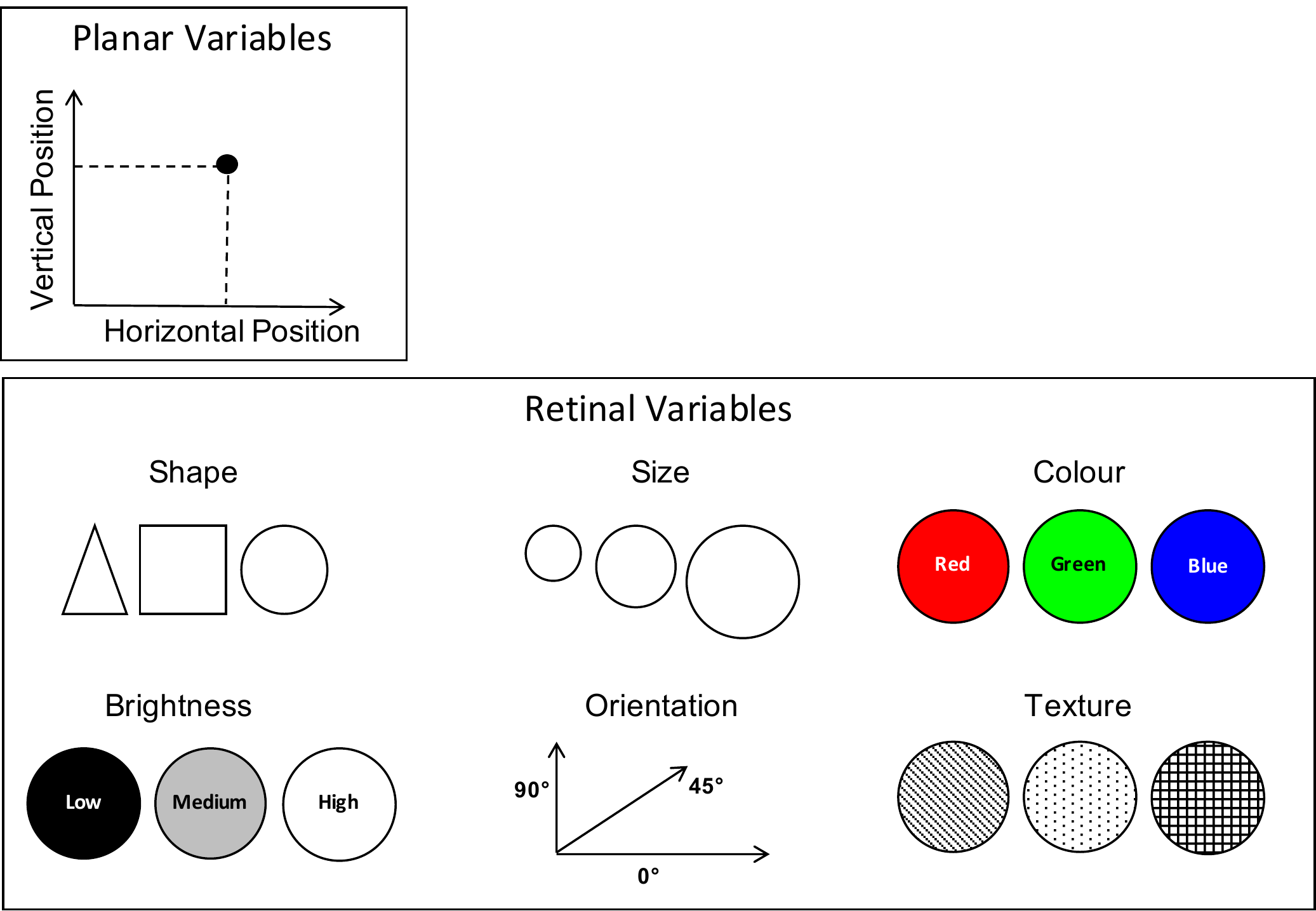}
\caption{Visual variables (~\citep{bertin1983}, adapted from~\citep{moody2009}.}
\label{bertin}
\end{figure}

Bertin~\citep{bertin1983} identified eight visual variables that can be used to graphically encode information as depicted in \autoref{bertin}. The decoding side is based on the human decoding processes, which can be divided into two phases: perceptual processing (seeing) and cognitive processing (understanding). As the perceptional processing system is much faster, it is more effective to move as much of the decoding work from the cognitive to it.

Now, based on these theories and empirical evidence Moody has developed a prescriptive theory for visual notations~\citep{moody2009}, which is formulated as nine principles for designing cognitively effective visual notations, summarized as follows:

\begin{itemize}
\item Semiotic clarity: there should be a 1:1 correspondence between semantic constructs and graphical symbols

\item Perceptual discriminability: different symbols should be clearly distinguishable from each other

\item Semantic transparency: use visual representations whose appearance suggests their meaning

\item Complexity management: include explicit mechanisms for dealing with complexity

\item Cognitive integration: include explicit mechanisms to support the integration of information from different diagrams

\item Visual expressiveness: use the full range and capacities of visual variables

\item Dual coding: use text to complement graphics

\item Graphic economy: the number of different graphical symbols should be cognitively manageable

\item Cognitive fit: use different visual dialects for different tasks and audience

\end{itemize}

As long as humans create and interpret models, we must be aware of the meaning triangle as all models are conceptualized in mind and exist there as cognitive models. Furthermore, any communication between humans is based on natural language and the need to establish a collaborative understanding through social interaction. One result of this fact is that missing or unclear information can be corrected by the human interpreter of a model; on the other side, it also can lead to misinterpretation of models. If machines interpret models (in our case typically workflow engines), they need to be semantically clear without any room for interpretation. Therefore, modeling languages intended to be interpreted by machines need an ontologically exact syntax and semantic.

The method of ontological analysis and the set of principles for designing cognitive effective visual notations, together with the understanding of the notion of model and semiotics assembles a full set of building blocks for a coherent and solid foundation for business process modeling notations. Finally, combining it with a corresponding formal model for business process execution leads to a complete and coherent theory of business processes as a solid foundation for business process management.

\section{Business Process (Systems) Models}
\label{businessprocesssystemsmodels}

For any analysis of languages, for the purpose to create and communicate business process models, we need a profound understanding of the underlying concepts. For this purpose, we shortly discuss several different proposed languages or notations. Furthermore, we will present a general set of capabilities for business process modeling languages amalgamed from this selected set of semi-formal or formal languages or notations. Even more, informal notations include in principle the same concepts but cannot create models which can be unambiguously interpreted by software.

\subsection{Foundational Concepts}
\label{foundationalconcepts}

\subsubsection{Discrete Event Systems (DES)}
\label{discreteeventsystemsdes}

\begin{figure}[htbp]
\centering
\includegraphics[keepaspectratio,width=\textwidth,height=250pt]{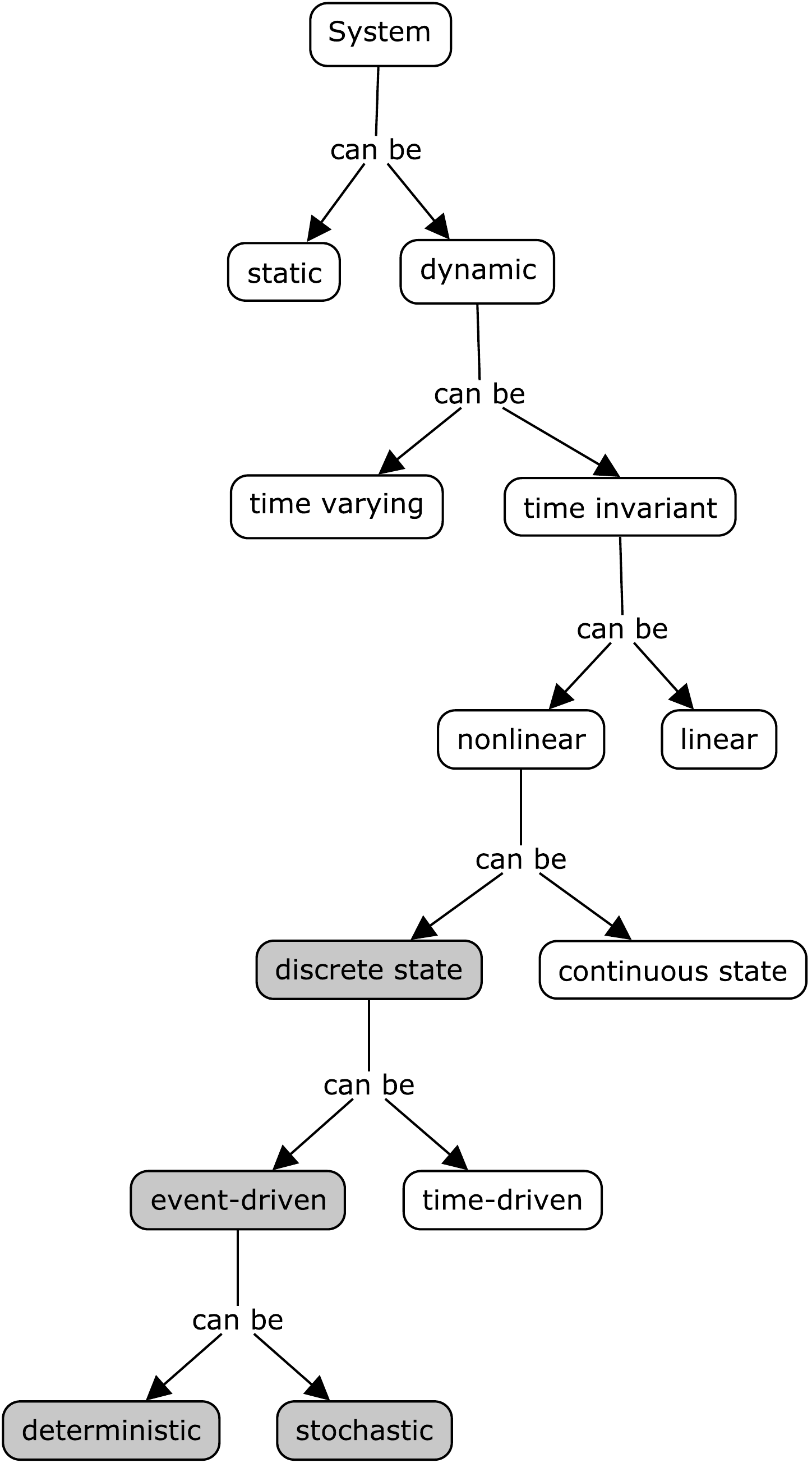}
\caption{Concept map: system classifications.}
\label{systemclassifications}
\end{figure}

In this section, we will discuss a specific class of systems, which are an integral part of our world and especially any firm. As we are interested in business processes and its support by information technology we focus on discrete state systems which are event-driven and deterministic~\citep{cassandras2008}. A typical classification of systems is depicted in \autoref{systemclassifications}.

A model is defined by a set of measurable variables associated with a ``system under study'' (SUS). By measuring these variables over a period of time $[t_0, t_f]$ we may then collect data. If we vary such a set of variables over time, we can define a set of time function as input variables $\{u_{1}(t),\ldots,u_{p}(t)\}$, where $t_{0}\leq t\leq t_{f}$. Then, we can select another set of variables which we can directly measure while varying the input variables $u_{1}(t),\ldots,u_{p}(t)$, and thus define a set of output variables $\{y_{1}(t),\ldots,y_{m}(t)\}$. These sets of input and output variables can be rewritten as vectors $\vec{u}(t)=[u_{1}(t),\ldots,u_{p}(t)]^{T}$, and $\vec{y}(t)=[y_{1}(t),\ldots,y_{m}(t)]^{T}$.

It is now reasonable to postulate that there exists some mathematical relationship between input and output. Thus, we can define functions.

\begin{align*}
y_1(t)&=g_{1}\left(u_{1}(t),\ldots,u_{p}(t)\right)\\
\vdots\\
y_{m}(t)&=g_{m}\left(u_{1}(t),\ldots,u_{p}(t)\right)
\end{align*}

and get the system model in the mathematical form

\begin{align}
\vec{y}&=\vec{g}(\vec{u})\\
&=\left[g_{1}\left(u_{1}(t),\ldots,u_{p}(t)\right),\ldots,g_{m}\left(u_{1}(t),\ldots,u_{p}(t)\right)\right]^{T}
\end{align}

where $g(.)$ denotes the column vector whose entries are the functions $g_{1}(.),\ldots,\allowbreak g_{m}(.)$. Now, if we know the set of functions $\vec{g}$ we have a so-called white-box model if we do not know how the value of the output variables depends on the values of the input variables we have a so-called black-box model. Additionally, we assume that $g(.)$ does not depend on time, that means the same input always gives the same output (time-invariant system). We also want a model which can predict the behavior of a system.

In the domain of business process management, the notion of business process orchestration (a synonym for business process, represented as process pool in BPMN) and the notion of business process choreography (collaborative business process orchestrations, whereby collaboration is realized through the exchange of messages) are familiar concepts~\citep{weske2012}. Process orchestrations need to be white-box models as they always define a logical order of activities; Process choreographies can be a mix of white- and black-box models, as the interaction between particular business processes has to be modeled and whatever the intention of the model is.

Now, the state of a system at time $t_0$ is the information required at $t_0$ such that the output $y(t)$, for all $t\geq t_0$, is uniquely determined from this information and from $u(t), t\geq t_0$. As the input $u(t)$ and the output $y(t)$, the state is also generally a vector, which shall be denoted by $x(t)$. The components of this vector, $x_1(t), ..., x_n(t)$, are called state variables. The modeling process then consists of determining suitable mathematical relationships involving the input $u(t)$, the output $y(t)$, and the state $x(t)$~\citep{cassandras2008}. These relationships define the dynamics of a system. The set of equations required to specify the state $x(t)$ for all $t\geq t_0$ given $x(t_0)$ and the function $u(t), t\geq t_0$, are called state equations. The state space of a system, denoted by $X$, is the set of all possible values that the state may take.

If the state space $X$ is a continuum consisting of vectors of real numbers it usually leads to differential equations and associated techniques for analysis. In discrete-state models, the state space is a discrete set. In this case, a typical sample path is a piecewise constant function.

The state transition mechanism is generally based on simple logical statements of the form ``if something specific happens and the current state is $x$, then the next state becomes $x'$.''

This mechanism is precisely the case for business processes where we always move from one state to the next; for example, an invoice changes its state from unpaid to paid or purchased material changes the state of the available material from $a$ to $a+n, n\in\mathbb{N}$.

Furthermore, business process modeling means the \emph{design} of white-box models, because we want to define how the input into a business process produces a specific output of that business process. Therefore we have to design the state space (states and the state transitions) of a business process to define the behavior of that process. If a business process is a so-called system workflow, that means without any human interaction, it is evident, that all states have to be modeled. However, in many human interaction processes it will not be possible or even needed to define all possible states; this is a result from complexity theory, as we know, that organizations are complex socio-technical systems. Nevertheless, as long as a workflow system is involved in the execution of a business process, we have to define those states which are supported by the system.

\subsubsection{Labeled Transition System (LTS)}
\label{labeledtransitionsystemlts}

The input-output model discussed in the previous chapter is a general modeling concept. In the case of a business process we have states and state transitions to describe the transformation from input to output; in the case of business processes input and output can be immaterial (data, such as an order) and material entities (delivered products). Such entities are often entitled business objects (BO) in the context of business process management.

Although transition systems are not suitable for modeling industrial information systems and business processes in practice, they illustrate the essence of modeling~\citep{vanderaalst2011}. A labeled transition system (LTS), or an automaton, consists of a set of states, a set of labels (or actions) and a transition relation $\rightarrow$ describing changes in process states; if a process $P$ can perform an action $\alpha$ and becomes a process $P'$, we write $P\xrightarrow{\alpha}P'$

We can define a labelled transition system (LTS) as a triple $\text{Proc}, \text{Act}, {\xrightarrow{\alpha}\mid \alpha\in \text{Act}}$, where $\text{Proc}$ is a set of states (or processes), $\text{Act}$ is a set of actions (or labels), and $\xrightarrow{\alpha}\subseteq\text{Proc}\times\text{Proc}$ is a transition relation for every $\alpha\in \text{Act}$.

An LTS is finite if its sets of states and actions are both finite. For every action, a transition relation $P\xrightarrow{\alpha}P'$ is defined. This means $P$ evolves to $P'$ by performing the action $\alpha$.

\subsubsection{Petri- and Workflow-Net}
\label{petri-andworkflow-net}

In computer science, Petri-nets are a well-known method to define business process models in a formal and abstract way and are an essential conceptual foundation for all business process languages. To use Petri-nets to define business process workflows has been proposed by van der Aalst and van Hee ~\citep{vanderaalst2004}~\citep{vanderaalst2011} and has led to intensive research about workflow patterns~\citep{russell2016a} and business process automation~\citep{yawl2010}. A comprehensive overview of the use of Petri-nets for business process management can also be found in in~\citep{weske2012}. They have three specific advantages:

\begin{itemize}
\item formal semantics despite the graphical nature,

\item state-based instead of event-based, and

\item an abundance of analysis techniques.

\end{itemize}

For practical use so-called colored Petri-nets~\citep{jensen2009} have to be used to define typed attributes for token and to define guards depending on values of the attributes. Nevertheless, colored Petri-nets have similar deficits as other languages as some information is not visible in the visual representation anymore; that means, business process models cannot be studied based on a print-out of the model alone. A tool for editing, simulating, and analyzing colored Petri-nets can be found on the Web\footnote{http:\slash \slash cpntools.org}.

Definition~\citep{weske2012}: a coloured Petri-net is a tuple $(\Sigma,P, T,A,N,C,G,E,I)$ such that

\begin{itemize}
\item $\Sigma$ is a nonempty finite set of types, called color sets

\item $P$ is a finite set of places

\item $T$ is a finite set of transitions

\item $A$ is a finite set of arc identifiers, such that $P\cap T=P\cap A=T\cap A=\emptyset$

\item $N:A\rightarrow(P\times T)\cup(T\times P)$ is a node function that maps each arc identifier to a pair (start node, end node) of the arc

\item $C:P\rightarrow\Sigma$ is a color function that associates each place with a color set

\item $G:T\rightarrow BooleanExpr$ is a guard function that maps each transition to a predicate

\item $E:A\rightarrow Expr$ is an arc expression that evaluates to a multi-set over the color set of the place

\item $I$ is an initial marking of the colored Petri-net.

\end{itemize}

Workflow nets are an approach to enhance traditional Petri nets with concepts and notations that ease the representation of business processes. At the same time, workflow nets introduce structural restrictions that prove useful for business processes~\citep{yawl2010}~\citep{weske2012}. Like Petri nets, workflow nets focus on the control flow behavior of a process. Places represent conditions and tokens represent process instances. Activities of a business process are represented by transitions in the workflow net. Because tokens represent business process instances, tokens hold application data including process instance identifiers, that is, the tokens are colored.

Definition~\citep{weske2012}: a Petri-net $PN = (P, T, F)$ is called workflow net if and only if the following conditions hold.

\begin{itemize}
\item There is a distinguished place $i\in P$ (called initial place) that has no incoming edge, that is, $\bullet i=\emptyset$.

\item There is a distinguished place $o\in P$ (called final place) that has no outgoing edge, that is, $o\bullet=\emptyset$.

\item Every place and every transition is located on a path from the initial place to the final place.

\end{itemize}

The inability of standard Petri nets and workflow nets to directly capture the notion of cancelation within a business process spurred the use of reset nets for this purpose. Reset nets can explicitly depict notions of cancelation within a process definition~\citep{yawl2010}.

Furthermore, for the ease of use some special symbols have been defined (syntactic sugaring), for example, to represent Xor- and And-join and -split patterns. Workflows can be modeled and executed with Yet Another Workflow Language (YAWL)~\citep{yawl2010}, for example; YAWL has been developed to have a language which directly supports the control workflow patterns. Nevertheless, while it uses workflow nets as a primary ingredient, the execution semantics of process instances is specified by state transition systems and not by Petri nets. In total, YAWL directly supports 31 of the 43 control-flow patterns and provides partial solutions for a further three patterns~\citep{yawl2010}.

\subsubsection{Abstract State Machine (ASM)}
\label{abstractstatemachineasm}

Abstract State Machines (ASM) were first postulated by Yuri Gurevich in 1985 and later axiomatized in~\citep{gurevich2000}. Gurevich's motivation for the new computational model was to improve Turing's thesis to understand better what algorithms are. The ASM method is a framework for designing hardware and software systems in a structured way. It does so by building up a ground model, which can be seen as the ``blueprint'' of a system and further refines the ground model until it is detailed enough to be put into code. It is a method for transforming the human understanding of a system from an application perspective into compilable code.

Technically ASMs can be defined as a natural generalization of Finite State Machines (FSM) by extending FSM-states to Tarski structures. Tarski structures also called first-order or simply mathematical structures, represent truly abstract data types. Therefore, extending the particular domains of FSM-computations to these structures turns Finite State Machines into Abstract State Machines~\citep{borger2003}.

Over time ASM experienced a shift in its notion from ``simultaneous parallel actions of a single agent,'' to a more general definition of ``multiple agents act and interact in an asynchronous manner'' {[ibid]}. ASMs can be understood as a virtual machine executing pseudo-code operating on abstract data structures. The ASM-method is suited for procedural single-agent and asynchronous\slash synchronous multiple-agent distributed systems. The intention is to bridge the gap between a human understanding and formulation of real-world problems and the deployment of their algorithmic solutions (implementation as a software and\slash or hardware machine). That is precisely what is done when a business process model is uploaded for \emph{interpretation} by a workflow engine.

Definition of Basic ASMs---basic ASMs are single-agent machines with a finite set of so-called transition rules of the form

\textbf{\texttt{if}} \emph{Condition} \textbf{\texttt{then}} \emph{Updates}

which transforms abstract states. The \emph{Condition} (or \emph{guard}) is an arbitrary predicate logic without free variables which evaluates to \emph{true} or \emph{false}. \emph{Updates} is a finite set of assignments of the form

$$f(t_1,\ldots,t_n):=t$$

which is to be understood as changing (or defining) the value of the functions $f$ according to the provided arguments and values. The notion of ASM states is the classical notion of mathematical structures where data come as abstract objects, i.e., as elements of sets which are equipped with basic operations (partial functions in the mathematical sense) and predicates (attributes or relations). A state can be understood as a ``database of functions'' instead of predicates~\citep{borger2003}.

Additionally, to the basic rules, there are further constructs to implement common conditional logic. The \texttt{forall} construct lets you express the simultaneous execution of rules. Where you execute each rule $R$ for every element of a certain set or type which suffices the guard $\phi$:

\textbf{\texttt{forall}} $x$ \textbf{\texttt{with}} $\phi$

 $R$

Non-determinism can be modeled by the \texttt{choose} construct. Where the user of the machine can choose which rule shall be used.

\textbf{\texttt{choose}} $x$ \textbf{\texttt{with}} $\phi$

 $R$

In summary, to define an ASM $M$ one has to indicate its signature, the set of declarations of functions and rules, the set of its initial states, and the main rule identified by the machine $M$. A comprehensive description of ASMs can be found in the book of Börger and Stärk~\citep{borger2003}.

Kossak et al. have used ASMs to define the signature of the ground model of BPMN~\citep{kossak2015}. This analysis allows a comprehensive analysis of the semantics of BPMN and---as others~\citep{borger2012}---various inconsistencies, as well as ambiguities in the BPMN standard, could be identified. A summary of the complete ASM model can be found on the Web\footnote{http:\slash \slash h--bpm.scch.at}. As ASMs define automata, they describe a semantic for executable business process models.

A subject-oriented---or resource-based---modeling language for business processes is S-BPM~\citep{fleischmann2012}~\citep{borger2012a}, which is discussed later. As the modeling language is based on process-calculus, it can be easily defined via an ASM ground model~\citep{fleischmann2012}, and models can be executed on a corresponding ASM interpreter~\citep{fleischmann2017}. Furthermore, the S-BPM syntax is defined via an ontology and the semantics via ASM~\citep{borgert2019}.

\subsection{Process-Calculi}
\label{process-calculi}

Process algebras are prototype specification languages for reactive systems. A short history of the ideas can be found in~\citep{baeten2005}. A discussion of the role that algebra plays in process theory is discussed in~\citep{luttik2006}.

A crucial observation at the heart of the notion of process algebra is that concurrent processes (that means, business process collaborations or choreographies) have an algebraic structure. For example, we can create a new process combining two separate processes $P$ and $Q$ sequentially or in parallel. The results of these combinations will be a new process whose behavior depends on that of $P$ and $Q$ and the operation we have used to compose them~\citep{aceto2007}. The description languages are algebraic, as they consist of a collection of operations for building new process descriptions from existing ones.

As these languages specify parallel processes that may interact with one other, they need to address how to describe communication or interaction between processes running at the same time. The crucial insight was that we need not distinguish between active components, such as sender and receiver, and passive ones such as the communication media---they may all be viewed as processes (i.e., systems that exhibit behavior). All these processes can interact via message-passing models as \emph{synchronized communication}~\citep{aceto2007}. Furthermore, data exchange is only possible via the exchange of messages that means processes do not share data with other processes (i.e., there is no shared memory space).

Some years ago there has been a discussion about a possible revolution in business process management, starting with a paper from Smith \& Fingar titled ”Workflow is just a Pi process”~\citep{smith2003}~\citep{smith2004}. The article was inspired by the $\pi$-Calculus developed around 1990 by R. Milner, J. Parrow and D. Walker~\citep{milner1989}~\citep{milner1992}. This formalism belongs to the family of process algebras (or process calculi) and can be seen as an enhancement of the Calculus for Communicating Systems (CCS)~\citep{milner1980}. While CCS could be used to describe concurrent communicating processes, the $\pi$-Calculus allows the formal description of so-called mobile processes. It is designed to model systems with dynamically changing structures in which the links between different components vary during the evolution. Another process algebras developed around the same time is Communicating Sequential Processes (CSP), published by Hoare~\citep{hoare1978}~\citep{hoare1985}.

In the following section, we will discuss the main concepts of these process calculi very shortly, as we think that---even if we do not want to create models in these formal notations---these concepts are very relevant for actual challenges when we want to execute parallel and distributed business processes.

\subsubsection{Calculus of Communicating Systems (CCS)}
\label{calculusofcommunicatingsystemsccs}

The primary objective of CCS is to provide a mathematical framework to describe communicating systems in a formal way. This formalization allows for verification of properties like checking for two processes being equivalent. Observation and synchronized communication are the two central ideas of the CCS. Observation aims to describe a concurrent system accurate enough to determine the behavior seen by an external observer. If two systems are indistinguishable from the observer’s point of view, they show the property of observation equivalence. Milner further states that every interesting concurrent system is built from independent agents which communicate in a synchronized way. So the objects whose behaviors are modeled are called agents. An agent can be seen as a term for a locus of activity, a process, or a computational unit. The action defines the agent's behavior it can perform and represented using algebraic expressions.

The basic capabilities of an agent are sending a message, receiving a message and performing an unobservable action $Act=N\cup\overline{N}\cup\tau$. $N$ represents a set of names, $\ overline{N}$ the set of corresponding co-names, and $\tau$ stands for an unobservable or so-called silent action. The ability to receive a message is denoted by using lowercase letters like $a,b,c,\ldots$ (names) whereas overlined lowercase letters like $\overline{a},\overline{b},\overline{c},\ldots$ (co-names) are used to denote the ability to send a message. A simple example of an employee asking his boss for a few days off could be described as the process $E=\overline{request}.date.\overline{answer}.(accept.E+deny.E)$.

This process can be read in the following way: an employee ${E}$ sends a $\overline{request}$ for leave, get questioned about the $date$, gives $\overline{answer}$ and finally he or she receives an $accept$ or $deny$. Afterward, the process is reset by invoking ${E}$ again (recursion) to make the process runnable again. The semantics of the CCS is defined by a labeled transition system (LTS), as previously discussed.

\subsubsection{Communicating Sequential Processes}
\label{communicatingsequentialprocesses}

Communicating Sequential Processes (CSP) provides a formal framework to describe the communication interaction between processes. As in CCS the number of processes is static and cannot be changed during runtime. As with other process calculi, the processes coordinate their behavior exchanging messages.

A process sends a message executing the command $Q!(expr)$ and the corresponding receiving process executes $P?(vars)$ to get the message and to map the content on to variables. Similar to CCS and $\pi$-Calculus message exchange is based on unbuffered message exchange so that the sender- and receiver process have to be explicitly stated by name in the corresponding commands. If $expr$ and $vars$ are empty a message without data is sent which typically is called a signal. The receiving command can be enhanced by logical expressions to get so-called guarded commands which are only executed if the logical expression evaluates to true. It is worth to mention, that---for example---the functional programming language Erlang has built-in capabilities which are very similar to these concepts; any program is called a process and processes coordinate their behavior through the exchange of messages (even the syntax is identical).

\subsubsection{$\pi$-Calculus}
\label{pi-calculus}

The $\pi$-Calculus can be understood as CCS enhanced with the so-called link passing mobility~\citep{milner2004}. The essential elements of the $\pi$-Calculus are, like in the CCS, names, and agents. Agents perform actions (sending, receiving, execution of an unobservable action, performing a match between two names), and names are a collective term for concepts like links, pointers, references, identifiers, and channels. As a consequence of the concept names can act as both transmission medium and as transmitted data~\citep{smith2004}. Agents (which can be seen as processes) interact with other agents by sending and receiving messages identified by a name. As contents of messages are also channels, at the end of a communication the recipient is capable of using the received channel for further communication (link passing mobility).

\begin{figure}[htbp]
\centering
\includegraphics[keepaspectratio,width=150pt,height=0.75\textheight]{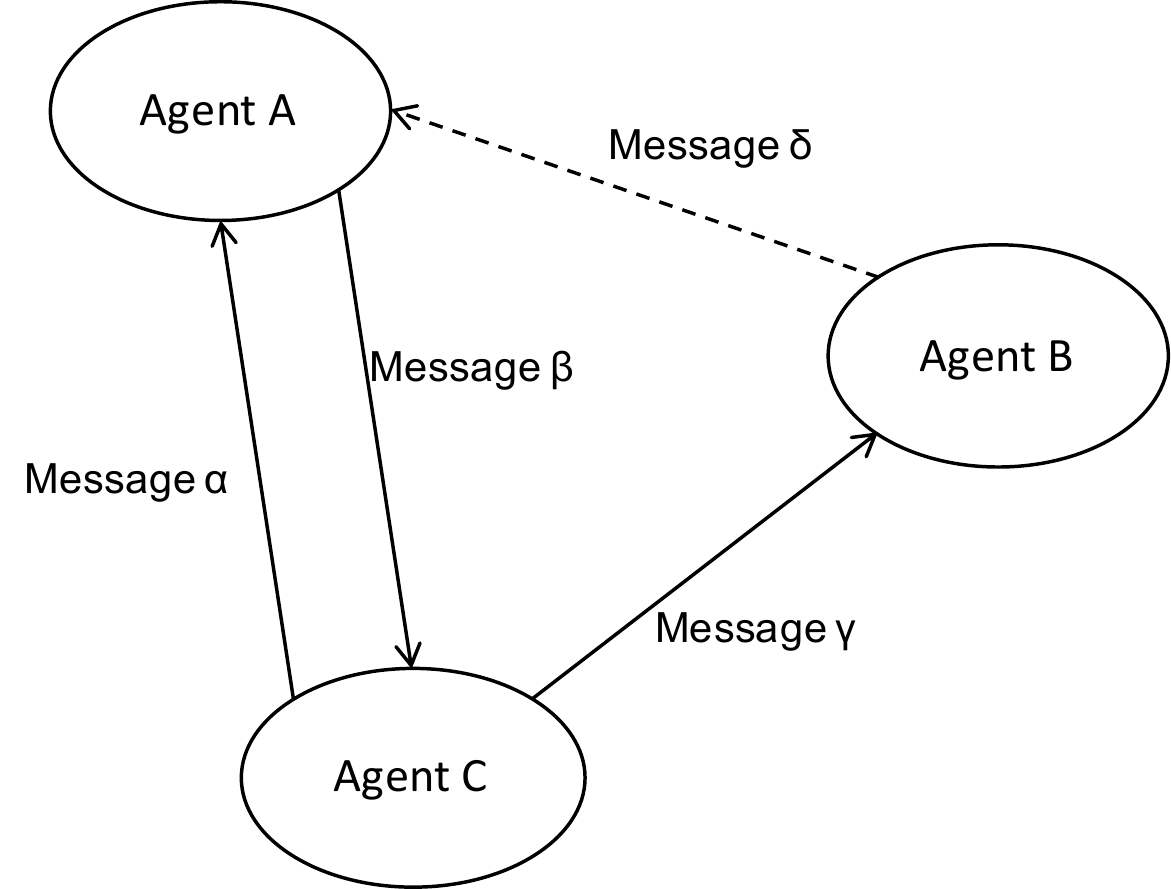}
\caption{The concept of link passing.}
\label{picalculus}
\end{figure}

For example: A process $A$ (a synonym for agent $A$) uses the link $\beta$ (a synonym for message $\beta$) to send the value $x$ (the business object as content of the message) to process $C$ (as can be seen in Figure \autoref{picalculus}). Now assume that instead of sending the answer as message $\alpha$ directly, $C$ delegates this action to $B$. To achieve this $C$ sends the name of the link $\alpha$ and the value $x$ to $B$ via $\gamma$. Now, $B$ knows how to communicate with $A$ via the link $\delta$.

The principles of $\pi$-calculus can also be discussed with the help of the well-known concept of e-mail~\citep{singer2012}. Consider electronic mail as a process. We can send an e-mail to another person, this one, for example, forwards the e-mail to a third party, and this one is then able to communicate and collaborate with me as the initiator of the email. How does this happen? By receiving e-mail, or more specifically by receiving an e-mail address, directly or indirectly, interchange the capability to communicate with others linked to that e-mail address. This is what makes e-mail work. We give a name, in the form of an e-mail address, to others, and this gives them the ability to communicate with yet other participants in the thread of the conversation---continuously extending the conversation over time, involving new participants that contribute value to the process. Through this simple model, a dynamic way of conversation becomes possible---a new business process. Another advantage in informal business processes is the possibility to send any type of business object (data) without the need to define a rigorous data model in advance (agent 1 transmits a spreadsheet to agent 2, agent 2 adds a column and forwards to agent 3, etc.).

Nevertheless, in this way it is not possible to work in a more structured way (a workflow). One way to do this is to send a link to, for example, a database to work with; the problem of role-based data views remains to be handled. A workflow system which is capable of handling link pass mobility would solve the problem to enhance workflows during execution, for example, to involve additional roles which are not considered in the process model. This concept could be named as structured communication as a foundational concept of modern workflow systems, as discussed in ~\citep{singer2015}, for example.

$\pi$-calculus is a formal and mathematical foundation of computer languages such as WS-BPEL, for example. This technical underpinning provides the foundation for business process execution to handle the complex nature interactions. Given the nature of WS-BPEL, a complex Business Process could be organized in a potentially complex, disjointed, and unintuitive format that is handled very well by a software system~\citep{omg2013}.

\subsection{Actor System}
\label{actorsystem}

The Actor System~\citep{hewitt1973} (or Model), proposed in the 1970s, has a somewhat different view on systems. In the Actor Model, all objects are independent, computational units. These units only respond to received messages and do not share a common state. Actors change their state only when they receive a stimulus in the form of a message. So, an actor is a computational entity that, in response to a message it receives, can concurrently~\citep{vernon2015}:

\begin{itemize}
\item send a finite number of messages to other actors

\item create a finite number of new actors

\item designate the behavior to be used for the next message it receives

\end{itemize}

There is no assumed sequence to these actions, and they could be carried out in parallel. In a fully enabled actor system, everything is an actor. Actors implement finite state machines which change state by internal or external events; the only allowed external events are messages.

There is a broad range of software tools to support the realization of actor systems. The most prominent nowadays is the Akka\footnote{https:\slash \slash akka.io} framework (available in Java and Scala). Nevertheless, the Erlang\footnote{https:\slash \slash www.erlang.org} Virtual Machine (VM) is designed on the principles of the Actor Model and has its roots in the 1990s. Enterprise systems, designed as actor systems, are message driven reactive systems and therefore responsive, resilient, and elastic~\citep{vernon2015}.

\subsection{Business Process Model and Notation (BPMN)}
\label{businessprocessmodelandnotationbpmn}

This \emph{de facto} industry standard provides a semantic to define business processes. It is maintained by the Object Management Group (OMG) and also adopted as ISO\slash IEC standard~\citep{omg2013}. The standard has been released by the OMG in its actual version (besides minor adaptions) in 2011. The strength of the standard document is that it is not a simple graphical notation, but it also offers the possibility to serialize visual models into a computer readable format, i.e., into XML. A business process even can be defined purely in XML without an associated graphical model; that means, it is possible to generate process models by computer algorithms without human interaction.

As we want to transform BPMN semantics, serialized as XML-files, into OWL files, we shortly discuss some important aspects of the standard regarding serialization in the following sections.

\subsubsection{Metamodel}
\label{metamodel}

The standard document defines a metamodel for BPMN, a formal specification of the semantic elements containing a BPMN model and their relationships to each other. A valid BPMN model must conform to the specification of the metamodel, and its components are defined as object classes with defined requirements and optional attributes. Some classes (Root Element, Base Element) are purely abstract and not directly used in BPMN models.

\begin{figure}[htbp]
\centering
\includegraphics[keepaspectratio,width=150pt,height=0.75\textheight]{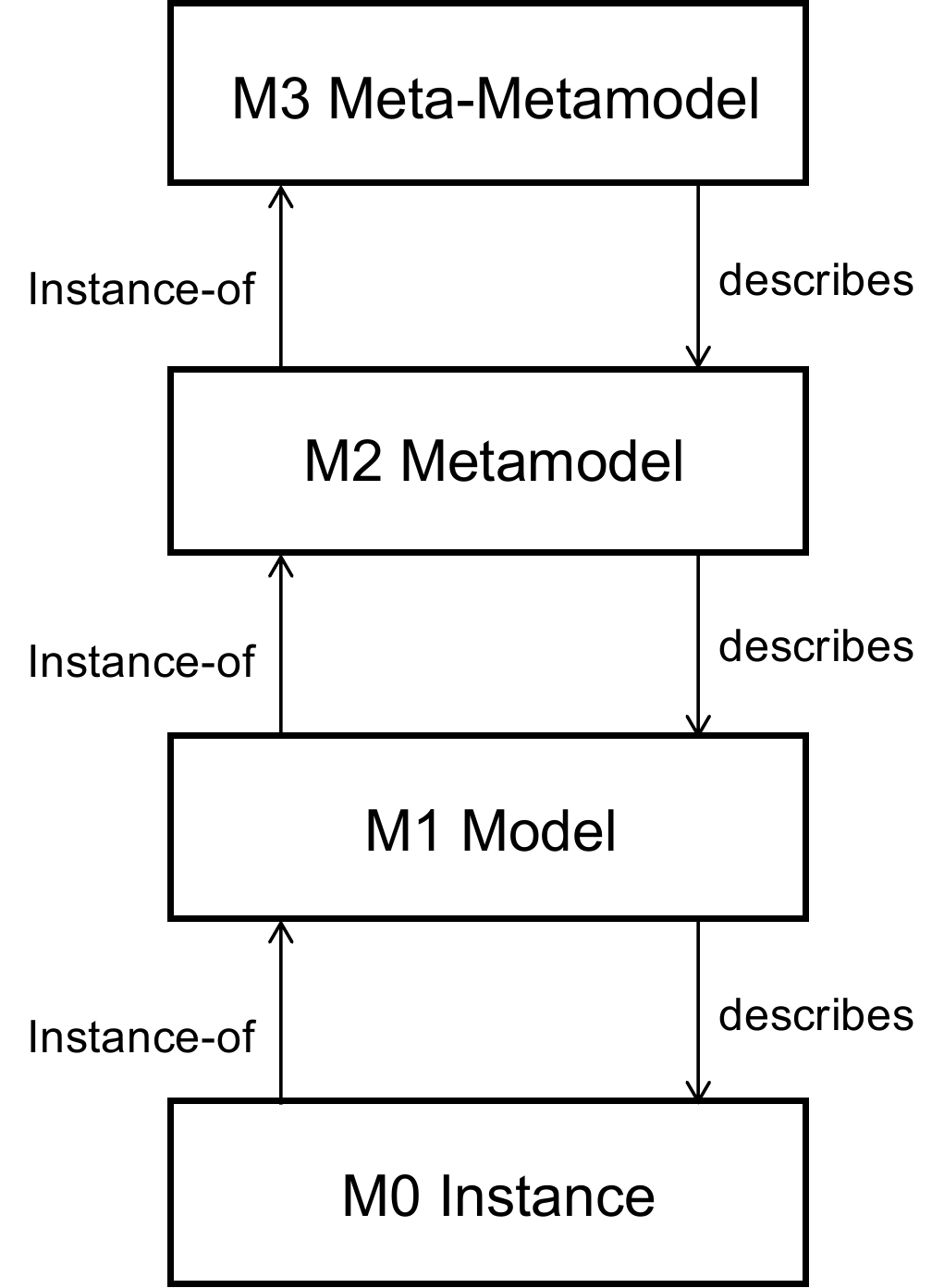}
\caption{OMG's Meta Object Facility (MOF).}
\label{mof}
\end{figure}

The BPMN standard is based on OMG's four-layer MOF modeling structure~\citep{weske2012}~\citep{henderson-sellers2012a}, as depicted in \autoref{mof}. The lowest layer M0 represents Data, layer M1 a particular model, layer M2 the metamodel, and M3 the metametamodel. Each layer is an instance of the layer of above. In case of BPMN layer, M2 contains the BPMN syntax and semantic, whereby M3 includes UML as the meta-language to define the standard as class diagrams. That means the standard document assumes a direct (or flat) relationship between metamodel and model, as depicted in \autoref{binarymeaning}. A critical discussion on the restrictions of the MOF architecture can be found in~\citep{henderson-sellers2012a}, for example.

The metamodel is defined by UML class diagrams, enhanced by tables and text. The metamodel is also published as OMG's XML Metadata Interchange (XMI) and W3C's XML Schema Definition (XSD). In this work, we will focus on XSD, as it is a widely accepted standard. Nevertheless, XSD cannot represent certain relationships of the UML, such as multiple inheritance; in XSD an element can only have one so-called \emph{substitutionGroup}, whereas in UML an element may be a subclass of more than one other class.

Now, a BPMN model is---by definition---not correct, if it is not a valid instance of the BPMN schema. The BPMN models itself are stored as XML files, which then can be validated against the XSD (supported by software tools). It is essential to understand that not all rules of BPMN are enforced by schema validation, but passing schema validation is an absolute minimum requirement to get valid BPMN models.

The BPMN schema is defined as a set of five XSD files: the file ``BPMN20.xsd'' file is the top level, which includes all the other ones. BPMN models also store graphical information---such as the position and size of each element in the graphical model---as defined in the BPMNDI schema, which can be omitted entirely.

Most elements in the XSD have an \emph{id} attribute; within an XML instance (a model) the values of the \emph{id} must be unique. The \emph{id} attribute is used to define relationships between the model elements.

\subsubsection{Modeling Conformance Subclasses}
\label{modelingconformancesubclasses}

The BPMN standard defines four types of conformance classes, namely

\begin{itemize}
\item Process Modeling Conformance

\item Process Execution Conformance

\item BPEL Process Execution Conformance

\item Choreography Modeling Conformance

\end{itemize}

The implementation claiming conformance to Process Modeling Conformance type is not required to support Choreography Modeling Conformance type and vice-versa. Similarly, the implementation claiming Process Execution Conformance type is not necessary to be conformant to the Process Modeling and Choreography Conformance types.

As an alternative to Process Modeling Conformance, the standard defines three Process Modeling Conformance subclasses:

\begin{itemize}
\item Descriptive

\item Analytic

\item Common Executable

\end{itemize}

The conformance subclasses define a limited set of elements and a limited set (as a minimal requirement) of attributes for each modeling element. This conformance sub-class Common Executable is intended for modeling tools that can emit executable models, whereby data type definition language must be an XML schema, service interface definition language must be WSDL, and data access language must be XPath. Implementations which claim Process Modeling Conformance are not expected to support the BPMN execution semantics.

\subsubsection{Serialization}
\label{serialization}

One of the main goals of this International Standard is to provide an interchange format that can be used to exchange BPMN definitions.

The top-level element of any BPMN model instance document is \emph{definitions}. The \emph{rootElement} children of \emph{definitions} represent reusable elements of the BPMN semantic model. These include the basic model types \emph{process}, \emph{collaboration}, and \emph{choreographie}, plus any other globally reusable elements, such as global activities, and messages. As some of the classes of the metamodel are defined as an abstract class, they will not show up in the BPMN instance; therefore, there is---for example---never a \emph{rootElement} in the instance file visible. Concrete root elements that designate \emph{rootElement} as its \emph{substitutionGroup} automatically inherit the properties of the root element class in the metamodel.

A process model with more than one participant constitutes a \emph{collaboration model}, whereby the participants interact via message flows. Each participant is a separate root element. A participant may only be associated with precisely one process (a pool in the diagram). The \emph{processRef} attribute of the \_participant\_element is omitted if it is a black box pool.

A data object in BPMN represents a local instance variable. In executable models, the \emph{itemDefinition}, a root element, points to a datatype. In the model itself, data is modeled via \emph{dataObjects}; the attribute \emph{itemSubjectRef} points to an \emph{itemDefinition}.

\begin{figure}[htbp]
\centering
\includegraphics[keepaspectratio,width=250pt,height=0.75\textheight]{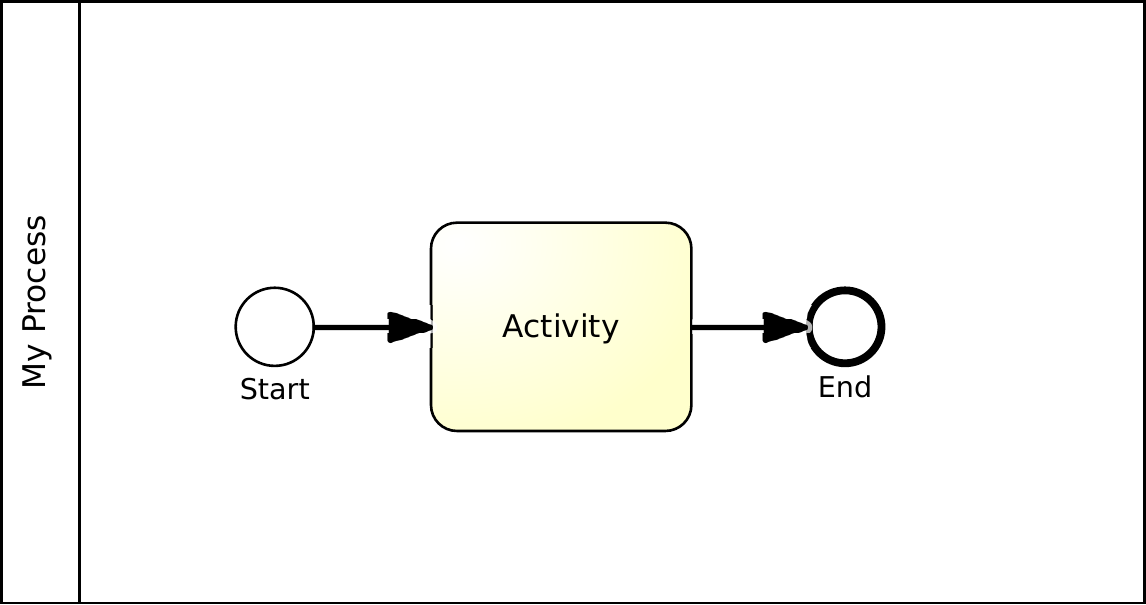}
\caption{An elementary BPMN process model.}
\label{averysimpleprocess}
\end{figure}

\begin{figure}[htbp]
\centering
\includegraphics[keepaspectratio,width=250pt,height=0.75\textheight]{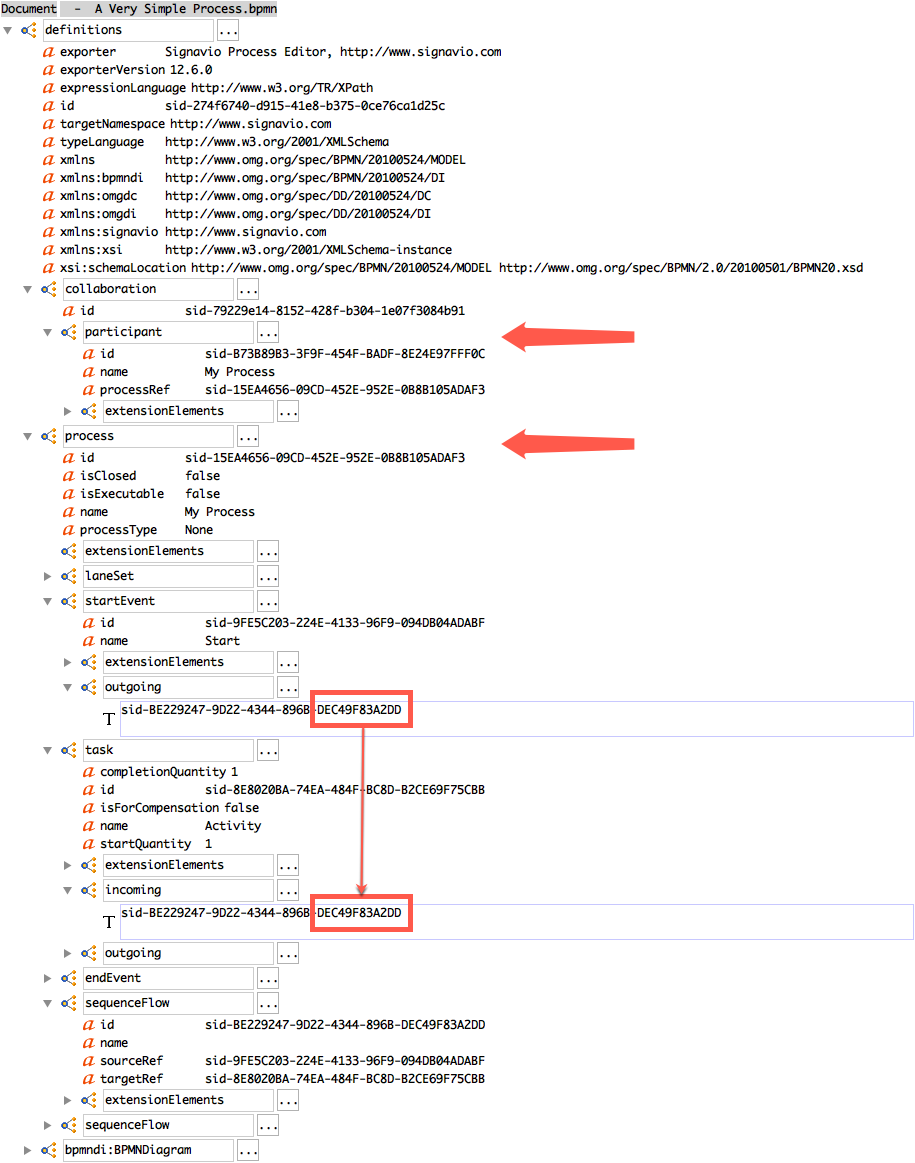}
\caption{An elementary BPMN process model as XML structure.}
\label{averysimpleprocessxml}
\end{figure}

Figure \autoref{averysimpleprocess} shows a very simple process in BPMN notation and figure \autoref{averysimpleprocessxml} shows the corresponding XML file. It can be seen, for example, that a process consists of a collaboration with one or more processes, which have (in this example) exactly one start event, one activity, and one end event; the elements are connected via their \emph{id} attributes.

\subsubsection{Critique (BPMN)}
\label{critiquebpmn}

Regarding syntax, the standard only provides a semi-formal definition of the BPMN metamodel in the form of class diagrams, corresponding tables specifying the attributes and model associations, as well as XML schemas. However, the definition of an element in the class diagram is partly overlapping with the refined specification in the corresponding table and redundant to the XML schema. Due to this redundancy, the description of the metamodel is in several cases inconsistent and contradictory~\citep{kossak2015}. Additionally, further syntactical rules are defined within natural text descriptions, also containing deviating information. Nevertheless, some of the critique~\citep{borger2012}~\citep{kossak2015} is based on the assumption that BPMN models should be directly executable on a workflow engine. This is not correct as this is only true under consideration of the part of the standard, which discusses the BPMN Execution Semantics; not all BPMN elements are considered to be executable on a workflow engine (non-operational elements). Nevertheless, the standard further states that ``the execution semantics are described informally (textually), and this is based on prior research involving the formalization of execution semantics using mathematical formalisms.'', but this is not documented or referenced in the standard document.

\subsection{Subject- or Agent-based BPM (S-BPM)}
\label{subject-oragent-basedbpms-bpm}

There is already a long history of the idea of interacting agents. The application of the agent concept into the domain of BPM has emerged from the field of distributed software~\citep{fleischmann1994} by Albert Fleischmann, who developed the Subject-oriented BPM (S-BPM) methodology in the early 2000s based on his Parallel Activities Specification Scheme (PASS) language. All language constructs of PASS can be transformed into pure CCS~\citep{aitenbichler2011}. The S-BPM methodology enhances the process algebra languages by graphical representations and adds some technical feature definitions.

Any collaboration contains more than one subject, so it is per definition a multi-agent system, which is a subclass of concurrent systems~\citep{wooldridge2009}---this is an important fact as it has consequences for possible technical implementation.

The problem of synchronizing multiple processes is not trivial and has been widely studied through the 1970s and 1980s~\citep{ben-ari1990}. It is helpful to consider the way that communication is treated in the object-oriented programming paradigm to understand the problem~\citep{wooldridge2009} of coordinating the behavior of several agents. In the field of object-orientation communication is realized as a method invocation. The crucial point is, that an object does not have control over the execution of its own public methods---any other object can execute the object's public methods whenever they want.

\begin{figure}[htbp]
\centering
\includegraphics[keepaspectratio,width=200pt,height=0.75\textheight]{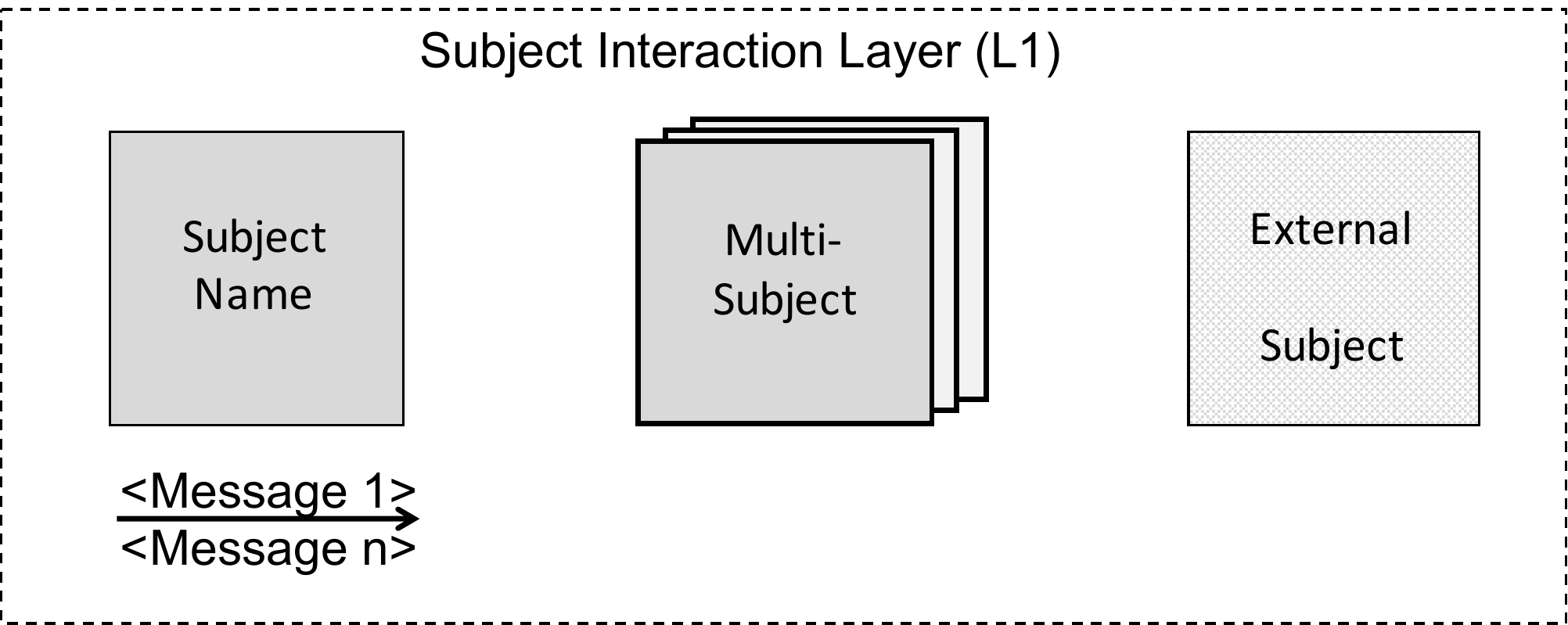}
\caption{Elements of a Subject Interaction Diagram. Multisubjects represent a modeling element which can be instantiated more than once. External Subjects represent other process participants where the SBD is not available, for example}
\label{sid}
\end{figure}

\begin{figure}[htbp]
\centering
\includegraphics[keepaspectratio,width=200pt,height=0.75\textheight]{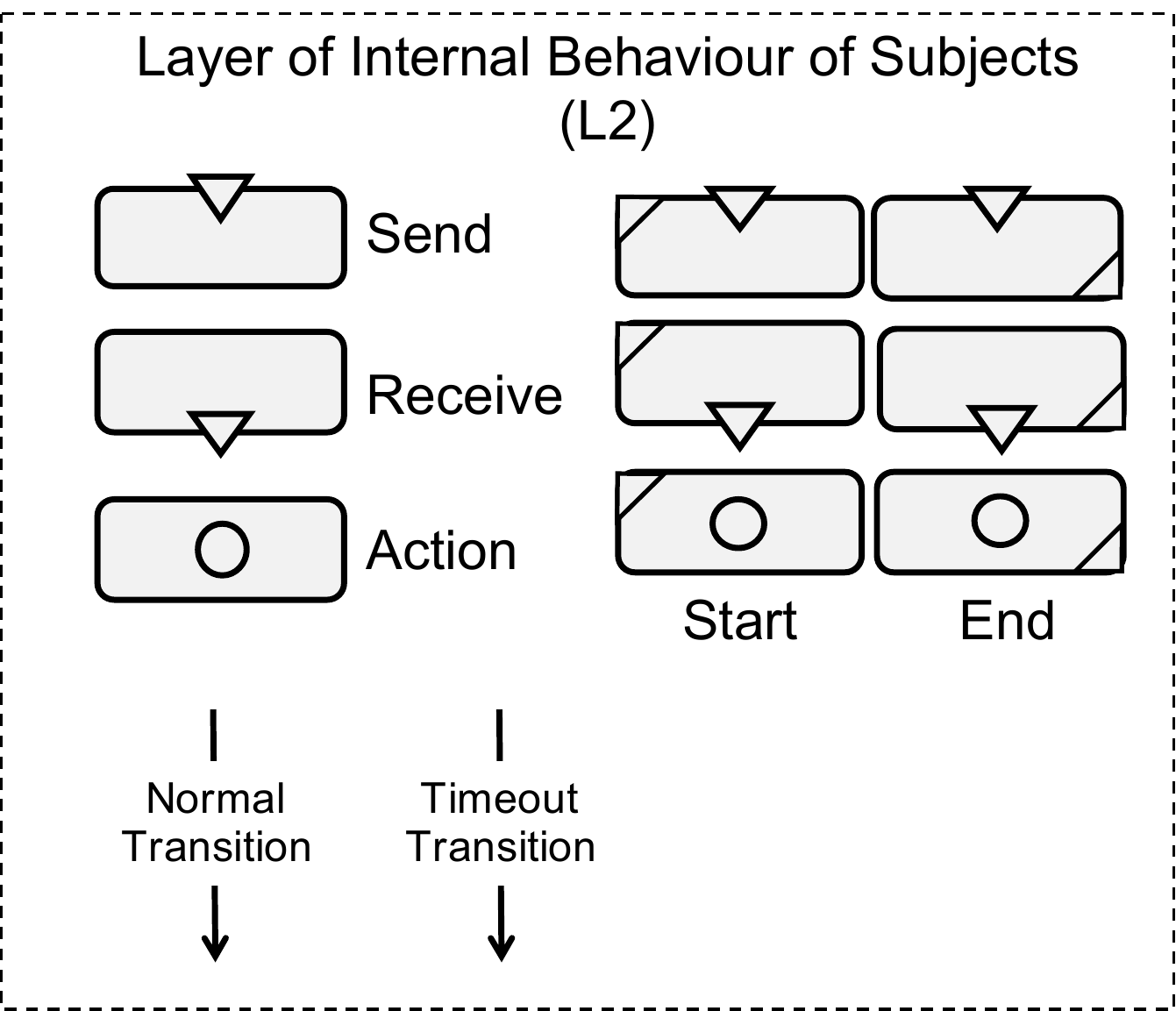}
\caption{Elements of a Subject Behaviour Diagram}
\label{sbd}
\end{figure}

An S-BPM process is defined via the communication exchange between subjects (actors are instantiated subjects in this context, or the other way round---subjects are generalizations of actors). Additionally, each subject has a defined (but invisible to the outside world) internal behavior, which is determined as a process flow using states for receiving or sending a message (to another subject), and states in which the subject is doing some work. States can be flagged as starting or ending states and are connected using directed arcs. In the context of BPM, actors define who is doing what, as they are mapped to a resource for execution (organizational roles). Typically, S-BPM models consist of two types of representations: a Subject Interaction Diagram (SID) and a set of Subject Behavior Diagrams (SBD). The SID includes the subjects, and the messages exchanged between the subjects and the business objects attached to the messages as depicted in Figure \autoref{sid}. The SBD includes all possible state sequences of a subject: a finite set of send, receive and function states as depicted in Figure \autoref{sbd}. There are more special elements, which will be introduced during the discussion if needed.

It can be shown, that with these elements all workflow patterns and all service interaction patterns can be modeled and also executed on an appropriate workflow engine~\citep{singer2014}~\citep{graef2009}.

\begin{figure}[htbp]
\centering
\includegraphics[keepaspectratio,width=150pt,height=0.75\textheight]{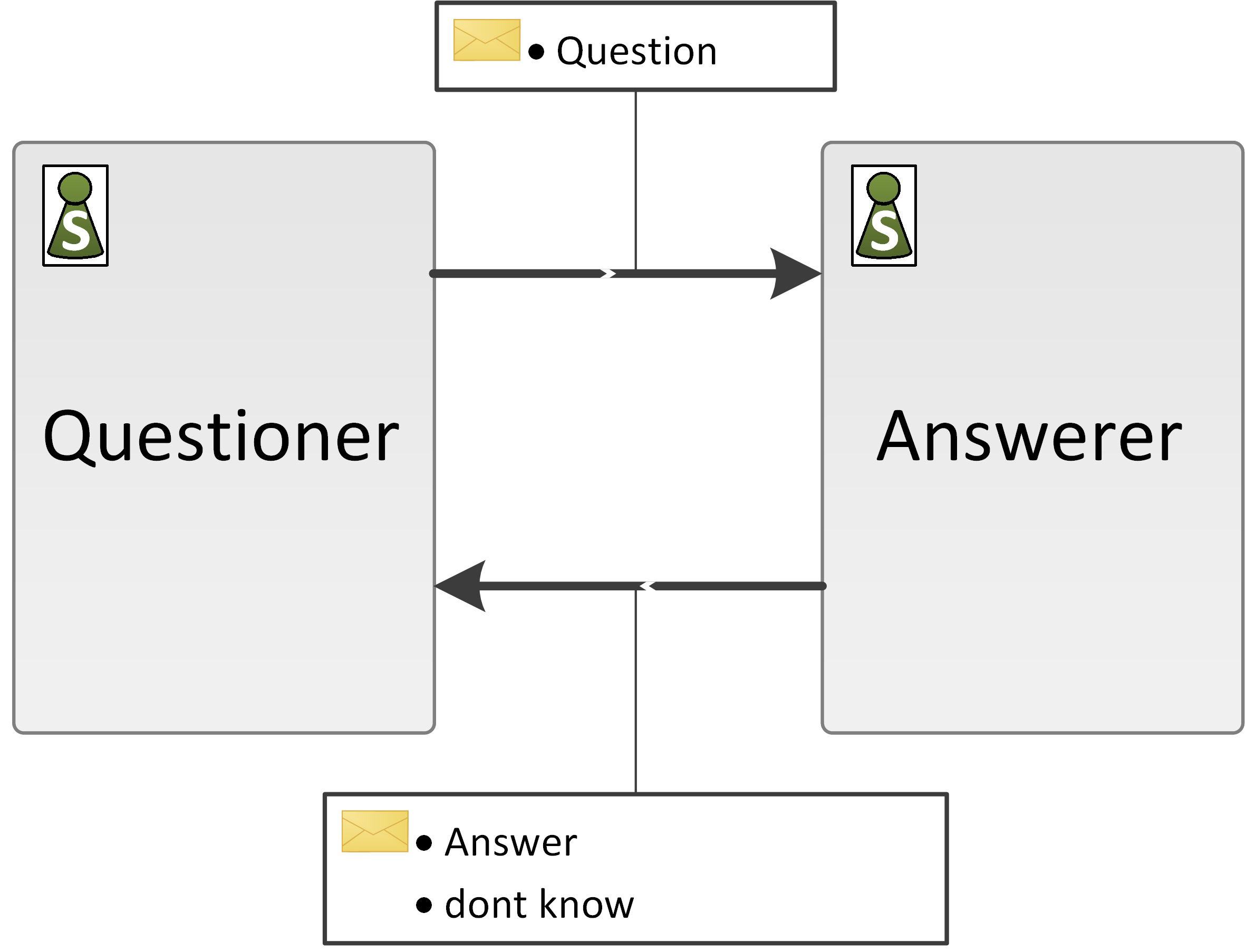}
\caption{A very simple SID example.}
\label{sidexample}
\end{figure}

\begin{figure}[htbp]
\centering
\includegraphics[keepaspectratio,width=150pt,height=0.75\textheight]{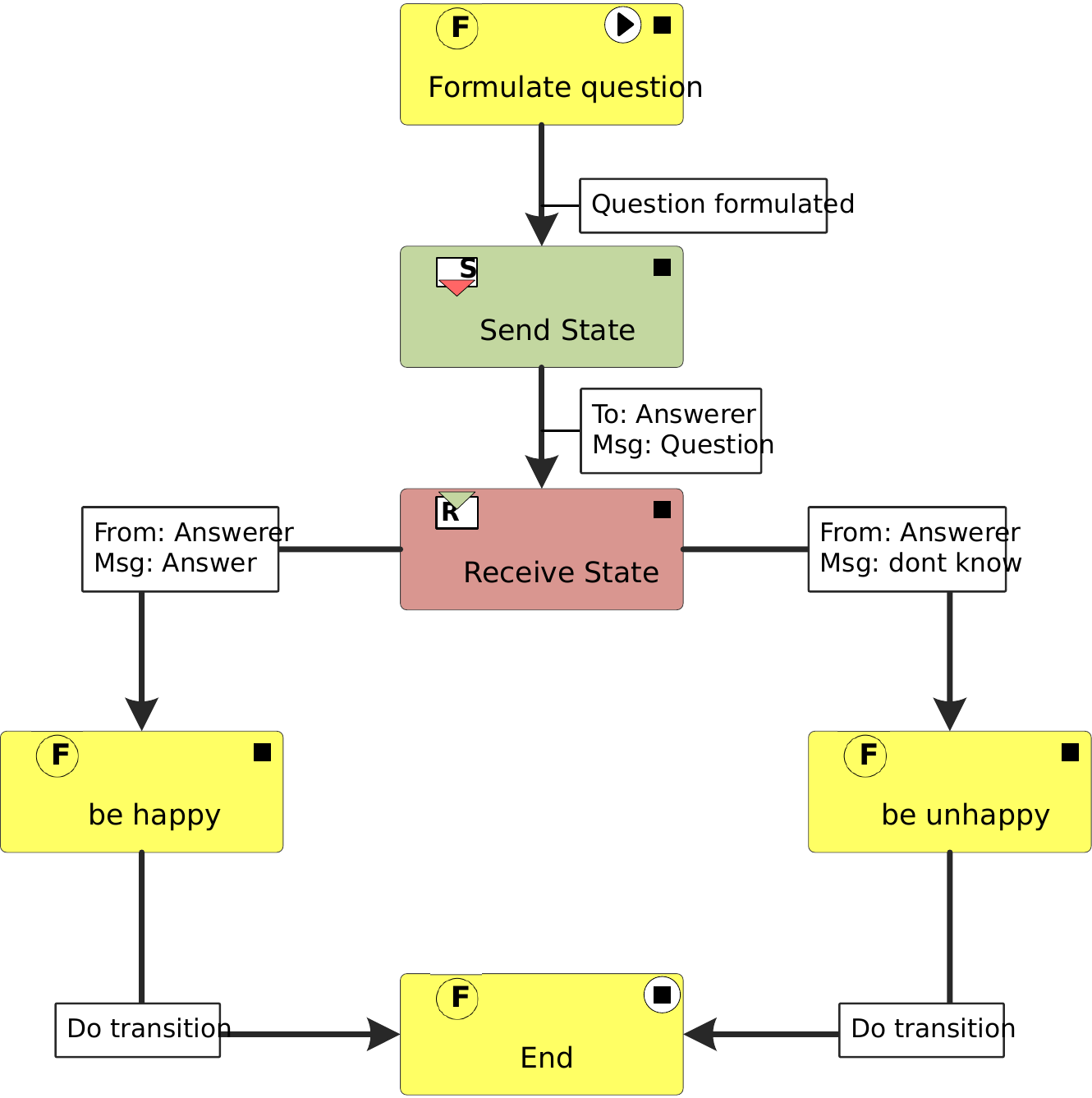}
\caption{SBD (questioner) diagram for the process depicted in \autoref{sid_1}.}
\label{sbdexamplequestioner}
\end{figure}

\begin{figure}[htbp]
\centering
\includegraphics[keepaspectratio,width=150pt,height=0.75\textheight]{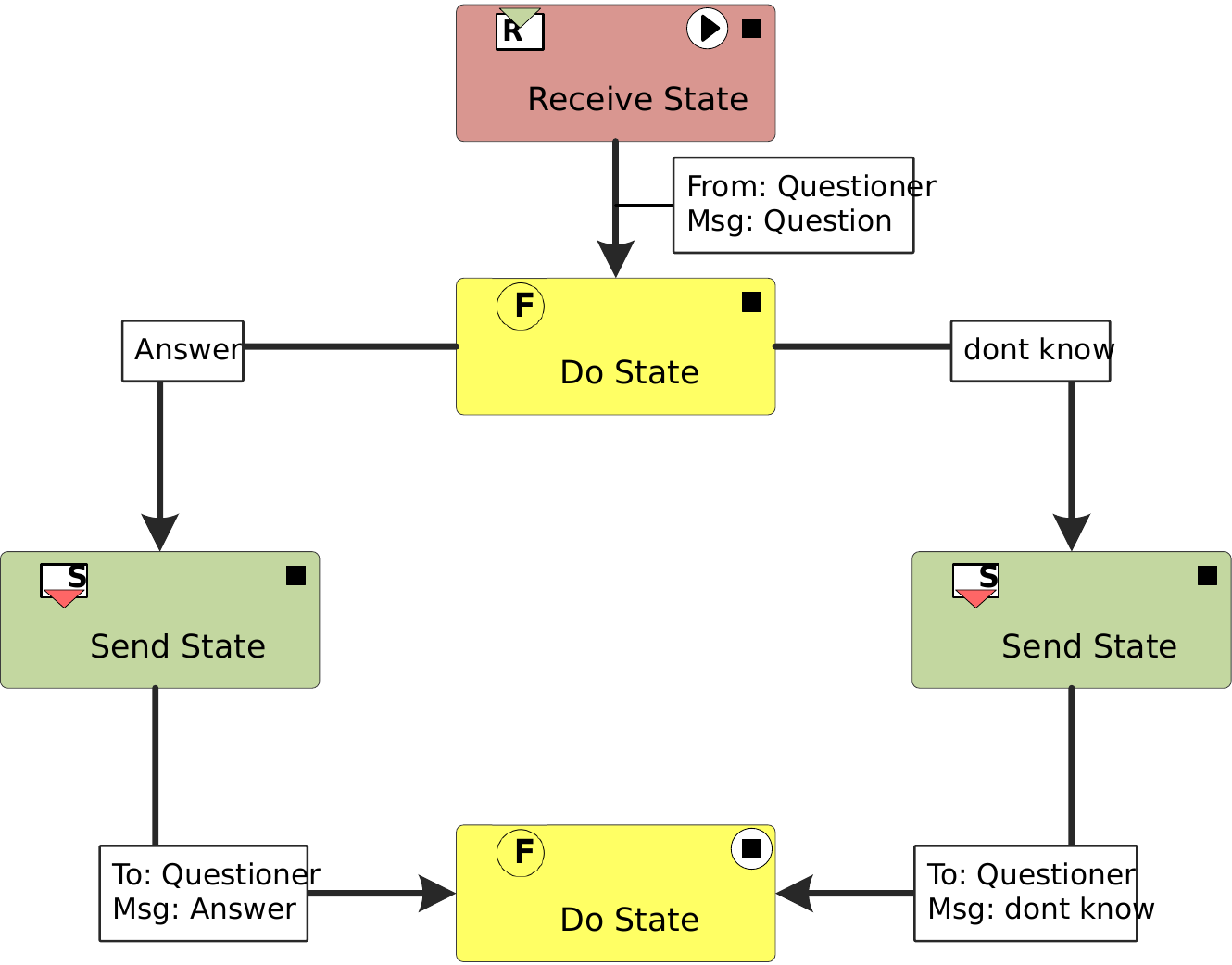}
\caption{SBD (answerer) diagram for the process depicted in \autoref{sid_1}.}
\label{sbdexampleanswerer}
\end{figure}

A simple process model for illustration is depicted in in Figures \autoref{sidexample}, \autoref{sbdexamplequestioner}, and \autoref{sbdexampleanswerer} as S-BPM model. The meaning of the models is: one person asks another person for a decision (or question). The decision is then communicated to the requester (positive or negative answer). We also have to mention, that the S-BPM notation has also some more enhanced concepts, which are not mentioned here as they are not important for the essential understanding (see ~\citep{fleischmann2012} for a complete reference). The process can easily be defined in BPMN with the same meaning.

\subsubsection{Text Based Modeling}
\label{textbasedmodeling}

Another possibility is to use the natural language capabilities of the S-BPM approach: we can express the process network as a finite set of formal sentences expressed as complete sentences with subject, predicate, and object~\citep{fleischmann2011b} which can be directly executed by information technology in the same way as graphically defined processes. Such an approach offers the possibility to involve more people in the modeling phase as they can use the language they already know---natural language sentences (but in a formal structure), such as, for example, ``Miller (subject) sends (predicate) the invoice (object) to the customer (subject)''. Based on this characteristic of the subject-oriented approach, it is easy to define a Domain Specific Language~\citep{hover2013}, for example.

Based on formal natural language sentences a process model can be defined merely using a general purpose spreadsheet application; afterward, process definitions can be exported as CSV-files for the input into an execution engine. For this purpose, we have recently created a multi-user web platform to define text-based business process models~\citep{singer2016}.

\begin{figure}[htbp]
\centering
\includegraphics[keepaspectratio,width=150pt,height=0.75\textheight]{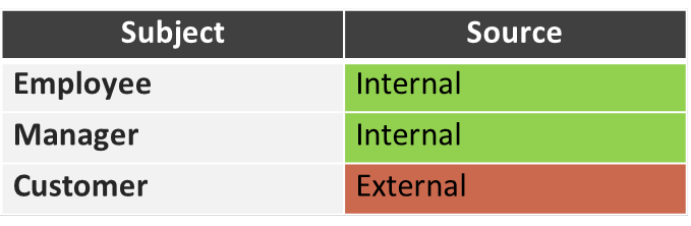}
\caption{Set-up internal or external subjects. Internal subjects are colored green and external subjects are colored orange-red.}
\label{textmodel1}
\end{figure}

As depicted in \autoref{textmodel1} all subjects, which take part in the business process have to be set up first.

\begin{figure}[htbp]
\centering
\includegraphics[keepaspectratio,width=150pt,height=0.75\textheight]{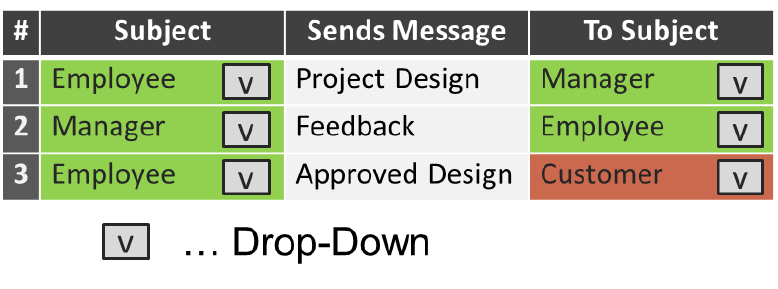}
\caption{Define interactions between the subjects.}
\label{textmodel2}
\end{figure}

The next step handles the interaction of the subjects, that means the message exchange as depicted in \autoref{textmodel2}. To get a clear overview of the message exchange sequence, it is required to define the sender and the receiver of a message as well as the purpose of the interaction.

\begin{figure}[htbp]
\centering
\includegraphics[keepaspectratio,width=200pt,height=0.75\textheight]{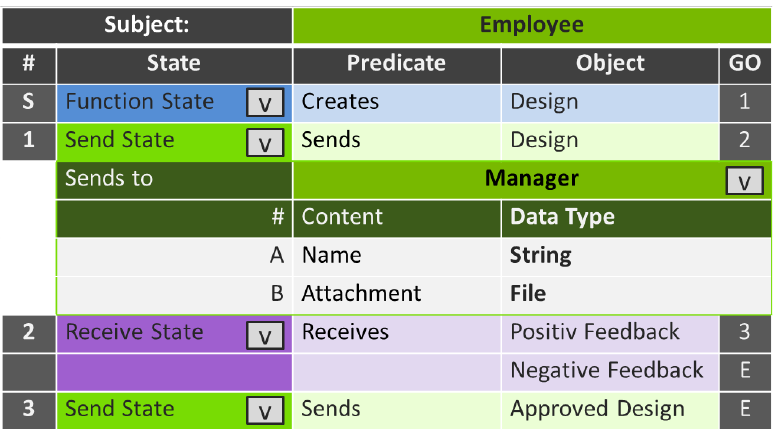}
\caption{Overview of defining the subject behavior. The function-state is colored blue, the send-state is green, and the receive-state is violet. S and E denote start- and end-states respectively.}
\label{textmodel3}
\end{figure}

The third and last step has to be done for every single subject in the process. This step is about modeling the internal behavior of all subjects successively as depicted in \autoref{textmodel3}, i.e., defining a finite sequence of send-, receive- and function-states. To distinguish the states, all states are colored. Within the state, always the object and the goto-number has to be selected. Especially when there is a branch point at a state, the goto-number defines the next states, if one of the possible cases occurs.

\begin{figure}[htbp]
\centering
\includegraphics[keepaspectratio,width=200pt,height=0.75\textheight]{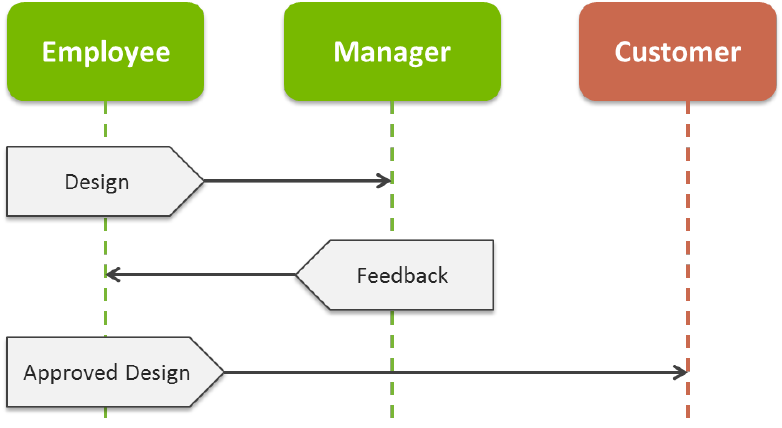}
\caption{Concept of a Subject Interaction Diagram. Arcs show the message exchange along a time axis from top to down.}
\label{textmodel4}
\end{figure}

\begin{figure}[htbp]
\centering
\includegraphics[keepaspectratio,width=200pt,height=0.75\textheight]{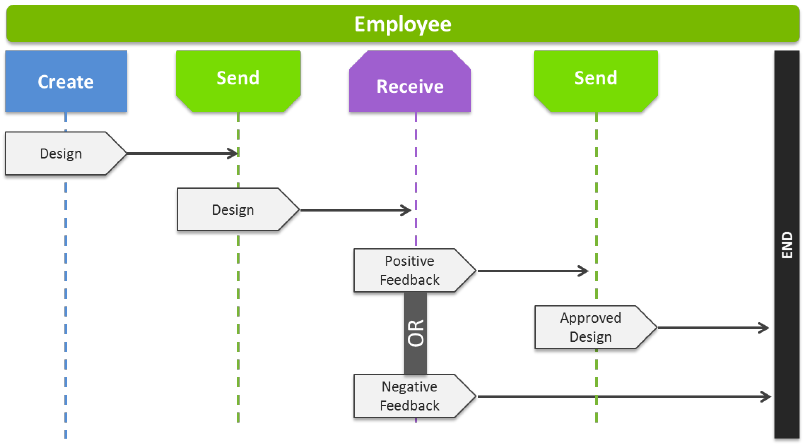}
\caption{Concept of a Subject Behavior Diagram. Arcs show state changes along a time axis from top to down. OR denotes a decision; in our case, it is an exclusive or, but as we use natural language we propose not to use XOR.}
\label{textmodel5}
\end{figure}

After entering all necessary process data, the graphic of the process model is generated automatically. UML sequence diagrams inspire our visual concept; an example for a Subject Interaction Diagram is depicted in \autoref{textmodel4}, and an example for a Subject Behavior Diagram is depicted in \autoref{textmodel5}.

It is evident that the generated process models can be exported as OWL files (based on the proposed S-BPM ontology, for example) for further use, for example as input in an appropriate WfMS.

\subsubsection{Critique (S-BPM)}
\label{critiques-bpm}

There are some advantages subject-oriented modeling of business processes: beside technical and architectural aspects it is more user-focused. Each process participant only needs to define the behavior of its role including the interaction with other roles. Additionally, this can be done in principle with only five symbols which is much easier to learn and train as using BPMN. Moreover, it also has a mathematical underpinning and corresponds with a modern view on software development which makes it much easier to develop a workflow engine.

Off course it is another proprietary approach, and from a business point of view, it would be wise to consider an industry standard as the first choice to reduce risk and to protect investments in modeling. Nevertheless, we will demonstrate that subject-orientation is more a general concept than a specific modeling language and it is rather easy to see, that this view can also be realized using BPMN as a modeling language. To unfold its full potential, the subject-oriented view has to be applied to the design of workflow engines of course.
\#\#\#

From a business point of view, it seems not to matter which modeling language to use~\citep{recker2007} as long as we are aware that people need adequate training to be able to design useful on correct business process models~\citep{recker2010}.

There are some advantages subject-oriented modeling of business processes: beside technical and architectural aspects it is more user-focused. Each process participant only needs to define the behavior of its role including the interaction with other roles. Additionally, this can be done in principle with only five symbols which is much easier to learn and train as using BPMN. Moreover, it also has a mathematical underpinning and corresponds with a modern view on software development which makes it much easier to develop a workflow engine.

Off course it is another proprietary approach, and from a business point of view, it would be wise to consider an industry standard as the first choice to reduce risk and to protect investments in modeling. Nevertheless, we will demonstrate that subject-orientation is more a general concept than a specific modeling language and it is rather easy to see, that this view can also be realized using BPMN as a modeling language. To unfold its full potential, the subject-oriented view has to be applied to the design of workflow engines of course.

\subsection{Business Process Modeling}
\label{businessprocessmodeling}

As there is (yet) no coherent and \emph{general accepted} theory of business processes and business process management (BPM)~\citep{sanz2013}~\citep{russell2016a}, any way to define process models is the right one---it depends on the purpose of the model; for this purpose domain-specific languages (e.g. notations such as BPMN or UML) are defined, most of them are rooted in the information systems domain. That is a consequence of the fact, that information systems engineers need formally defined models without any semantic ambiguity. Additionally, modeling is typically conducted by experts, that means business analysts or requirements engineers, for example. But studies~\citep{caire2013} and experience from teaching show that end users and learners understand such expert models very poorly. One of the reasons for this is that it is hard for experts to think like novices, a phenomenon called \emph{the curse of knowledge}. There are well-known differences in how experts and novices process diagrams~\citep{cheng2001}.

It is good practice to involve ``users'' in the analysis and design of business processes; this also works in developing software systems (e.g., user-centered design) or in developing new products. Why not involve them in the design process of notations? Caire at al.~\citep{caire2013} have done this for example in a research study regarding requirements engineering (RE) notations.

\begin{figure}[htbp]
\centering
\includegraphics[keepaspectratio,width=250pt,height=0.75\textheight]{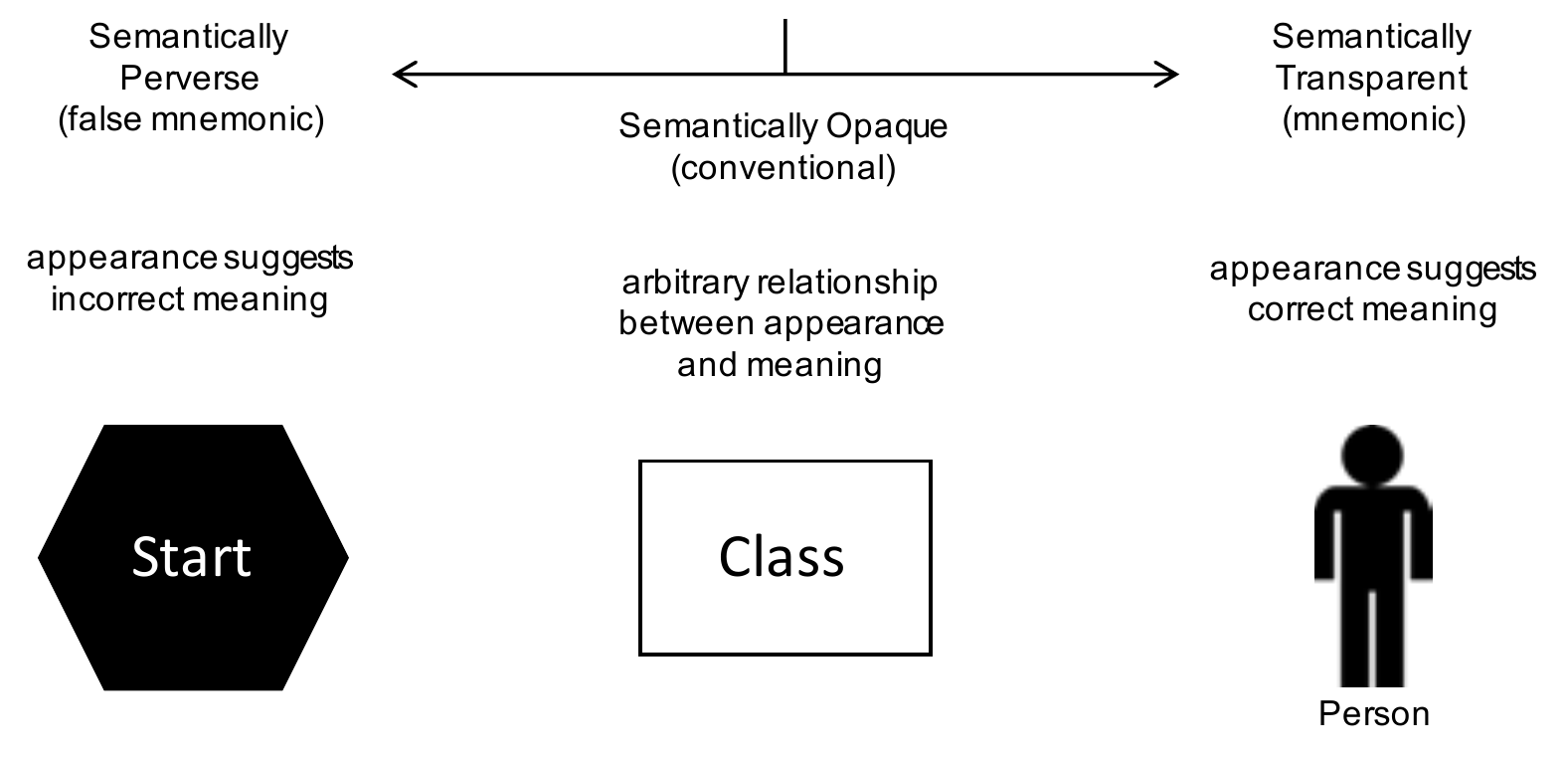}
\caption{Semantic transparency is a continuum (adapted from ~\citep{caire2013}.}
\label{semtrans}
\end{figure}

The key to designing vial notations that are understandable to  users is a property called semantic transparency. This means that the meaning (semantics) of a symbol is clear (transparent) from its appearance alone; Semantically transparent symbols reduce cognitive load because they have built-in mnemonics~\citep{petre1995} (see \autoref{semtrans}).

However, semantic transparency is typically evaluated subjectively: experts (researchers, experts from software vendors) try to estimate the likelihood that novices will be able to infer the meaning of particular symbols. Even when notations are specifically designed for communicating with business stakeholders, members of the target audience are rarely involved. For example, BPMN 2.0 is a notation intended for communicating with business stakeholders, yet no business representatives were engaged in the notation design process, and no testing was conducted with them before its release~\citep{recker2010}~\citep{caire2013}.

Business process models are needed to facilitate a shared understanding in the organization; therefore the process of creating and documenting the model includes employees unfamiliar with the chosen process design method. Typical workshops on process design use design tools such as a whiteboard, flip charts and post-its to capture knowledge about a current or future process. Informal sketches and diagrammatic drawings were found to be vital to any design activity, as they serve as an externalization of one's internal thoughts, and assist in idea creation and problem-solving.

There is a clear difference how novice and expert modelers conceptualize essential domain elements as reported by Wang and Brooks~\citep{wang2007}, who found that novice modelers conceptualize in a reasonably linear process in contrast to experts, who have better analysis and critical evaluation skills. Also based on inexperienced modelers, Recker et al.~\citep{recker2010} developed a range of typical process design archetypes; they found out, that ``moderate use of graphics and abstract shapes to illustrate a process is more intuitive and would aid the understanding on the concept of process modeling.'' There is some research to develop modeling guidelines based on an empirical evaluation, as the 7PMG rules from Mendling at al.~\citep{mendling2010}, for example.

Another point to consider is the practical quality of a model: it is evident that the syntactic and the semantic quality must fit the purpose of the model, but also the pragmatic (from semiotics) of the model must be considered. This is not an easy task, and few frameworks have been proposed to define evaluation criteria to measure the quality of business process models, as the 3QM framework~\citep{overhage2012}, for example.

Another interesting measurement value could be the complexity of a business process model, for example, which could be done evaluating the information entropy as discussed by Jung~\citep{jung2008}~\citep{jung2011}.

\section{Workflow Management Systems}
\label{workflowmanagementsystems}

One crucial aspect and challenge in the field of business process aware systems is the alignment of the software with requirements from business. This alignment has always been driven by business process modelers and is based on ambiguous process models. The predominant view on process modeling is still based on flow diagrams, a tool for software developers to define algorithms in the fifties of the last century. This thinking also includes implicit concepts such as ``goto statements''~\citep{dijkstra2001} and leads to what can be called ``spaghetti process diagrams.'' Now, what can business process modelers learn from software developers~\citep{gruhn2007} should, therefore, be the question.

Over the last couple of years, new concepts have emerged or got more attention, as, reactive~\citep{blackheath2015} and flow based programming~\citep{morrison2011}. There is also an increasing interest in microservices, functional programming, and actor based systems to support the need to develop solutions that are responsive, resilient, elastic and message driven---the core requirements stated in the Reactive Manifesto~\citep{boner}. Reactive systems respond on time (usability), stay responsive in the face of failure, stay responsive under varying workload, and are message driven; relying on asynchronous message-passing means to establish a boundary between components that ensures loose coupling, isolation, location transparency, and provides the means to delegate errors as messages. This all also supports the smooth integration of machines and devices into business processes.

Furthermore, large systems are composed of smaller ones and, therefore, depend on the reactive properties of their constituents. This leads to the concept of microservices~\citep{newman2015} as a design pattern to build reactive systems meeting the requirements mentioned before. The microservice architectural style is an approach to understanding any application as a collection of small services, each of them running in its process environment and communicating with lightweight mechanisms; state changes of a service can only be triggered by receiving certain message types and business objects (data).

It seems worth to learn from the software community and think about concepts such as microservices or reactive programming, for example. Therefore, we propose to enhance the method pool of business process modeling and execution by defining business process models based on modern architectural perspectives and knowledge from software development. It is evident that the concepts of the modeling language induce the architecture of a workflow engine.

As a model should help to understand, we think, we should use languages which inherently incorporate real-world concepts such as message exchange between interaction partners, for example. In some aspects, we propose to rethink the dominant logic of business process management and the architecture of workflow management systems. There is nothing wrong with the industry standard BPMN, for example, but practical application scenarios show severe problems defining correct process models which are understandable and also executable.

Therefore, in this section, we discuss the well-known workflow reference architecture proposed by the WfMC, review the state of affairs with actual industry implementations, and present a reference architecture for the distributed execution of agent-based (S-BPM) processes. The analysis should lead to a critical discussion about the general architecture of workflow systems.

\subsection{Reference Architecture}
\label{referencearchitecture}

\begin{figure}[htbp]
\centering
\includegraphics[keepaspectratio,width=250pt,height=0.75\textheight]{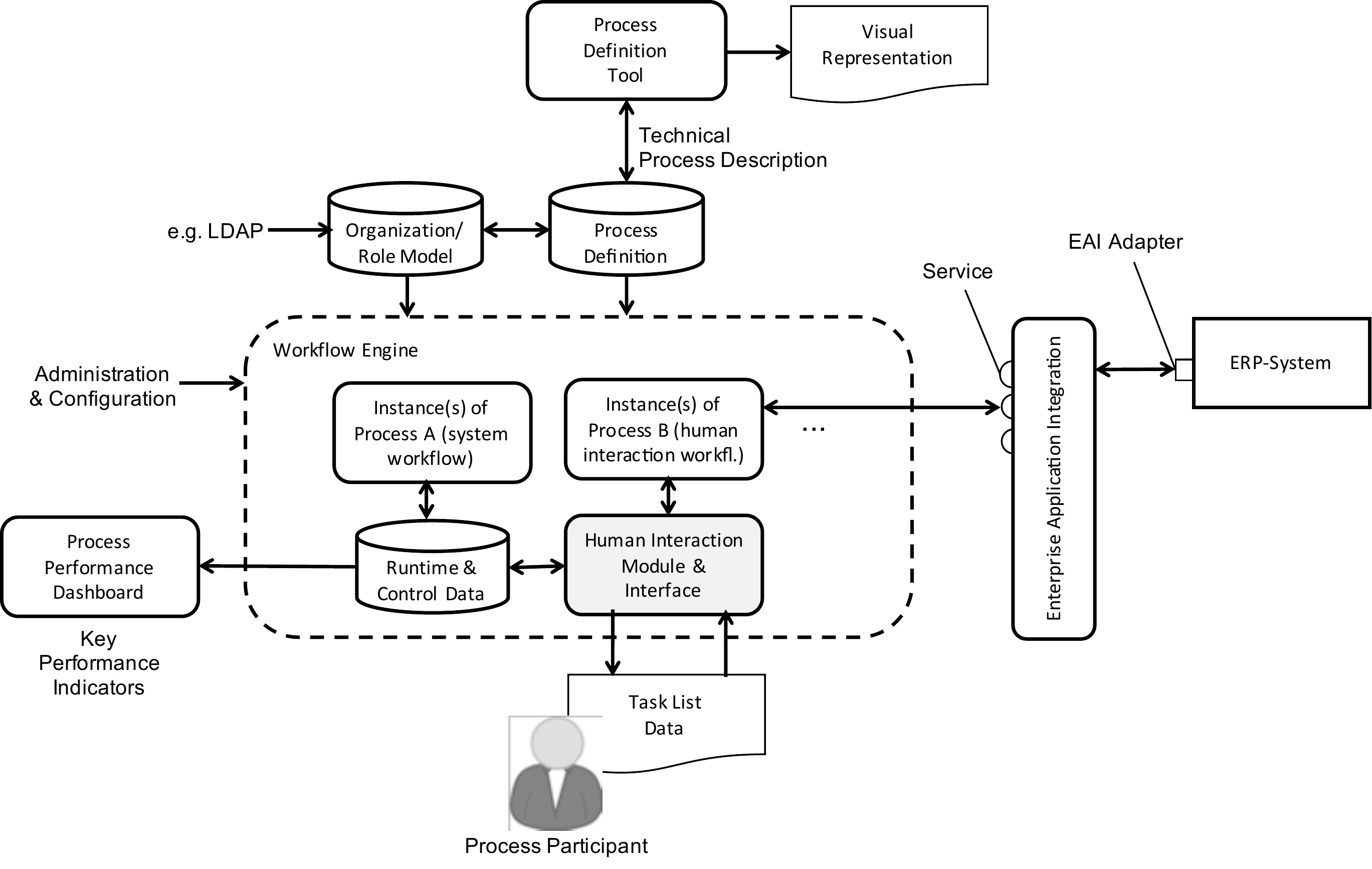}
\caption{Typical workflow system components, adapted from~\citep{hollingsworth1995} and~\citep{singer2016}.}
\label{wfm1}
\end{figure}

A typical conceptual architecture of a workflow system, as part of a business process management system, is depicted in \autoref{wfm1}. The Workflow Management Coalition (WfMC) defines~\citep{hollingsworth1995} workflow as ``the computerised facilitation or automation of a business process, in whole or part.'', and workflow management system as ``a system that completely defines, manages and executes `workflows' through the execution of software whose order of execution is driven by a computer representation of the workflow logic.''

Hillingsworth~\citep{hollingsworth1995} is often referred to as a reference model for workflow system and apparently had significant influence in the development of such systems. The reference model describes a standard model for the construction of workflow systems and identifies how it may be related to various alternative implementation approaches:

\begin{itemize}
\item the Build-time functions, concerned with defining, and possibly modeling, the workflow process and its constituent activities

\item the Run-time control functions related to managing the workflow processes in an operational environment and sequencing the various activities to be handled as part of each process

\item the Run-time interactions with human users and IT application tools for processing the various activity steps

\end{itemize}

A short description of the depicted architecture is as follows: business process models (including business object data definitions) are stored in a repository; we also need a formal model of the organization, so we can link organizational groups, roles and individual persons with the activities of the process model (who is doing what). Process models can be uploaded to and started by the workflow engine; they are interpreted by the software logic of the application (process execution). Furthermore, it is essential to understand that there must be some mechanism to interact with human process participants.

Firstly, the application needs to create and maintain a task list to distribute work~\citep{russell2016a}. Secondly, a task requires some input (data) from a human actor. Typically, a task is presented as a form-based window to the process participant, which includes some read-only data and offers some interactive elements to enter or change data; the forms have to be designed and developed manually, or, can be generated automatically from business objects. Finally, we also need to interact with other systems in the enterprise, which nowadays typically is done via (web) service calls.

This conceptional architecture is used in principle in nearly all commercially available BPMS, and it is a logical imperative of the way how business process models are defined. The challenge lies in the automatic translation from human-readable process model to strictly formal computer readable and executable models. In the following section, we will also critically address which parts of the reference architecture is typically not realized in a commercial software solution.

\begin{figure}[htbp]
\centering
\includegraphics[keepaspectratio,width=250pt,height=0.75\textheight]{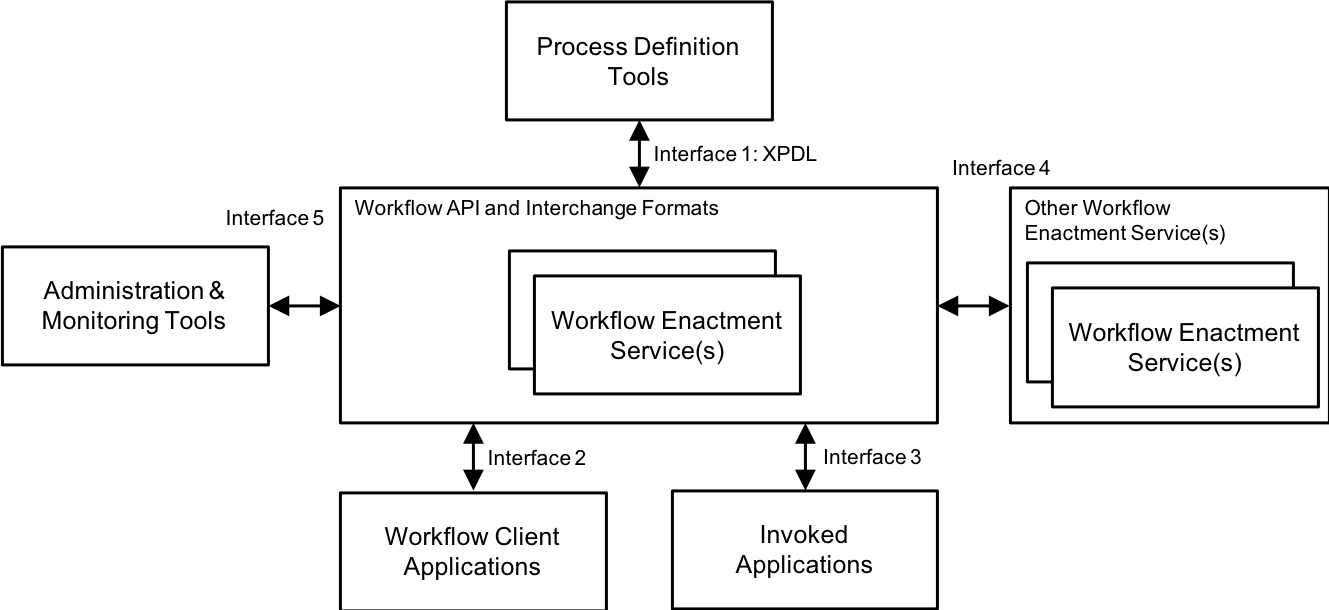}
\caption{Workflow reference model---components and interfaces~\citep{hollingsworth1995} and~\citep{russell2016a}.}
\label{wfm2}
\end{figure}

Now, the full range of so-called interfaces being defined by the WfMC covers (see \autoref{wfm2}):

\begin{itemize}
\item specifications for process definition

\item interfaces to support interoperability between different workflow systems

\item interfaces to support interaction with IT applications

\item interfaces to support interaction with process users

\item interfaces to provide system monitoring and process metrics

\end{itemize}

The WfMC proposal also considers a centralized or distributed workflow enactment service as implementation alternatives: the workflow enactment software consists of one or more workflow engines, which are responsible for managing all, or part, of the execution of individual process instances. Such a system may be set up as a centralized system with a single workflow engine responsible for managing all process execution or as a distributed system in which several engines cooperate, each managing part of the overall process execution.

A workflow enactment service is defined as a software service that may consist of one or more workflow engines to create, manage and execute workflow instances. Applications may interface to this service via service calls, for example. The reference architecture further states that the workflow enactment service may be physically either centralized or functionally distributed. In a distributed workflow enactment service, several workflow engines each control a part of the process enactment and interact with that subset of users and application tools related to the activities within the process for which they are responsible. A software service or ``engine'' that provides the run time execution environment for a workflow instance is named a workflow engine, which is part of the workflow enactment service.

In its simplest view, the workflow enactment service may be considered as a state transition machine, where an individual process or activity instances change states in response to external events (e.g., completion of an activity) or to specific control decisions taken by a workflow engine (e.g., navigation to the next activity step within a process).

A key objective of the coalition is to define standards that will allow workflow systems produced by different vendors to pass work items seamlessly between one another. Four possible interoperability models have been identified:

\begin{itemize}
\item Chained services model: This model allows a connection point within process A to connect to another point within process B. This model supports the transfer of a single item of work (a process instance or activity) between the two workflow environments, which then operates independently in the second environment with no further synchronization.

\item Nested subprocesses model: This allows a process executed in a particular workflow domain to be encapsulated entirely as a single task within a (superior) process performed in a different workflow domain.

\item Peer-Peer model: This model allows a thoroughly mixed environment; the diagram indicates a composite process C, which includes activities which may be executed across multiple workflow services, forming a shared domain. Activities C1, C2, and C5 could be coordinated by server A (or even several homogenous servers within a common domain) and activities C3, C4 and C6 co-ordinated by server B.

\item Parallel synchronized model: This model allows two processes to operate substantially independently, possibly across separate enactment services, but requires that synchronization points exist between the two processes.

\item Gateway operation: If different workflow environments do not share the same process definition, they may interoperate based on some exchange mechanism.

\end{itemize}

\subsection{BPMN Workflow Systems}
\label{bpmnworkflowsystems}

Most of the rare scientific publications on workflow management systems cite the reference architecture of the WfMC, and it is implicitly assumed that a typical commercial Workflow Management Systems is built on the WfMC reference architecture as presented in the previous section.

A review of typical commercial WfMS shows that there are some common architecture patterns (we have only BPMN based systems in mind as all others are legacy\footnote{From a management view the core business processes are assets and modeling induces cost and is, therefore, an investment. Therefore it is better to use an industry standard to protect the investment and to be more independent from a specific software vendor.}, in our opinion)

\begin{itemize}
\item upload of process models into an execution system (the WfMS)

\item the data model, including data views and access rights, need to be directly designed on the execution platform

\item the organizational role model has to be created on the execution platform (this can be done by linking LDAP roles to process activities, for example)

\item human interaction form-based dialogs have to be designed and programmed on the execution platform

\item all business process models are executed on one singular workflow engine

\item a WfMS is executed on one singular Web-Server

\end{itemize}

The WfMC standard states, that ``the process definition contains all necessary information about the process to enable it to be executed by the workflow enactment software.''; this includes references to applications which may be invoked, definitions of data, and ``the process definition may refer to an Organisation\slash Role model which contains information concerning organizational structure and roles within the organization.'' These concepts constitute an architecture which is based on a clear distinction between process modeling and execution. Actual architectures have an unclear conceptual mismatch between modeling and execution phase. Furthermore, any process definitions done on the execution platform needs substantial knowledge in software development which foils the idea of business process modeling by the people involved.

Another missing architectural component is the concept of more than one workflow engine which can interact via several concepts. Therefore, actual implementations hardly support interaction between process instances. The minimally needed capability of a WfMS would be to interact with another WfMS to constitute collaborative behavior as is required, for example, in business process choreographies. As proposed in the WfMC standard interaction can be constituted between several workflow engines running on one WfMS, or between workflow engines running on different WfMS (i.e., cooperation between different firms). Furthermore, such a collaboration has to be built on the concept of message exchange which constitutes a technical safe but loose coupling of business processes.

\subsection{S-BPM Workflow System}
\label{s-bpmworkflowsystem}

Previously we have presented an entirely functional S-BPM workflow solution using the Microsoft Windows Workflow Foundation functionality as discussed in~\citep{singer2014} and~\citep{singer2015}. The work has the intention to work out and to define a WF reference architecture based on a concrete implementation. Another aim was to prove the hypothesizes that all practical workflow implementations share a typical software architecture, but the architecture of the workflow engine strongly depends on the underlying modeling concepts. Finally, a reference implementation of the S-BPM methodology should be developed.

We present an architecture which overcomes restrictions of traditional workflow architectures using concepts from software development and the idea of subject-orientation as modeling paradigma. We have started with a more conventional approach focusing on cloud architectures, and after that, we used a micro service-oriented approach~\citep{geisriegler2017}. Beside such proven architectures we also propose to investigate the possibility to build a compiler to translate business process models into executable code~\citep{singer2016}, which has also been proposed by Prinz and Kretzschmar~\citep{prinz2015a}.

\subsubsection{Architecture 1: Windows Workflow Foundation on Azure}
\label{architecture1:windowsworkflowfoundationonazure}

Previously we have presented an entirely functional S-BPM workflow solution using the Microsoft Windows Workflow Foundation functionality as discussed in~\citep{kotremba2013}~\citep{rass2013}~\citep{singer2014}~\citep{singer2015}, and~\citep{kotrembagit}. The designed architecture has full enterprise functionality and supports the execution of inter- and intra-company business processes (i.e., collaborations, choreographies). Nevertheless, the developed prototype is still based on the concepts of traditional workflow systems---e.g., one web server for all process instances---and needs many infrastructure components as a functional backbone.

\begin{figure}[htbp]
\centering
\includegraphics[keepaspectratio,width=100pt,height=0.75\textheight]{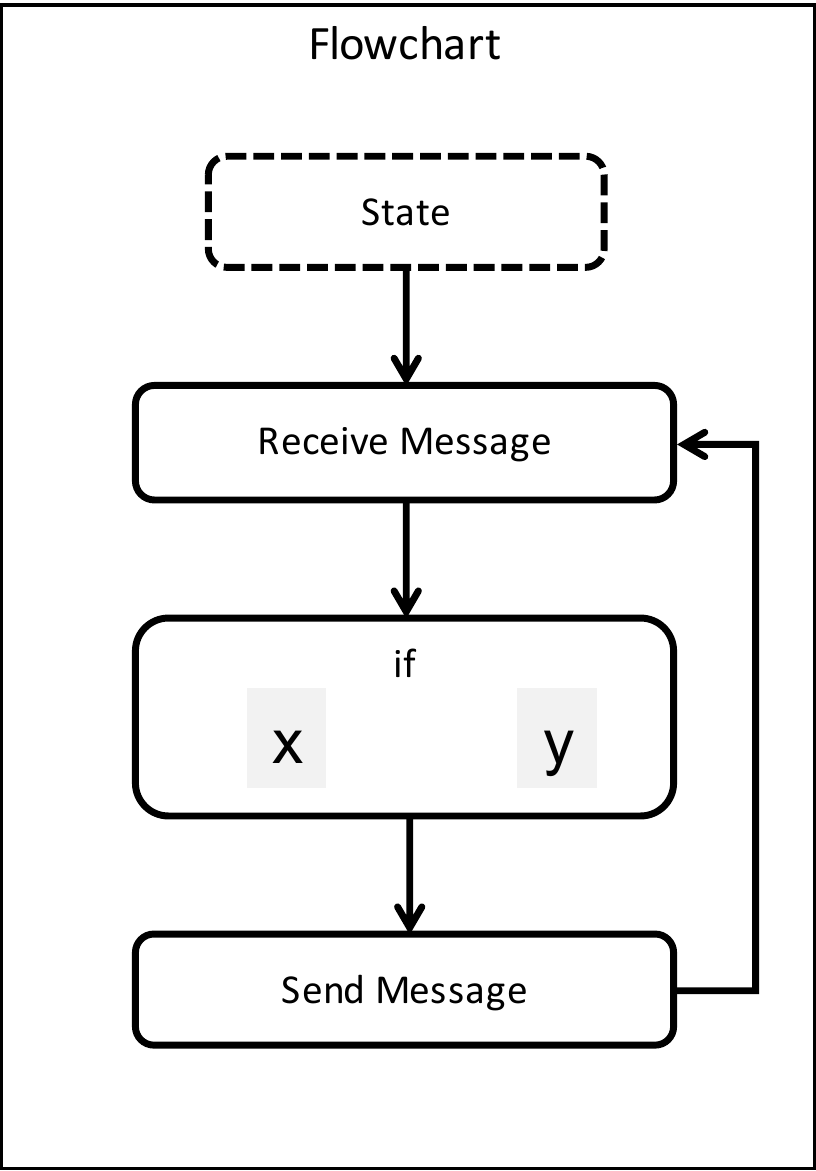}
\caption{The structure of a WF workflow; all work is done by activities (modified from {[4]}), and process flow can be routed back to previous activities.}
\label{dotnet}
\end{figure}

We have chosen Microsoft Windows Server as a platform that provides an easy install of the Microsoft .NET framework, which includes the Microsoft Windows Workflow Foundation (WF)~\citep{chappel2009} as a workflow engine. A WF workflow provides functionality to maintain state, gets input from and sends output to the outside world, provides control flow, and executes code---this is done by so-called activities. Each workflow has an outer activity that contains all of the others (a Sequence or a Flowchart). A Flowchart, like other Activities too, can include variables that maintain its state and can contain other Activities. Each Activity in a WF workflow is a class. The WF runtime performs the execution of the workflow. The runtime does not know anything about the internal structure of an Activity but knows which Activity to run next. It can be easily seen, that WF is based on the Actor concept and directly supports a subject-oriented (resource based) approach, as discussed in previous sections (see \autoref{dotnet}).

Furthermore, we need an infrastructure which can be used by more than one company to define and execute integrated business processes crossing organizational boundaries. That means we have to create an architecture which does not run on only one company's server; from a technical point of view, this means that processes running on the infrastructure of one company need to interact with processes running on the infrastructure of another company. Other requirements are: the platform needs to be scalable; that means it must be capable of handling processes with a small and a large number of instances and transactions per time frame; there must be a security concept which allows fine granular steering of user rights and visibility of business process models or instances, and access to data (business objects), for example.

\begin{figure}[htbp]
\centering
\includegraphics[keepaspectratio,width=250pt,height=0.75\textheight]{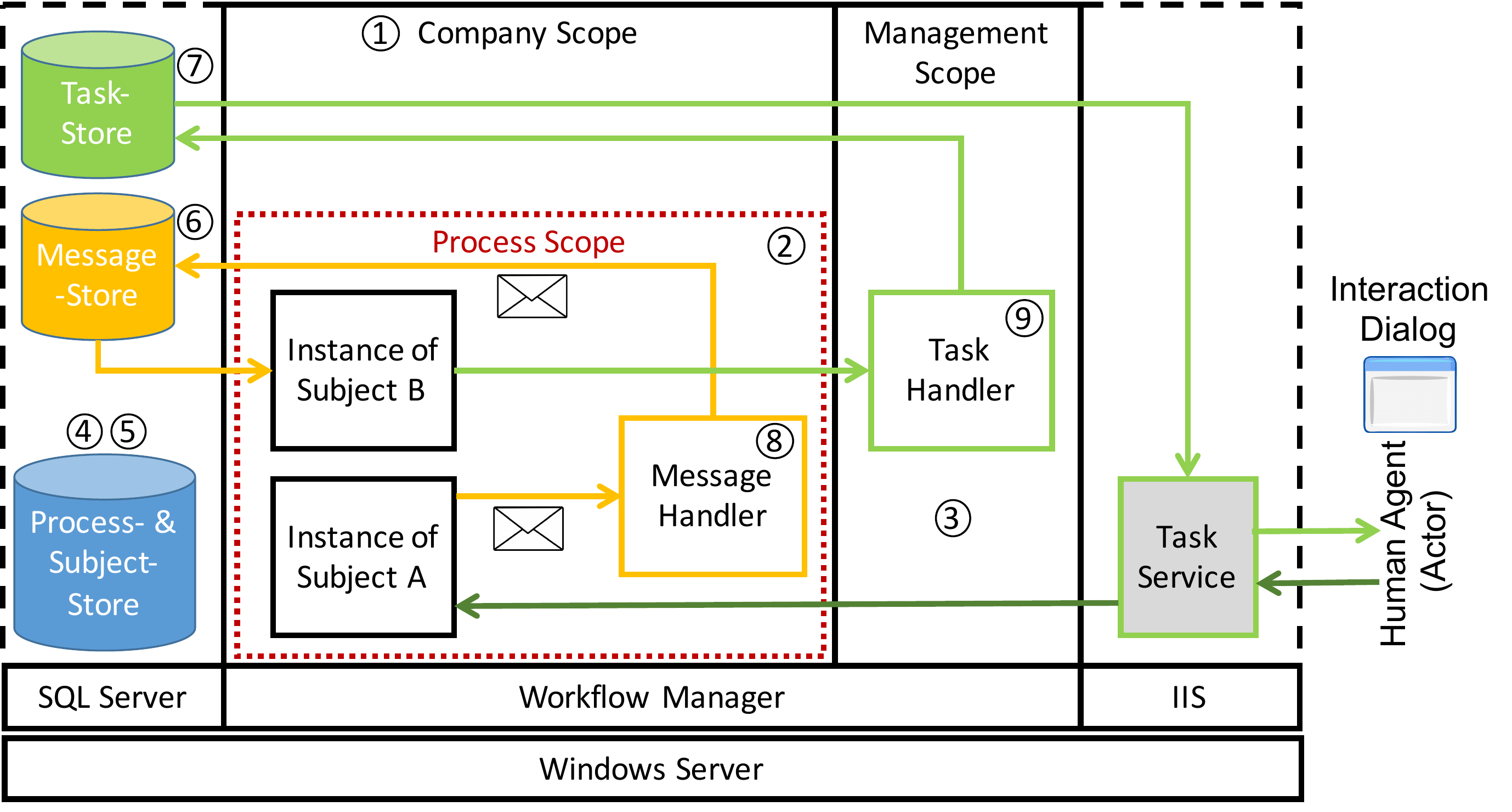}
\caption{An S-BPM WfMS based on the Microsoft technology stack~\citep{singer2015}.}
\label{inflow_architecture}
\end{figure}

We believe that the only way to implement a Multi-Enterprise Business Process Platform is to move a WfMS into the cloud~\citep{singer2015}. The architecture of a commercial implementation is depicted in Figure\autoref{inflow_architecture}. Processes are hosted on an instance of the Workflow Manager (WFM), which is responsible for the hosting, administration and configuration of the subjects based on so-called scopes (a mechanism of the Windows Workflow Manager to support multi-tenancy), such as a Company Scope (1) for the processes of one organization, a Process Scope (2) for each process and a Management Scope (3). Each company has its own Process Store (4) and Subject Store (5); the same for Message Store (6) and Task Store (7). Each company has Task Handler (9) instances to generate new tasks, and each process has Message Dispatcher (8) instances to manage message exchange. Task and Message Handler are implemented as workflows itself. The mechanism of Scopes ensures full encapsulation of one company or organization by the other. Further, it allows rights management on a very fine granular basis for each activity; depending on the rights of a role, activities can be visible or not, and activities can be executed or not.

\begin{figure}[htbp]
\centering
\includegraphics[keepaspectratio,width=250pt,height=0.75\textheight]{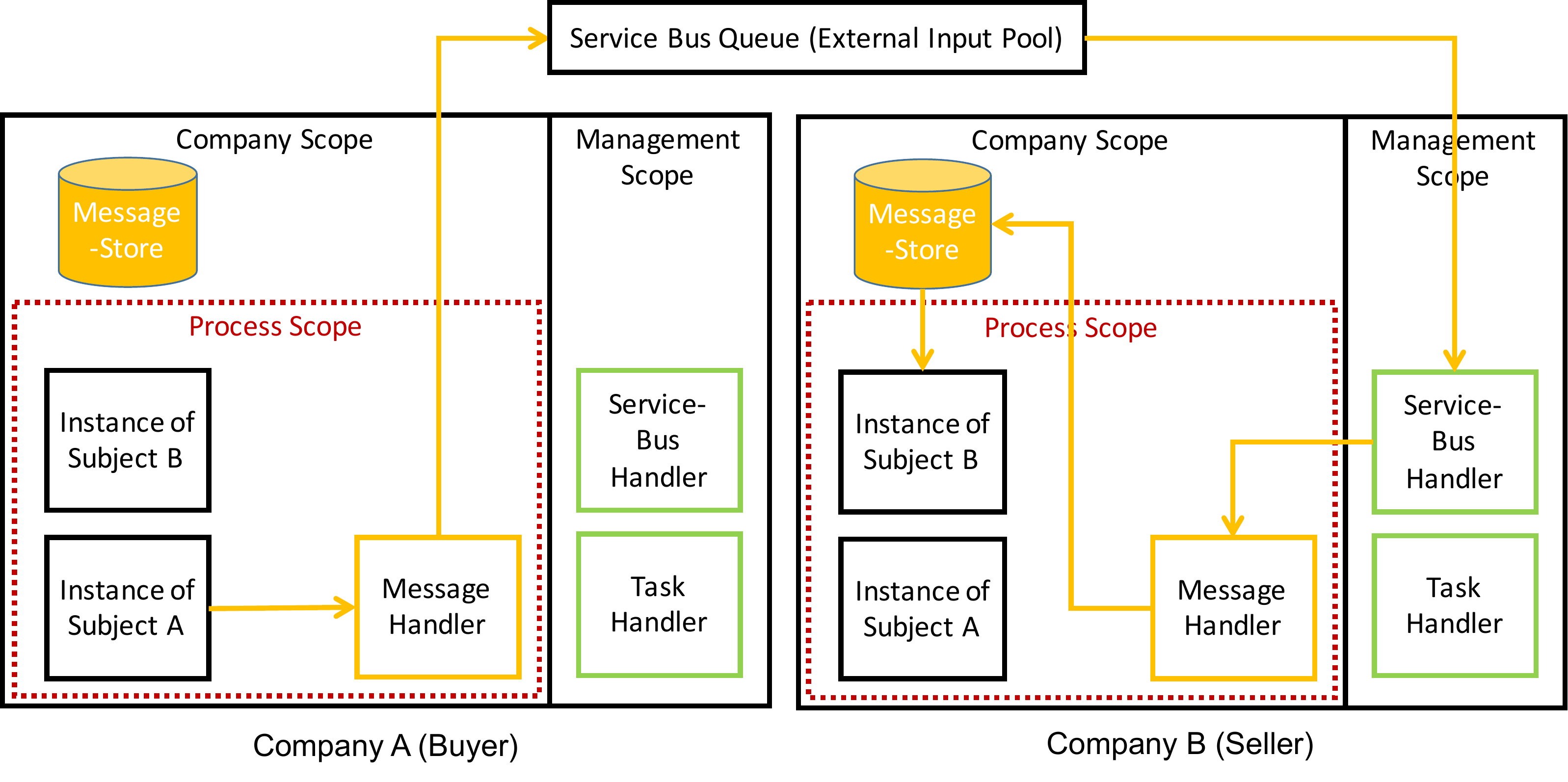}
\caption{Messages can be routed to external process partners via the functionality of a service bus.}
\label{inflow_servicebus}
\end{figure}

The proposed architecture heavily uses key functionality of the MS Workflow Manager (hosting of workflows) and the MS Service Bus (exchange of messages). The service bus provides relay and broker messaging functionalities that enable the exchange of messages between different services (see Figure\autoref{inflow_servicebus}). This architecture supports the exchange of Messages between S-BPM workflow nets.

Based on this architecture we could demonstrate, that it is possible to map any SID\slash SBD (that means all defined S-BPM constructs) on a WF workflow. We could further prove, that all Service Interaction Patterns~\citep{barros2005} can be modeled and executed with our S-BPM prototype~\citep{rass2013} (S-BPM is also capable of modeling and implementing, in general, all Workflow Control Patterns as discussed in~\citep{graef2009}).

\subsubsection{Architecture 2: Microservices and Akka}
\label{architecture2:microservicesandakka}

The architecture presented in the preceding section executes business processes defined as S-BPM model which are designed on the Microsoft Visual Studio development platform as Windows Workflows. Recently the S-BPM standard, which defines the S-BPM modeling syntax as an OWL ontology and the execution semantics via Abstract State Machine code, has been proposed~\citep{borgert2019}.

This section includes basic information about the developed execution platform to support such a business process standard. Furthermore, the presented architecture can be understood as reference architecture for WfMS; it will be shown that it is a general architecture which is principally independent of the modeling notation as the workflow engine is designed as microservice~\citep{ippr16}~\citep{ebusa19}.

In the developed platform process models are defined independently from the execution platform and can be uploaded for execution as OWL files. Nevertheless, one problem of all architectures remains in the actual prototype: business objects still need to be defined on the execution platform yet. This is an implementation detail but not a fundamental restriction of the architecture itself.

Firstly, the utilized technologies for the creation of the various services within the execution platform are shown, and secondly, the single services are described in more detail.

\begin{figure}[htbp]
\centering
\includegraphics[keepaspectratio,width=250pt,height=0.75\textheight]{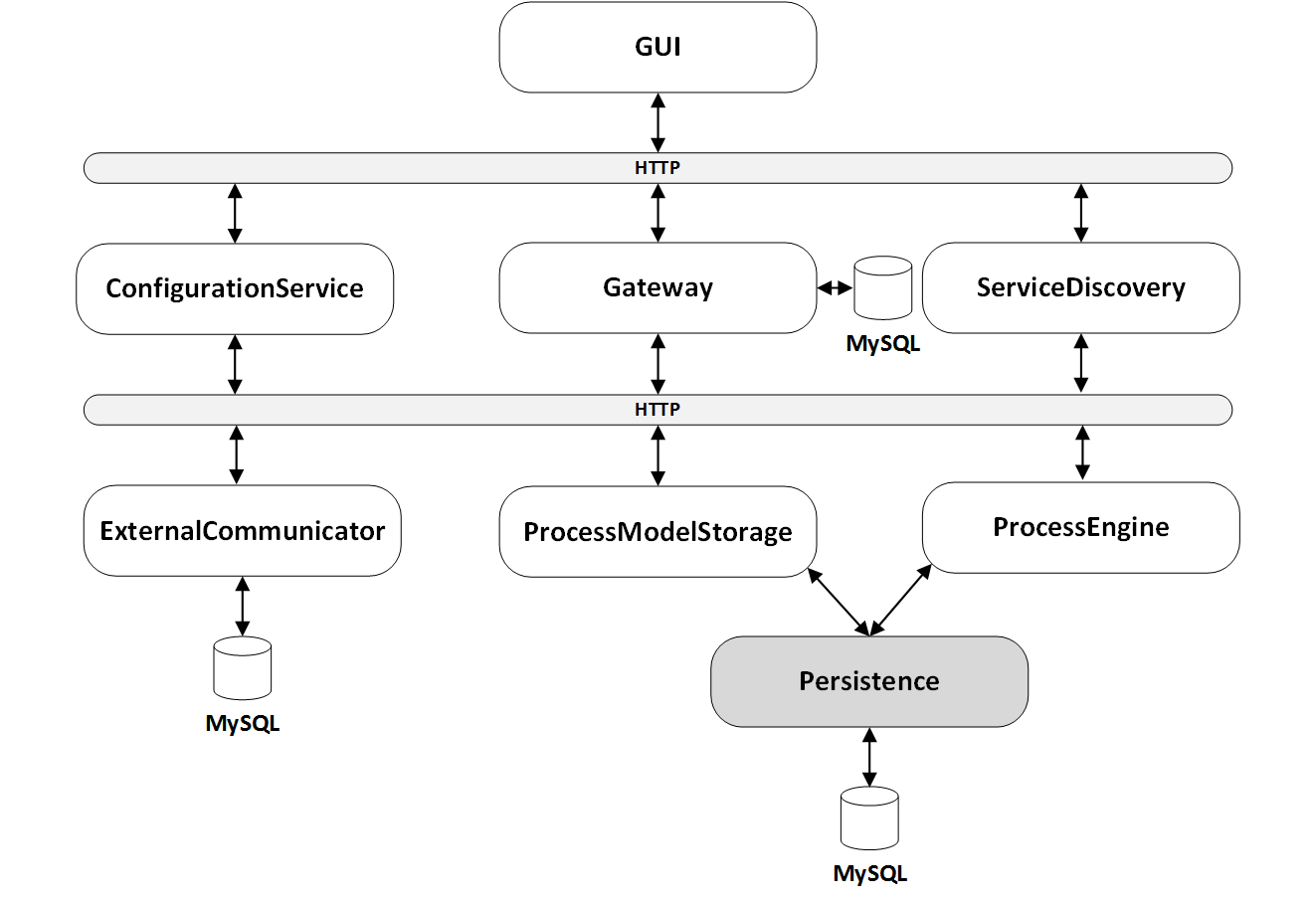}
\caption{The architecture of the prototype: a collection of microservices~\citep{geisriegler2017}.}
\label{architecture_big_picture}
\end{figure}

The core of the execution platform is built on the following technologies:

\begin{itemize}
\item Spring Boot is the dominant technology in the execution platform. This technology enables the easy creation of standalone applications based on the Spring technology stack. Furthermore, it includes an embedded web server, e.g., Tomcat in this case. So, the application can just run standalone. Due to Spring Boot, the creation of separate services is simplified. The data interfaces between the services are based on Representational State Transfer (REST) and JavaScript Object Notation (JSON). Therefore, platform independent communication between different services and external systems is possible. Since the User Interface (UI) only depends on the JSON data, that is retrieved via Asynchronous JavaScript (AJAX) calls on the REST interfaces, it is possible for everyone to build an own graphical interface for the developed S-BPM process engine.

\item For the storage of the process data, a MySQL database is used. In general, the designed database model can be divided into two parts. (1) The repository stores the process model. The process model contains all information (For S-BPM: subject models, states, transitions, business object models, for example) to invoke a process. (2) The repository further persists the data of a running process. For example, a user chooses a process, which is stored as a model, to start. After that, all the instances, e.g., process instance, business object instances, to run a process, is generated.
-The mapping of the database tables and the Java objects is done by using Hibernate. This object\slash relational mapping (ORM) framework is using metadata that describes the mapping between the classes of the application and the schema of the SQL database.

\end{itemize}

The underlying architecture depends mainly on four different services as depicted in \autoref{architecture_big_picture}: (1) Gateway, (2) Process Model Repository, (3) Process Engine and (4) Management UI, which provides a graphical user interface to the Process Model Repository and the engine. Persistence is a separate project, that is imported into Process Model Storage and Process Engine, to enable the database access. The explanation of the different services is part of the next sections. Such a design allows for the loose coupling of services.

\textbf{Gateway}---this service acts as a router to forward requests or responses to the correct recipient. Additionally, the Gateway provides an authentication service to use external authentication providers like Active Directory. Thus, this service maps and persists the users of the external authentication provider. Users can have one or more roles, and each role can be assigned to one or more rules (permissions). Our authentication principle follows an extended role based access control model (RBAC with rule groups). The assignment of subjects or subjects that can start processes relies on the allocated rules. All authentication requests are forwarded to an external provider, which accepts or rejects the authentication request. Since nearly all companies utilize authentication providers, this approach offers several advantages. Users can log in on the UI platform and authenticate themselves via so-called JSON Web Tokens (JWT). JWT is an open standard for securely transmitting information between parties as a JSON object. After providing the credentials to a REST interface, the user gets a JWT that will be included in every subsequent request. This token will be saved in the local storage of the browser, which enables a stateless authentication mechanism.

\textbf{ProcessModelStorage}---the primary responsibility of the Process Model Storage is the persistence of process models. Therefore, different parsers are part of this service. These parsers transform for example OWL files and store the results of this transformation in the database. Consequently, all information to run a process is persisted with this service. The Process Engine uses the data for the execution of processes to retrieve the next states, business object models, for example. An OWL file, which describes an S-BPM process model according to the semantic specifications of the S-BPM standard can be imported. In the first step the process model, the subjects, their behavior, the messages, and the message flows will be parsed, JSON serialized and sent to the user. In the second step, the user has to provide some information for the process model, that is not or cannot be designed in the modeling phase. Not included in the model is, e.g. the linking between and subject models and actual role groups, and the specification of the business objects (BO). In our prototypical solution, business objects can be designed with a maximum of customizability that includes multiple fields, different field types and read and write control for every state. In the last step, the process model is sent back to the Process Model Storage; a new process model is generated and stored in the database.

\textbf{ProcessEngine}---The ProcessEngine is the most extensive service of the execution platform. This service provides features to start, stop processes, and handles the complete workflow of processes. The Process Engine is based on Akka, which is one of the popular Actor Model frameworks. The Akka framework is based on the Actor Model concept. In general, it is an event-driven, middleware framework to build concurrent, scalable, distributed applications. Concurrent programs are split into separate entities that work on distinct subtasks. Each actor performs his quota of tasks (subtasks), and when all the actors have finished their individual subtasks, the bigger task gets completed. Actors can change their state and behavior based on the message passed. This capability allows them to respond to changes in the received messages. In S-BPM subjects change its state depending on the received message. So basically, the concepts of Actor Models and S-BPM are quite similar.

\begin{figure}[htbp]
\centering
\includegraphics[keepaspectratio,width=200pt,height=0.75\textheight]{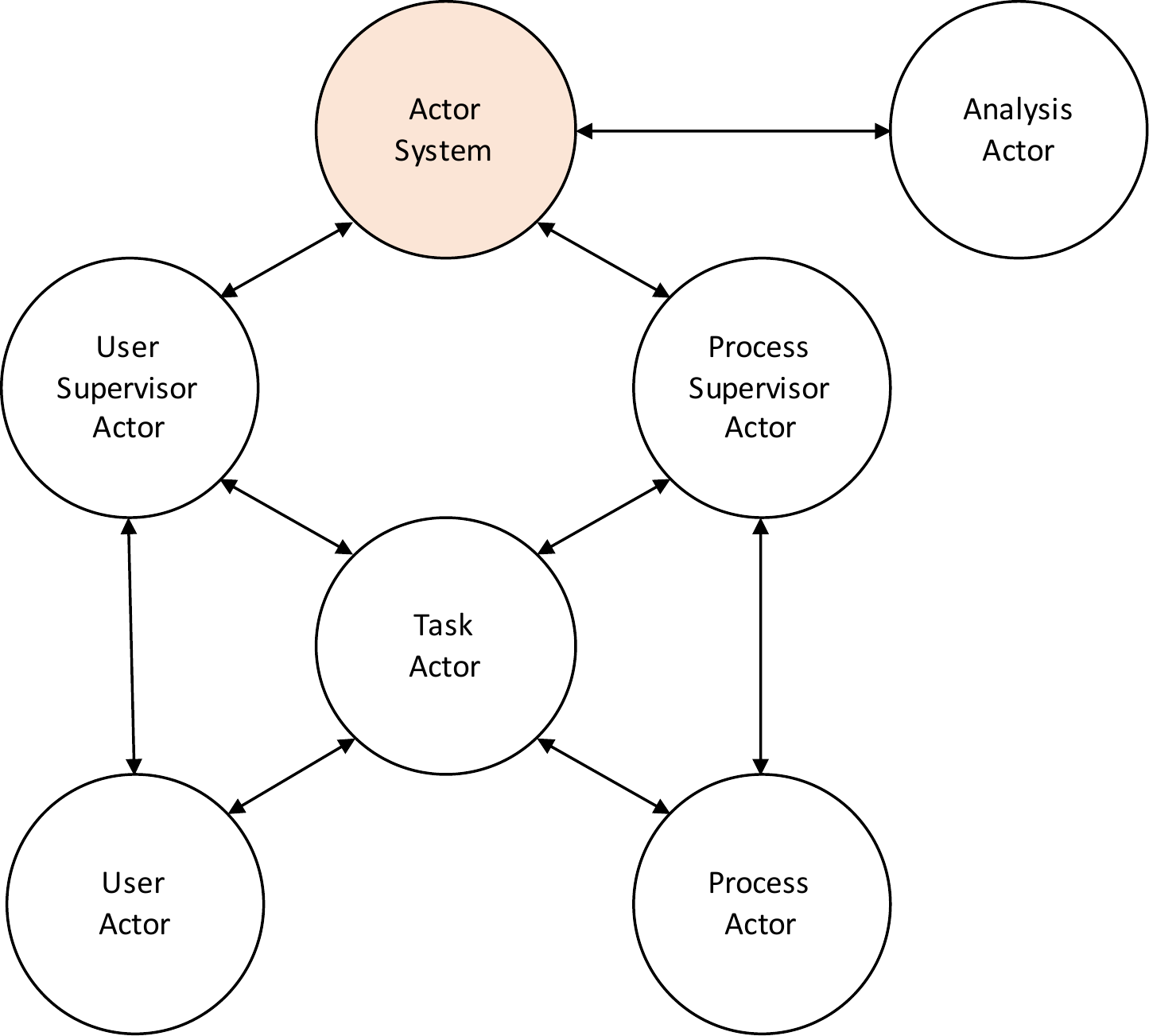}
\caption{The actor system of the workflow engine.}
\label{actorservices}
\end{figure}

Figure \autoref{actorservices} shows the actor system as implemented in the Process Engine. The system can be split into two major parts: (1) The User Supervisor Actor that contains any number of user actors and (2) the Process Supervisor Actor which consists of process actors. This approach implements the Akka supervisor strategy and provides a clear separation of duties between process and user responsibilities. Furthermore, tasks (also an actor) can be executed by supervisor-, user and process actors to outsource complex activities to an independent actor to ensure that other actors are not affected by other tasks. Lastly, the Analyze Actor provides different key performance indicators (KPIs), e.g. some processes finished in a specified time frame.

The responsibilities of each actor are defined as follows:

-Process Supervisor Actor: It is responsible for starting and stopping processes. For each new process, a process actor is created or stopped. Furthermore, it handles process wake up messages, to restart the running processes after a system restart. All messages to the process actors are routed through the Process Supervisor Actor.
-Process Actor: This actor handles all messages to retrieve the current state (e.g., running, finished, canceled) of a process. It provides functions to check whether a process is still active.
-User Supervisor Actor: Initializes new actors for each user based on its user identifier. So, for each user, one user actor is created. This actor also handles wake up messages and acts as a router for the underlying user actors.
-User Actor: It returns the open tasks of one specific user. Additionally, the actor provides state objects for a user in a running process. A state object contains all information what can be done by a user. This information is process specific. So, a state object consists of business objects and returns the next possible states. Furthermore, it handles state object change messages, where the state of one user is changed. The actor also handles retrieved messages from other users.
-Tasks: Task provides a way to outsource complex activities like changing the subject state or storing business objects. User Supervisor Actor, Process Supervisor Actor, User Actor, Process Actor can execute tasks.
-Analyze: Actor which executes SQL queries to retrieve KPIs of the Process Engine.

The process engine is the most extensive service of the execution platform. This service provides features to start processes, stop processes, and handles the complete workflow of processes. The process engine is based on Akka~\citep{vernon2015}, which is an implementation of the actor model framework.

In the Actor Model, all objects are independent, computational units. These units only respond to received messages and do not share a common state. Actors change their state only when they receive a stimulus in the form of a message. So, an actor is a computational entity that, in response to a message it receives, can concurrently~\citep{munish2012}:

\begin{itemize}
\item send a finite number of messages to other actors

\item create a finite number of new actors

\item designate the behavior to be used for the next message it receives

\end{itemize}

Akka is an event-driven, middleware framework to build concurrent, scalable and distributed applications. Concurrent programs are split into separate entities that work on distinct subtasks. Actors can change their state and behavior based on the messages passed.

The actor system can be split into two major parts: (1) The User Supervisor Actor that contains any number of user actors and (2) the Process Supervisor Actor which consists of process actors. This approach implements the Akka supervisor strategy and provides a clear separation of duties between process and user responsibilities. Furthermore, tasks (also an actor) can be executed by the supervisor, user, and process actors to outsource complex activities to an independent actor to ensure that other actors are not affected by other tasks. Lastly, the Analyze Actor provides different key performance indicators (KPIs), e.g. some processes finished in a specified time frame.

The responsibilities of each actor in the workflow engine service are summarized as follows:

\begin{itemize}
\item Process Supervisor Actor: It is responsible for starting and stopping processes.

\item Process Actor: This actor handles all messages to retrieve the current state (e.g., running, finished, canceled) of a process. It provides functions to check whether a process is still active.

\item User Supervisor Actor: Initializes new actors for each user based on its user identifier. So, for each user, one user actor is created. This actor also handles wake up messages and acts as a router for the underlying user actors.

\item User Actor: It returns the open tasks of one specific user. Additionally, the actor provides state objects for a user in a running process. A state object contains all information what can be done by a user. This information is process specific. So, a state object consists of business objects and returns the next possible states. Furthermore, it handles state object change messages, where the state of one user is changed. The actor also handles retrieved messages from other users.

\item Tasks: Task provides a way to outsource complex activities like changing the subject state or storing business objects. User Supervisor Actor, Process Supervisor Actor, User Actor, Process Actor can execute tasks. Tasks can be invoked by the usage of the task manager.

\item Analyze: Actor which executes SQL queries to retrieve KPIs of the Process Engine.

\end{itemize}

The main features of the supporting graphical user interface (UI) are:

\begin{itemize}
\item User authentication

\item Starting and executing processes

\item Visualization of KPIs

\item Import of process models via OWL files

\end{itemize}

The UI platform communicates and interacts with the Gateway via REST calls. This loose coupling allows an easy replacement of components, as long as the REST interfaces remain unchanged. On the one hand, the Process Engine or the Process Model Repository could be replaced without changing the UI platform. On the other side, the UI could be replaced easily as well.

\subsection{Conclusion}
\label{conclusion}

Subject-orientation is a changed view on how to model a business process, which is based on modern views on software development, as discussed previously, and foundational theories from social sciences. We define a business process as an entity which determines how we plan to provide services or products to customers. Furthermore, if more than one acting person (the subject or actor) is involved in a process, we have to coordinate the work. Coordination is done by communication, either between humans and machines.

That means any business process can be seen as an exchange of messages between actors. No doubt, this can be modeled in BPMN; nevertheless, the standard does not provide a practical communication concept, such as messages which can be sent to more than one actor, or a message pool concept to define how to handle asynchronous messaging. Well known and understood concepts in software development.

Subject-orientation, which is the actor concept applied on business process management, inherently supports an enterprise architecture designed on reactive principles as stated in the Reactive Manifesto~\citep{boner}; the reactive principles are not restricted to the implementation of the enterprise architecture, but it also supports a reactive design on all levels of an organization: the so-called business-IT gap dissolves.

Now, building a WfMS based on the actor system concepts prooves, that there are some architectural concepts independent from the modeling notation as depicted in \href{}{}\footnote{\href{}{}}. Furthermore, based on modern design principles a WfMS can be designed which is independent how business processes are defined as long as the workflow engine as its core execution mechanism can be changed what is the case if the architecture is based on microservice principles, for example. A precondition is that the full business process model can be defined independently from the WfMS as required by the reference architecture of the WfMC. This is not the case in all known WfMS as the design of the data models representing the business objects generally is done within the workflow engine; that also means that all form (however implemented) have to be defined within the workflow engine as they rely on the underlying data model.

In the subject-oriented architectures presented in the previous chapters the data model is also defined in the workflow engine, but it is not a requirement of the architecture but just more comfortable to code and to get a more compact prototypical implementation with less effort.

\section{Business Process Ontology}
\label{businessprocessontology}

Several researcher have proposed process related ontologies~\citep{haller:2008aa}~\citep{hofferer:2007aa}, and to transform process models to ontologies~\citep{coskuncay:2017aa}~\citep{thomas:2009aa}, as is also the topic of this work.

An ontology is a graph-based model, which describes data objects (also called resources) in more detail because of their relationships to each other. Because of this structure, an ontology can also be seen as a Resource Description Framework\} (RDF). RDF is a standardized model, which, due to the linked structure, links resources with each other to enable data exchange (on the Web). Ontologies extend the concept of RDF through a more detailed description of the links.

Since ontologies are based on the RDF concept, the information can be read out of an ontology using the standardized query language SPARQL. The syntax for data access via SPAQRL is similar to that of RDF and identifies individual resources via their IRI (Internationalized Resource Identifier) which are an abstraction of URIs (Uniform Resource Identifiers)---and therefore conform to URIs and URLs.

That means that the developed BPMN 2.0 ontology can also be used in the course of process modeling as a machine-workable reference guide for the correct use of BPMN symbols. The query language SPAQRL allows information to be queried directly for specific symbols and does not need to be read out manually from the standard. With the individually definable query via SPARQL, for example, in addition to the correct syntax (restrictions), the formal description of the class can also be loaded directly. With regard to BPMN, a query can be defined for a symbol which, in addition to the description, also returns all restrictions and thus shows the correct use of the symbol.

In addition to the use of ontology as a machine-processable reference work, additional information on process models can also be obtained via the graph-based data structure. For example, a process model can be tested for the use of a specific compliance class if it is properly stored in the ontology (based on the properties of the class). As a result, additional information about an existing process model can be generated and thus a higher quality with regard to the information content in process models can be achieved.

By depositing the standard as ontology, the resulting machine-processable data structure can be read in from a wide variety of applications and used for specific purposes (review, analysis, knowledge processing, {\ldots}). For example, the information from the ontology for process modeling can be integrated via an application and used for automated documentation of the process model. Another possibility would be to use the BPMN 2.0 ontology as a source of information to answer text-based questions for process modeling in BPMN.

Another aspect of using ontology as a source of information to increase the quality of process modeling is the use of ontology as a verification tool. Since process models are stored in XML as BPMNs, syntactic validation using an XML schema is possible, but requires the schema to match correctly and to fully contain the syntactic constraints with the documented BPMN standard. Due to the technical specification, this is not possible with the help of XSD, as only a simple inheritance of restrictions is allowed.

Unlike an XSD schema, an ontology allows retrieval of multiple inheritance restrictions, allowing more accurate syntactic verification of the process model. In addition, in the event of an error, a detailed error message can be reported back through the stored semantic description, as a result of which the error can be eliminated more quickly.

Another advantage of ontology is the possible combination of syntactic restrictions and formal description into a single data structure. For BPMN the PDF contains the formal description and the XML schema the implemented rules. By providing and maintaining the OMG, it can be assumed that both files correlate with each other, but in the case of customization, it must be ensured that both files receive the changes. In an ontology both information (description and rules) are contained in a file, whereby only one-sided adaptation of the ontology is necessary.

\subsection{Ontology}
\label{ontology}

Ontology is the philosophical study of being. It deals with what we call ``reality.'' This is congruent with the traditional terminology of general metaphysics. The study of reality or metaphysics has a long philosophical tradition; a concise overview can---for example---be found in the introduction of Bunge~\citep{bunge1977}.

Bunge {[ibid]}, furthermore, writes that the modern version of a general theory of objects, namely \emph{general systems theory}, is constructed as a mathematical theory dealing with the explanations of observed phenomena or conceptual constructs in terms of information processing and decision-making concepts. Indeed, ontology has gone mathematical and is being cultivated by engineers and computer scientists. Now, computer scientists are interested in ontologies which are both exact and scientific. Based on this view and his definition of materialism~\citep{mahner2004} Bunge has developed a formal ontology of the world which has been very influential in computer sciences.

Another very influential definition of ontology is the one from Gruber~\citep{gruber1995}, a work in the context of artificial intelligence (AI) and knowledge sharing. Gruber {[ibid]} writes

\begin{quote}
A body of formally represented knowledge is based on a conceptualization: the objects, concepts, and other entities that are assumed to exist in some area of interest and the relationships that hold among them. Conceptualization is an abstract, simplified view of the world that we wish to represent for some purpose.
\end{quote}

And furthermore, he writes

\begin{quote}
An ontology is an explicit specification of a conceptualization.
\end{quote}

Wand and Weber~\citep{wand1990} start their seminal work ``\emph{An Ontological Model of an Information System}'' with the following---and, we think, still valid---remarks:

\begin{quote}
``The computer science (CS) and information systems (IS) fields are replete with fundamental concepts that are poorly defined.''
\end{quote}

A similar discussion about this topic---for example---can be found in ``The FRISCO Report''~\citep{falkenberg1998}. Based on these and other arguments, we think it is very fruitful to start the discussion with the notion of ontology. Doing this we can try to build a better understanding of a \emph{systemic view} on organizations; this leads us to a better understanding of the notion of model in general and process models in particular.

Business processes offer a \emph{dynamic view} on organizations as they describe the \emph{states} and \emph{state transitions} of a system or the corresponding world. States and state transitions also define the ontological model of a system, as elaborated by Dietz~\citep{dietz2006}: ``The ontological model of a world consists of the specification of its state space and its transition space.''

In the process of requirements engineering for information systems, we are confronted with the need to represent the requirements in a conceptual form. Often, however, they do not possess an underlying conceptual structure on which to base such models. As already discussed, real-world systems can be explained and described using ontology---the study of the nature of the world and attempt to organize and describe what exists in reality.

Wand and Weber~\citep{wand1993}~\citep{wand1995} suggest that the theory of ontology can be used to help define and build information systems that contain the necessary representations of real-world constructs, including their properties and interactions. Hence, they developed and refined a set of models based on an ontology defined by Bunge~\citep{bunge1977} for the evaluation of modeling techniques and the scripts prepared using such techniques. These models are referred to as Bunge-Wand-Weber (BWW) models.

The BWW representation model is one of three theoretical models defined by Wand and Weber~\citep{wand1995} that make up the BWW models. Its key constructs can be grouped into four clusters: things including properties and types of things; states assumed by things; events and transformations occurring on things; and systems structured around things. For a complete description of the BWW constructs, please refer, for example, to~\citep{weber1997}.

Weber~\citep{weber1997} suggests that the BWW representation model can be used to analyze a particular modeling technique to make predictions on the modeling strengths and weaknesses of the technique, in particular, its capabilities to provide complete and clear descriptions of the domain being modeled. He clarifies two main evaluation criteria in representational analyses that may be studied according to the BWW model: ontological completeness is indicated by the degree of construct deficit, i.e., the extent to which a modeling technique covers completely the constructs proposed in the BWW representation model. Ontological clarity is indicated by the degrees of construct overload, where one language construct covers several BWW constructs, construct redundancy, where one BWW construct maps to several language constructs and construct excess, where language constructs exist that do not map to any BWW construct.

\subsection{Ontology Web Language (OWL)}
\label{ontologyweblanguageowl}

In the early nineties, ontologies have been used to model knowledge to create knowledge-based systems. Afterward, ontologies have been used to develop ontology-based applications. The underlying idea is to gather all concepts of a particular area. In addition, all properties and relationships are also taken into account. Ontologies are nowadays used in the semantic web and typically are stored in a web ontology language; in this work, we use the Ontology Web Language (OWL) developed and maintained by the World Wide Web Consortium (W3C). All ontologies in this work are developed with the software protégé which can be downloaded from the Web\footnote{https:\slash \slash protege.stanford.edu}.

An ontology file starts with the declaration of the namespace (analog to an XML schema definition), and the essential ontology elements are classes, individuals and properties. Individuals are things of the real world and members of classes (i.e., instances of classes). To understand the composition of the presented ontologies a concise explanation as follows.

The root class is owl:Thing and all other classes are attached to this class. Every class is declared as owl:Class has a unique name and can have more than one class as a subclass, which is defined as rdfs:subClassOf. There are two main property categories, which are object and data properties. Object properties are used to link an individual to another one and data properties are used to describe data values of a specific individual. Data properties have unique names and data types. Properties can have constraints, which are defined by owl:cardinality. The cardinality constraint specifies how many relations a specific class has to another class. Relations are realized by an object property. Data properties present data values of classes and their occurrence can also be restricted with owl:cardinality.

\subsection{BPMN Ontology}
\label{bpmnontology}

There are only a few works based on an ontological analysis of BPMN. For example, Recker et al.~\citep{recker2006} presented an evaluation of BPMN 1.x based on the Bunge-Wand-Weber (BWW) ontology and interviews. The ontological evaluation reveals construction deficits, construct redundancies, construct excess and construct overload. One option to proceed with an ontological analysis of BPMN is to develop an ontology of the standard document. That means, to build a real ontology serialized as OWL-file which includes all classes, relations, and restrictions. Based on such a real ontology several analyses can be done. We will demonstrate how such an ontology can be used to compare a specific BPMN model with the standard (compatibility), and how the ontology can be used to transfer a concrete BPMN model into an S-BPM model and back. There are a few attempts to build a BPMN ontology, but as these are not available as OWL files, or are not complete, or are based on previous versions on the BPMN standard, we have decided to develop an own version, which is available on the Web for public reference and further research~\citep{reitergit1}~\citep{reitergit2}~\citep{hoedlgit}.

OWL ontologies are described by entities and individuals, where entities are called fundamental blocks of an ontology and in their sum are considered to be the signature of an ontology. Entities are represented as either class or property and are identified and addressed through IRIs (Internationalized Resource Identifier). So called individuals describe the actual data objects and are distinguished between named individual and anonymous individual. Unlike anonymous individuals, named individuals have an IRI and can be addressed directly. An OWL class possibly represents a lot of individuals. Via axioms, statements about the entities can be made, for example, the inheritance hierarchy of the class.

Properties define either the ontology (AnnotationProperty), individual objects (ObjectProperty), or data (DataProperty) in more detail. While data properties describe a data field in more detail, object properties reference and describe an entity more closely. The annotation property (AnnotationProperty) can be used to provide an ontology with additional information. Annotations, for example, add information about the creator of the ontology or describe the ontology itself.

\begin{figure}[htbp]
\centering
\includegraphics[keepaspectratio,width=250pt,height=0.75\textheight]{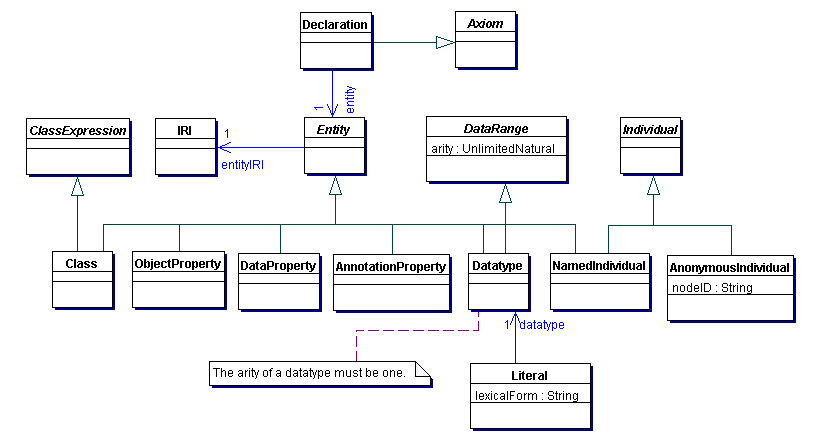}
\caption{The construction of an ontology.}
\label{aufbauontologie}
\end{figure}

In order to create a complete BPMN ontology, it is necessary to understand the relations between the components of BPMN models and OWL ontologies: BPMN elements, attributes, instances, inheritance, and relationships between classes or properties matches with the OWL concepts class, object and data property, named individual, hierarchy, and axioms respectively (see \autoref{aufbauontologie}).

\begin{figure}[htbp]
\centering
\includegraphics[keepaspectratio,width=200pt,height=0.75\textheight]{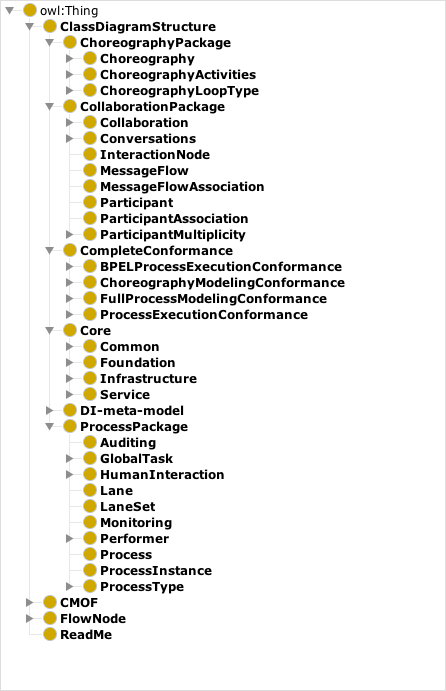}
\caption{Structure of the developed BPMN ontology.}
\label{bpmn_owl_1}
\end{figure}

The BPMN ontology has been developed manually by analyzing the standard document, the provided set of XML schema files, and analyzing serialized models from several modeling tools. Two teams conducted the analysis; each developed independently from the other group a complete BPMN ontology. Afterward, the two solutions were systematically compared, differences where discussed, and the final ontology was adapted accordingly. Later we will present further systematical checks and quality improvements of the ontology. The fundamental structure of the BPMN OWL-file is depicted in \autoref{bpmn_owl_1}.

\begin{figure}[htbp]
\centering
\includegraphics[keepaspectratio,width=250pt,height=0.75\textheight]{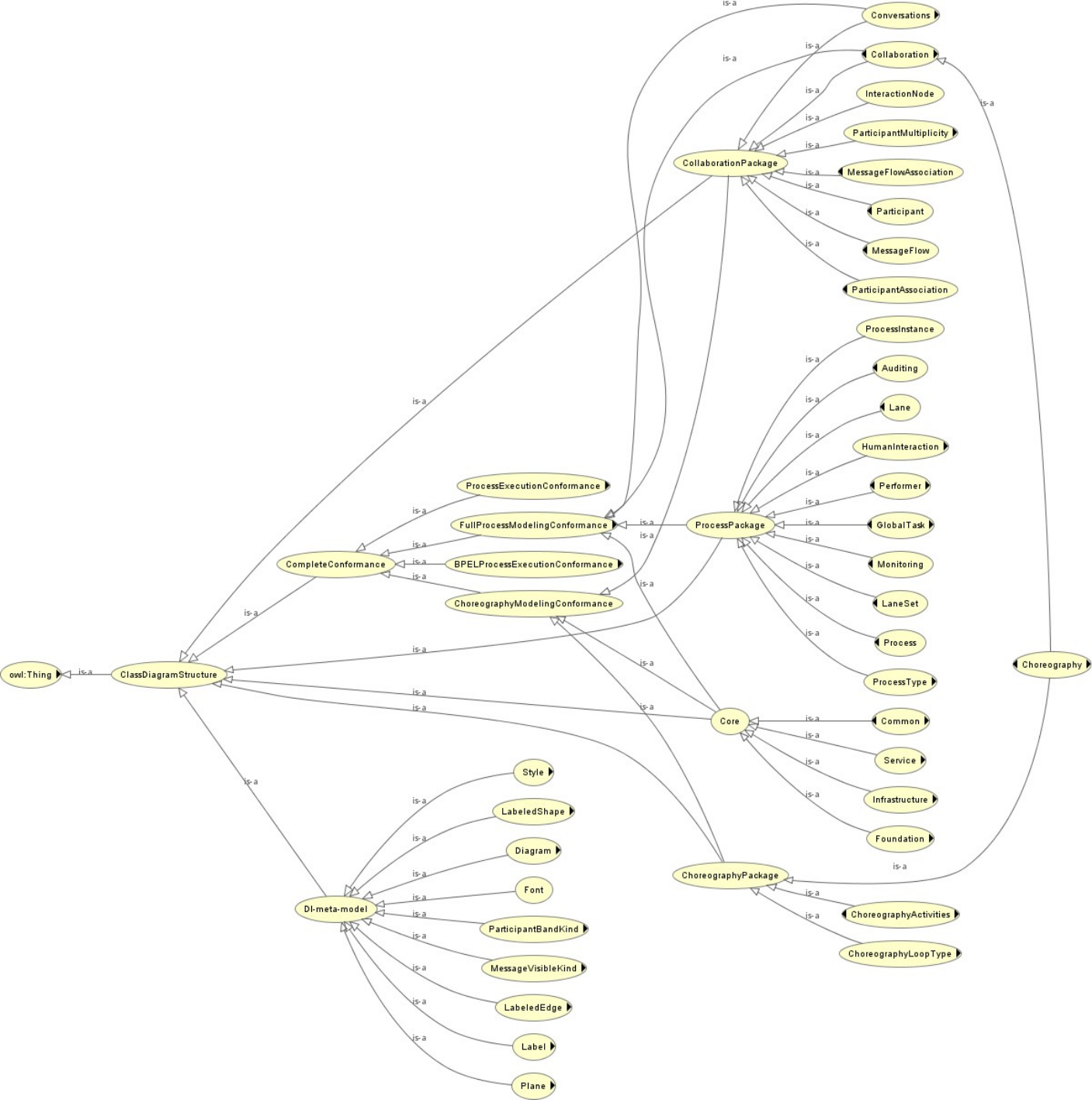}
\caption{Visual representation of the basic elements of the BPMN ontology.}
\label{bpmnvisual}
\end{figure}

\begin{figure}[htbp]
\centering
\includegraphics[keepaspectratio,width=250pt,height=0.75\textheight]{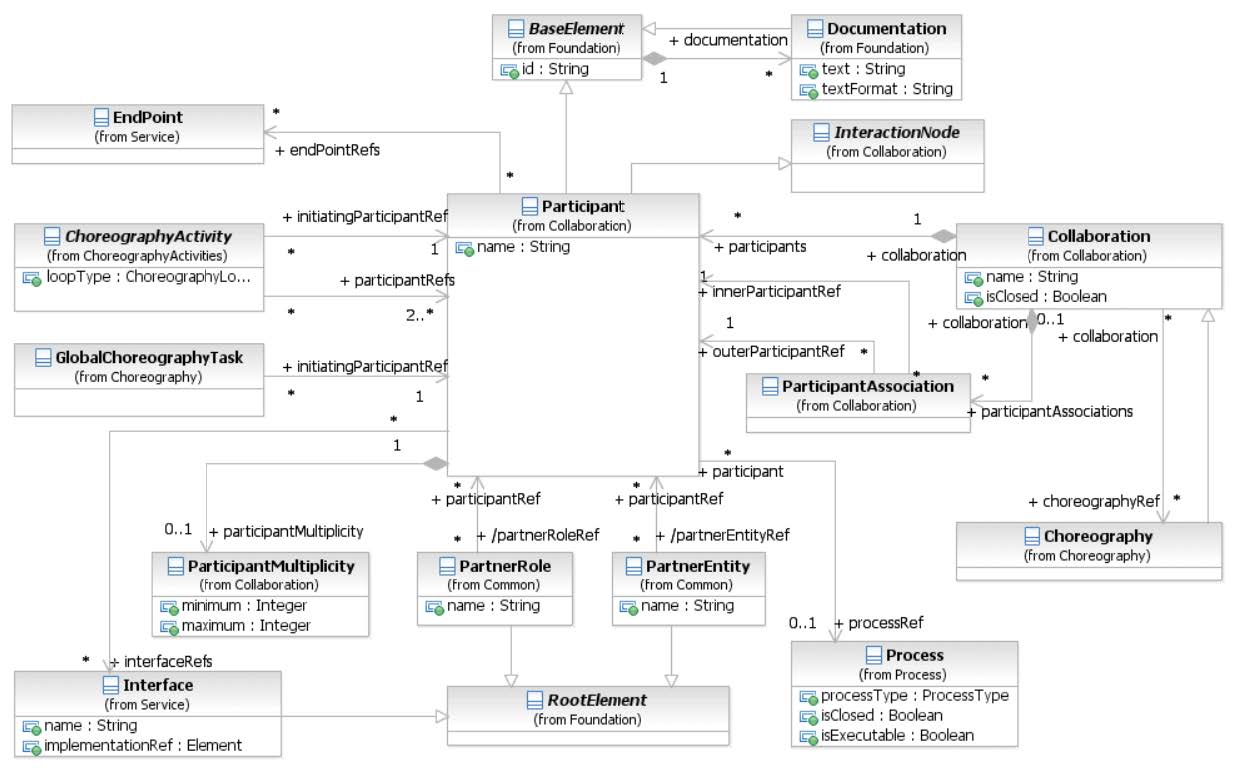}
\caption{The participant class diagramm in UML notation~\citep{omg2013}.}
\label{participantuml}
\end{figure}

All nodes, which are shown in \autoref{bpmn_owl_1} are classes. The top-level class is owl:Thing and contains the two child classes ClassDiagramStructure and CMOF. The ClassDiagramStructure contains the core packages of the BPMN 2.0 specification. The core packages are ChoregraphyPackage, CollaborationPackage, CompletePerformance, Core, DI-meta- model and ProcessPackage. All of these packages have a certain variety of different child classes. As an example, \autoref{participantuml} shows the class diagram of the participant class. A Participant belongs to the CollaborationPackage and has a name of type String and a Participant inherits all model associations and attributes from the BaseElement. The ontology is built of 270 classes, 176 object properties, and 70 data properties, for example. A visual impression is given in \autoref{bpmnvisual}.

\begin{figure}[htbp]
\centering
\includegraphics[keepaspectratio,width=250pt,height=0.75\textheight]{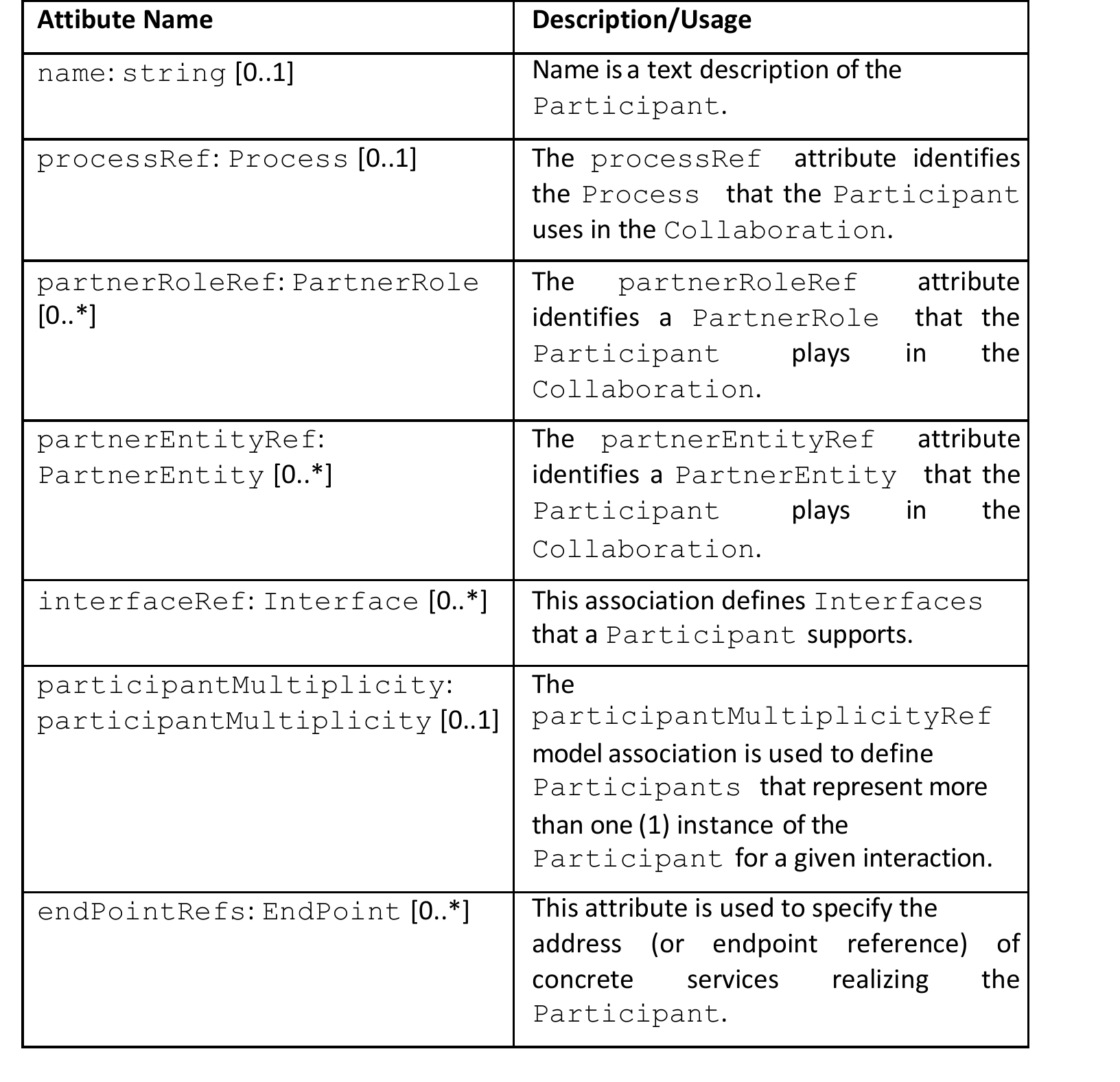}
\caption{Participant attributes and model associations~\citep{omg2013}.}
\label{participantbpmntable}
\end{figure}

\autoref{participantbpmntable} shows all attributes and model associations of the Participant element. Possible attributes for a Participant are name, processRef, partnerRoleRef, partnerEntityRef, interfaceRef and participantMultiplicity.

\autoref{bpmn_owl_1} shows the Participant class in the BPMN ontology hierarchy. The Participant class is part of the CollaborationPackage.

\begin{figure}[htbp]
\centering
\includegraphics[keepaspectratio,width=200pt,height=0.75\textheight]{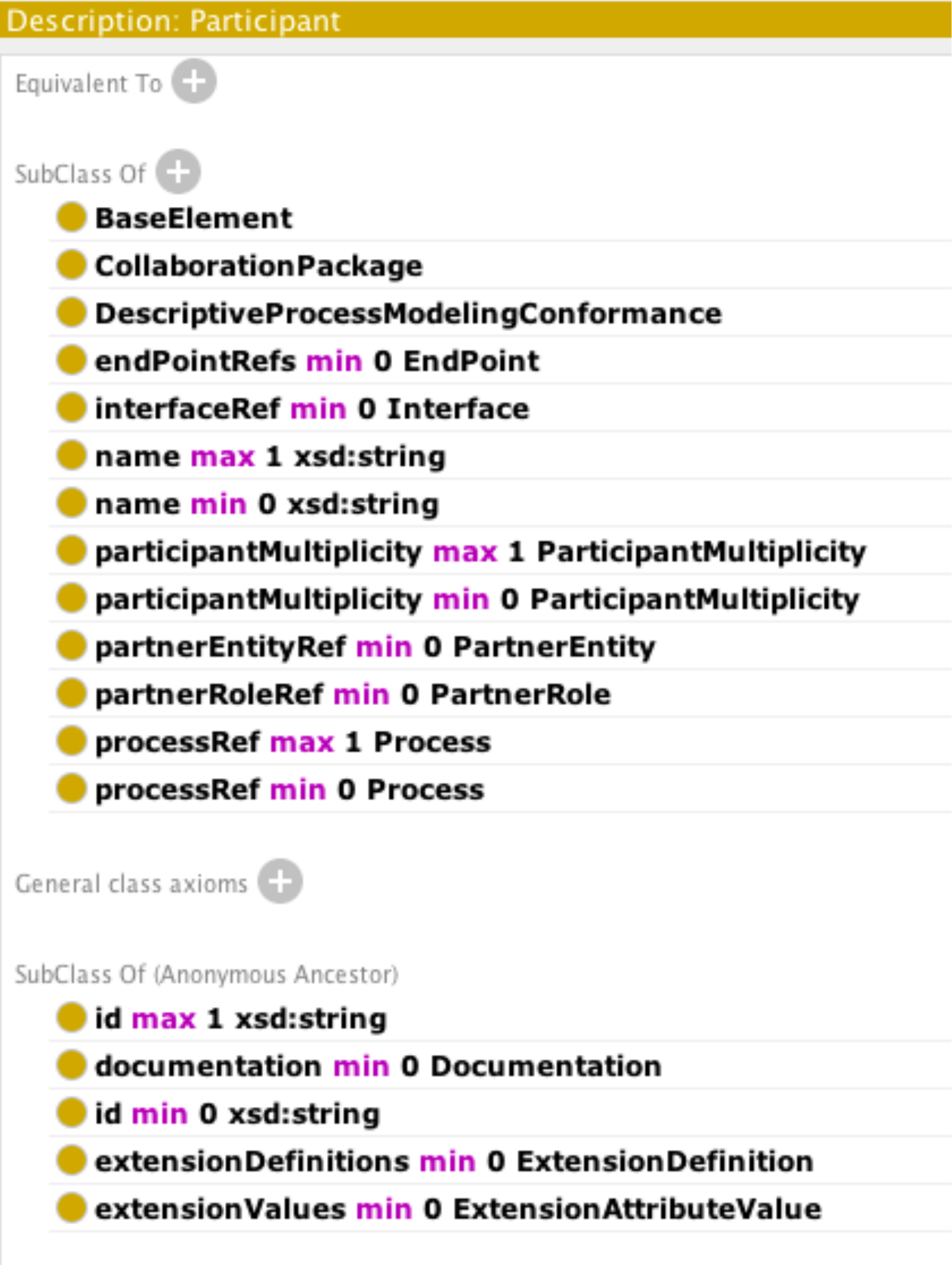}
\caption{BPMN ontology description of the participant.}
\label{ontparticipantdetail}
\end{figure}

\autoref{ontparticipantdetail} shows the description of the Participant class. A Participant can have 0 or 1 name of type string. A Participant can have 0 or more endpointRefs, which refer to EndPoints and 0 or more interfaceRef, which refer to Interfaces. A Participant can have 0 or 1 participantMultiplicity, which refer to ParticipantMultiplicity and 0 or 1 processRef, which refer to Process. A Participant can have 0 or more partnerEntityRef, which refer to PartnerEntity and 0 or more partnerRoleRef, which refer to PartnerRole.

The Participant class is one example of a BPMN element. The developed BPMN ontology contains all BPMN elements with their attributes and model associations. All references to other elements and all inherited attributes are included in the BPMN ontology.

Because of the OWA concept described, an ontology for the validity of data must define boundaries that are adhered to by data structures. For example, the valid name of a person consists of first and last name, while this statement is not correct for company names. Therefore, restrictions for classes. Relations and attributes depend on the semantic context.

To define these restrictions for OWL classes in an ontology, constraints must be defined. An ontology distinguishes limitations on the data type (datatype restrictions) and restrictions on classes and object properties (object property restrictions). Restrictions on a data type affect the content of the data object and restrict it (for example, to numbers). Restrictions on classes and object properties affect their intended use.

\begin{figure}[htbp]
\centering
\includegraphics[keepaspectratio,width=250pt,height=0.75\textheight]{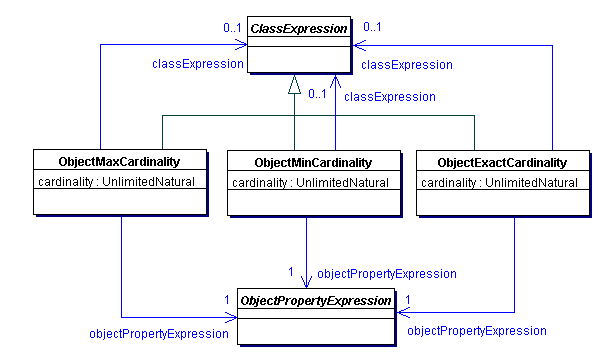}
\caption{UML diagram for restrictions on cardinalities.}
\label{aufbauowlrestriction}
\end{figure}

For restrictions of classes and object properties, the construct ``Class Expression'' is used. Using this construct, all constraints can be defined to limit data to concrete objects. In the BPMN 2.0 ontology, only restrictions related to the cardinality are used, the corresponding structure for this is shown in \autoref{aufbauowlrestriction}.

Cardinalities can be defined either in a range (minimum and maximum) or as an exact value. The restrictions specified in the BPMN 2.0 ontology are used for the verification of process models. Since every XML node of the process model corresponds to a class from the ontology, all (inclusive inherited) restrictions must be fulfilled by it. Only if all restrictions have been met is the assignment valid. Due to the OWA concept of ontology, however, the verification is not bound to XML nodes but is used accordingly in the concrete case of an application (BPMN models as XML pre-existing).

The classical and also logical way to test a BPMN model, serialized as an XML file, is via XML schema verification, but this has some limitations. The advantages of an ontology for the verification of an XML file lies in the independence of the data structure. While XML schema has been specified for the verification of XML documents, an ontology is independent of the structure of the data structure. The verification itself must be done via the application; the ontology serves only as a data source for the review. This flexibility allows versatile use of the ontology, but also requires the correct use of the provided information for verification by the application.

While XML schemas, due to their technical specification, can only perform limited validations of XML elements, an ontology allows a higher degree of detail due to its structure. This is reflected, for example, in the review of inherited restrictions. While an XML schema can only test for single inheritance, the ontology is unrestricted.

Regardless of the verification, one of the advantages of ontology lies in its standardized technical description. For example, data structures of a particular format can be transformed into a different data structure using an ontology. In terms of process modeling, ontology, for example, enables the transformation of process models in BPM notation into another notation language without loss of information as will be demonstrated later in the case of S-BPM.

The completeness and quality of the BPMN ontology have been thoroughly tested, as will be discussed later.

\subsection{S-BPM Ontology}
\label{s-bpmontology}

\begin{figure}[htbp]
\centering
\includegraphics[keepaspectratio,width=250pt,height=0.75\textheight]{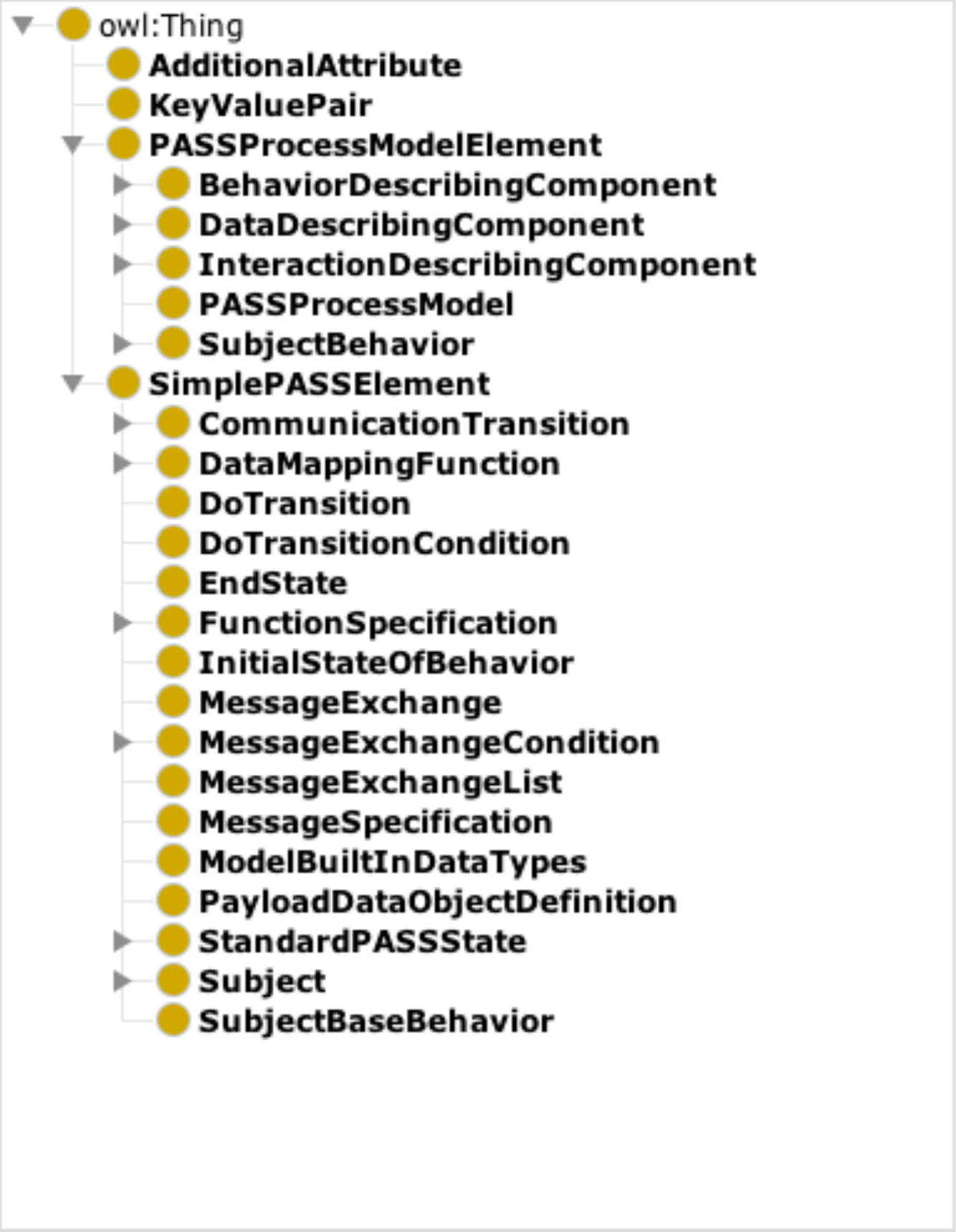}
\caption{Structure of the S-BPM ontology~\citep{elstermann2018}.}
\label{s-bpm_owl_1}
\end{figure}

The S-BPM Ontology (see \autoref{s-bpm_owl_1}), has been mainly developed by Matthes Elstermann~\citep{elstermann2017} based on proposals presented by~\citep{hover2014} and it now is a working draft for a standard document, which is used in order to store S-BPM processes based on the Abstract Layered PASS concept. PASS is the foundational concept of S-BPM and has been developed by Albert Fleischmann~\citep{fleischmann1994}, as discussed later. The S-BPM ontology contains all syntactical and semantical constructs, which are necessary to define a valid S-BPM model ready for execution.

The S-BPM ontology (as OWL file) and a detailed description and discussion can be found on the Web\footnote{https:\slash \slash github.com\slash I2PM\slash } and in~\citep{elstermann2018}.

\section{Implementation}
\label{implementation}

In the following sections, we first will present how the developed BPMN ontology has been validated and verified. Then we demonstrate how BPMN models, serialized as XML files from some modeling platform, can be transformed into valid OWL files. Based on this transformation we will discuss possible advantages of having models stored as ontologies. A concrete application, the conversion of BPMN into S-BPM models, is presented afterward. Finally, we discuss the execution of BPMN models on an actor based WfMS, which we introduced in a previous section. That leads to a general discussion about the architecture of workflow management systems and how such an architecture depends on the underlying concepts of how business processes are defined. Finally, we can discuss the fundamental concepts of business process models based on an ontological analysis of BPMN and S-BPM in the context of business process execution supported by WfMS. Based on the conclusions we will propose a reference architecture for workflow management systems.

\subsection{Transforming BPMN Models (BPMN to OWL)}
\label{transformingbpmnmodelsbpmntoowl}

\begin{figure}[htbp]
\centering
\includegraphics[keepaspectratio,width=250pt,height=0.75\textheight]{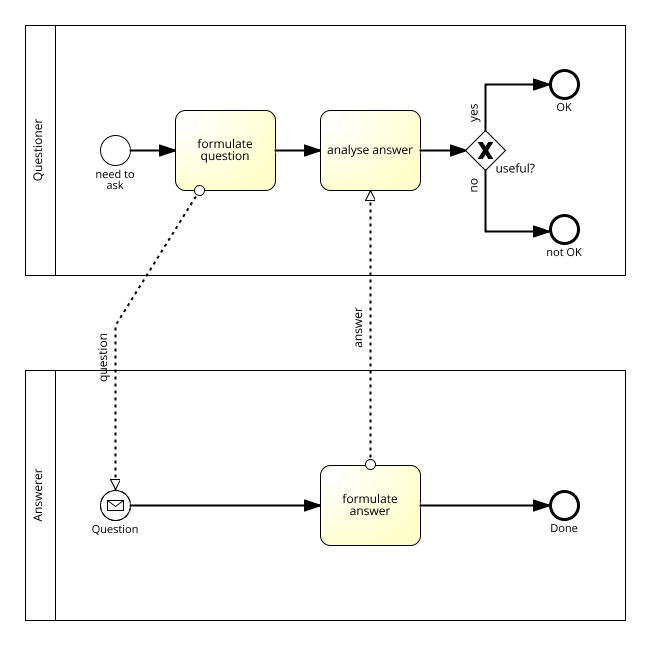}
\caption{A simple BPMN collaboration model.}
\label{question-answer}
\end{figure}

This section describes the transformation process from a serialized BPMN diagram to an OWL file, based on a simple example BPMN process as depicted in \autoref{question-answer}. The model is intended to be simple, but should also include gateways, and it should be a collaboration as defined in the BPMN standard document to support some first explorative testings. A more rigorous approach will be presented later.

For the transformation process, a prototypical software tool had to be developed, which transforms a BPMN-XML into a BPMN-OWL file~\citep{reitergit1}. The resulting file contains a partial ontology including classes and individuals of the input file. For some explorative tests, a selection of the well-known workflow patterns was used. Another testing strategy was to develop a round-trip; that means transforming a BPMN-file into an OWL-file and afterward transforming the resulting OWL-file back into a corresponding BPMN-file which should be equal to the original BPMN-file. For that reason all information from the input file is included in the OWL-file, even if not needed for an ontological analysis; so all vendor-specific attributes and graphical information is also transferred. This makes the OWL-file unnecessarily complicated, but such extra information could be stripped from the file with some lines of code. The reason for such a procedure is to practically test all assumption and the correctness of the software tools.

Some conceptional issues had to be resolved. For example, it is crucial that for each node in the OWL file a unique instance number is set, because every node in the BPMN diagram represents a single instance. In an OWL file, instances are stored as named individuals which must have a unique name.

There are a few steps necessary during the creation of named individuals: named individuals may have a node value, so it has to be tested if they have a node value or not. If they have a node value, the value of the node is set as text content in the OWL file. Named individuals may have attributes, which also need to be set in the OWL file. Named individuals belong to a specific class, which is their parent or superclass. The parent class defines their properties, and a parent class may have multiple instances. There is only one node, which has ``document'' as parent. This is the root node in the BPMN diagram. The root node of an ontology is always <owl:Thing>, and during the transformation, the document node is automatically set as a child of <owl:Thing>.

\begin{figure}[htbp]
\centering
\includegraphics[keepaspectratio,width=150pt,height=0.75\textheight]{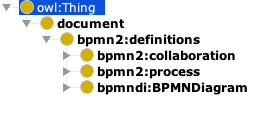}
\caption{Extract from the generated OWL file.}
\label{question-answer_owl_1}
\end{figure}

\autoref{question-answer_owl_1} shows that the document itself is added to the ontology node owl:Thing and has one child node, which is bpm2:definitions. The node bpmn2:definitions has one parent node named document and three child nodes, which are bpmn2:collaboration, bpmn2:process and bpmndi:BPMNDiagram.

Every node of the BPMN diagram represents a class and has a parent. It is necessary to get the parent node to find out to which parent a specific class belongs since the class hierarchy of the BPMN diagram is also transformed to the ontology to show the relationship between all nodes of the BPMN diagram.

Every node can have multiple attributes. BPMN diagrams only have a list of attributes, whereas ontologies have two categories of attributes. Attributes in an ontology are called properties, and they can belong to object properties or data properties. Data properties belong to the class itself and object properties refer to another class and represent a relationship. This is the reason why it is crucial to divide the attributes of each node into these two categories.

Data properties have a data type which has to be set accordingly. To find out if an attribute is an object property, it is necessary to check if the attribute value contains the name of another node. If that is the case, the attribute is a reference and should represent a relationship to another node.

Properties in ontologies are realized with the tag <owl:Restriction>. An OWL restriction of a data property contains three parts: <owl:onProperty> defines to which class the data property belongs, <owl:qualifiedCardinality> defines the exact cardinality of the data property, and <owl:onDataRange> defines the data type of the data property. An OWL restriction of an object property also contains three parts: <owl:onProperty> defines to which class the object property belongs, <owl:qualifiedCardinality> defines the exact cardinality of the object property, and <owl:onClass> defines the reference to another class and represents the relationship between them.

During the creation of instances, classes and properties they are immediately appended to different sets. Before the OWL file can be written, it is necessary to know how the OWL file has to be structured. It contains the following parts:

\begin{itemize}
\item the Ontology Tag includes the rdf:about property,

\item the RDF Tag defines the vocabulary for rdf and owl,

\item the Object Properties Set contains the names of all object properties.

\item the Data Properties Set includes the names of all data properties,

\item the Restrictions Set contains all restrictions for data and object properties,

\item the Named Individuals Set consists of all nodes from the BPMN diagram as named individuals,

\item the Class Set contains all names from all nodes from the BPMN diagram as class name, and

\item the owl file ends with a closing rdf tag.

\end{itemize}

These parts need to be in this order because some parts depend on other parts, which have to be already in place.

\begin{figure}[htbp]
\centering
\includegraphics[keepaspectratio,width=250pt,height=0.75\textheight]{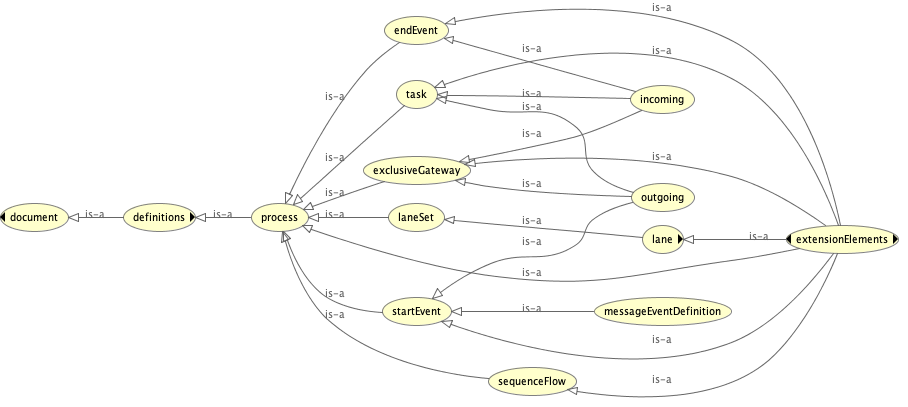}
\caption{Visual representation of the generated OWL file.}
\label{question-answer_owl_3}
\end{figure}

Now, \autoref{question-answer_owl_3} shows the ontology, which was generated from the example BPMN diagram (see \autoref{question-answer}). The ontology starts with owl:Thing, which is always the starting point of every ontology. The next class is document, which has only one child class named bpmn2:definitions. The class bpmn2:definitions consists of three child classes, as discussed before. The class bpmn2:process contains as child classes all BPMN elements which can be seen in \autoref{question-answer}. These elements are bpmn2:sequenceFlow, bpmn2:startEvent, bpmn2:task, bpmn2:exclusiveGateway, and bpmn2:endEvent. As mentioned above, there are sequence Flows between tasks and events. BPMN diagrams differ between ingoing and outgoing sequence flows. The ontology shows, which elements have ingoing and\slash or outgoing sequence flows. It can be seen, that the class bpmn2:startEvent has only a relationship to bpmn2:outgoing, whereas bpmn2:endEvent has only a relationship to bpmn2:incoming. The classes bpmn2:task and bpmn2:exclusiveGateway have a bpmn2:incoming as well as a bpmn2:outgoing sequence flow. The bpmn2:extensionElements are vendor specific extension which could be ignored for further analysis.

\begin{figure}[htbp]
\centering
\includegraphics[keepaspectratio,width=250pt,height=0.75\textheight]{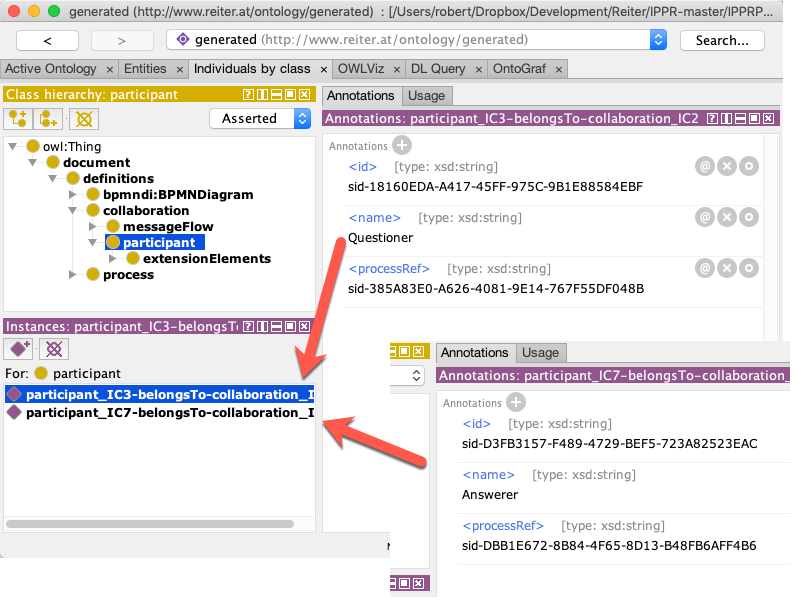}
\caption{Visualisation of the generated OWL file showing that a concrete model is based on individuals.}
\label{questioner-answerer_individuals}
\end{figure}

The class bpmn2:collaboration has two child classes, which are bpmn2:messageFlow and bpmn2:participant. The class bpmn2:participants stand for the previously mentioned labeled pools questioner and answerer (see \autoref{questioner-answerer_individuals}). The generated ontology shows all elements of the BPMN diagram and the relationship between them. It not only contains the elements, but also the concepts behind it. A closer look at the concepts allows the assumption that at least two pools indicate a collaboration; therefore the collaboration class has to be present.

In the end, the generated OWL file of a concrete BPMN model can be compared with the OWL file which contains the full BPMN standard. The comparison process between two ontologies is included in the developed application. The comparisons (based on comparing a set of strings) between the BPMN diagram (stored as an ontology) and the BPMN standard ontology file are automatically made for classes, object properties, data properties, subclasses, and are stored as a set of log files.

The next step is to parse the model ontology to find all restrictions and to compare it with the restrictions defined in the standard ontology. The program searches for the minimum, maximum and exact cardinalities, which are part of the standard but are missing in the model ontology. The model ontology has only exact cardinalities because it represents a concrete BPMN diagram with a specific number of pools, tasks, and gateways.

Now, it can be summarized that it is possible to convert any BPMN model into a OWL file based on an ontology reflecting the standard document. The OWL file inherently contains the information from the standard document, such as multiple inheritance, for example. We will use the OWL files for testing a process model for conformance with the standard and we will use it as a starting point to translate a process model from BPMN into S-BPM, i.e. a language translation from one notation into another one.

\subsection{Verification of the BPMN Ontology}
\label{verificationofthebpmnontology}

To use the BPMN ontology in a productive environment more tests and refinements of the ontology are needed. Therefore we have chosen the set of test cases of the BPMN Model Interchange Working Group\footnote{http:\slash \slash www.omgwiki.org\slash bpmn--miwg\slash } (BPMN-MIWG). The purpose of the BPMN MIWG is to support, facilitate, and promote the interchange of BPMN Models, as stated on their homepage. That means the models of the tests are developed to test the conformance of software tools for process modeling systematically. Because the BPMN ontology is intended to consistently represent the BPMN standard in a rigorous and machine-readable way this set can be used to test if the BPMN MIWG models can be transformed into OWL files without errors; this should also be a test of the completeness of the ontology. A reference of the used BPMN elements in the BPMN-MIWG model set is available via the corresponding Github repository.

To test specific software has been developed. The tool can be used to convert any BPMN model into an OWL file or test a set of BPMN models---such as the BPMN-MIWG set of models---against the BPMN ontology. The tool and all results from the conducted tests are available on the Web~\citep{hoedlgit}. During the test, the ontology was recursively refined, when needed.

The BPMN ontology is valid if the following statements can be proven:

\begin{itemize}
\item Each BPMN model element (XML-element in the file) from a particular model is defined as OWL class (including inheritance).

\item Each BPMN-attribute (XML-attribute in the file) from a specific model must be a documented OWL data or object property; ontology properties are related to classes, in XML they appear as attributes.

\item All in the ontology defined restrictions for a particular OWL class correspond with the BPMN model. Ontology restrictions refine class definitions as they specify constraints of how often a specific class or property may appear in a model, or they define the datatype of a property. The incidence of BPMN elements can be defined as a precise number, or as minimum and maximum values.

\end{itemize}

\begin{figure}[htbp]
\centering
\includegraphics[keepaspectratio,width=250pt,height=0.75\textheight]{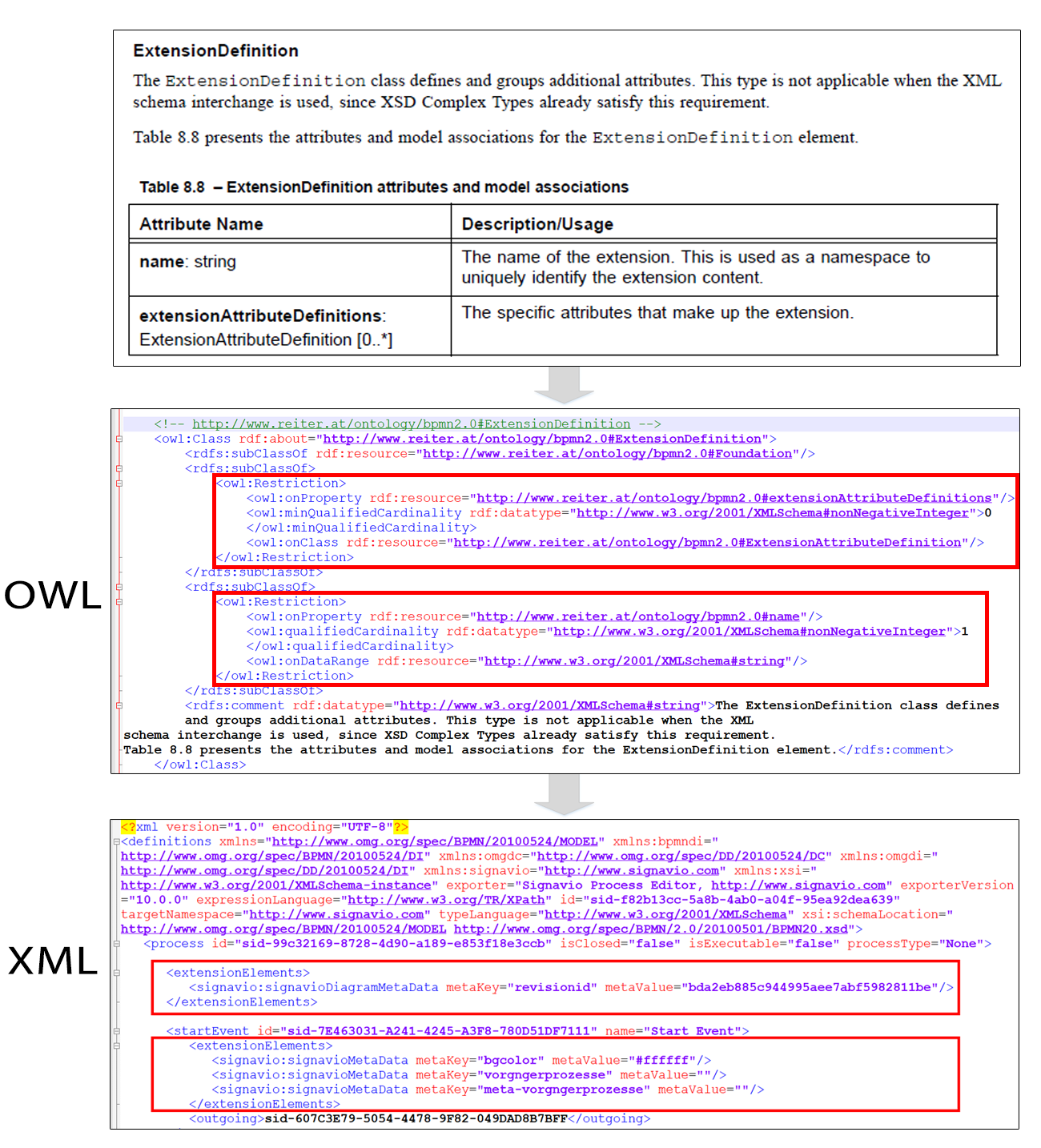}
\caption{Example of a naming conflict: ExtensionDefinition in the standard document and in the ontology vs. extensionElements in XML files.}
\label{namenszuordnungbeispiel}
\end{figure}

During the development of the ontology, some decisions had to be made according to the naming of ontology elements. In the ontology, all class must have a distinct name, but in the standard document, it can be that the same name is used for different things, as the meaning is clear from the context. Nevertheless, it seems that tool vendors also have enhanced or changed the standard in some way as they use names which are not defined in the standard document (for some practical reason); see \autoref{namenszuordnungbeispiel}, for example. To solve this optional naming conflict, a flexible translation table can be loaded into the test tool. Another issue is that all vendors include numerous specific attributes (which is explicitly allowed in the standard document) in their model serializations, but these extension attributes are not relevant to test the conformance of a particular model with the standard. Therefore, the test and conversion tool only considers the namespaces (URIs) defined in the standard document, i.e., http:\slash \slash www.omg.org\slash spec\slash BPMN\slash 20100524\slash MODEL and http:\slash \slash www.omg.org\slash spec\slash BPMN\slash 20100524\slash DI; the second one represents information for the graphical representation, such as x and y coordinates, for example.

Even after some fine tuning (test code and ontology) the testing of the BPMN-MIWG model set showed some discrepancies: on one side all XML elements (i.e. BPMN classes) of the test models could be matched with the corresponding OWL concepts (i.e. BPMN classes); on the other side several breaches of OWL restrictions (onProperty restrictions) were reported. The previous statements for the mapping of XML nodes and OWL entities do not permit a closer verification of the ontology, since the defined restrictions are not fulfilled exclusively via XML attributes. Constraints defined in the standard can also be accomplished by XML child elements. However, restrictions specified in the ontology are always assigned to a property of a class via the keyword onProperty and must, therefore, be fulfilled according to the chosen concept of XML attributes. In an ontology, a class is more closely defined by attributes (properties) and attributes are further refined by restrictions.

The third statement above---which is related to the restrictions defined in the standard---defines the verification process and contains a conceptual error that was made in the conversion of the standard because of the misinterpretation of the name \emph{Attribute}. The standard defines attributes of a class, but these do not necessarily have to be mapped as an XML attribute. Attributes of a class can also be fulfilled by other classes. In XML, this branch is represented as an XML-child element.

Because of this, constraints of an ontology must distinguish between restrictions on attributes (OWL properties) and classes (OWL class). If a restriction concerns an attribute, then the restriction itself and all associated (inherited) constraints must be satisfied by the class. In terms of XML, this means complying with the restriction of XML attributes of the XML element representing the class. However, if a restriction affects another class, the class containing the restriction must meet the frequency of occurrence. All inherited constraints must, yet, be achieved by the class that affects the restriction. In XML, this constellation would correspond to an XML child element. The XML parent element must meet the number of defined frequency of restriction, while the XML child element must satisfy all inherited constraints.

\begin{figure}[htbp]
\centering
\includegraphics[keepaspectratio,width=250pt,height=0.75\textheight]{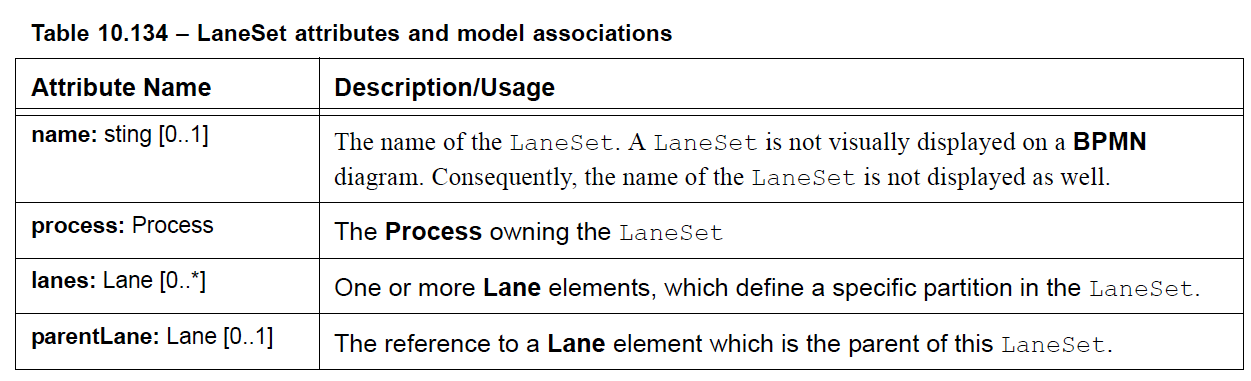}
\caption{Description of the class LaneSet from the standard document.}
\label{bpmn_laneset}
\end{figure}

Using the class LaneSet, the problem can be illustrated graphically. Figure \autoref{bpmn_laneset} shows the description of the class in the standard.

In the ontology, the class \texttt{LaneSet} was mapped according to the standard. The restriction points to an OWL property through the \texttt{onProperty} keyword but also uses the \texttt{onClass} keyword to reference its own class, which should satisfy the constraint of the assigned class. Due to the strict concept and the assumptions made, the test application expects an XML attribute instead of an XML child element \texttt{lane}. Also, the XML element \texttt{LaneSet} must satisfy all inherited restrictions of the class Lane.

Therefore, in addition to the adaptation of the statements for the verification, a concept for the ontology must be defined, which is capable of distinguishing between restrictions that are met by XML attributes and constraints that are met by XML child elements to ensure proper inheritance to the respective XML elements.

This distinction can be made using the optional OWL keywords of the restriction \texttt{onClass} and \texttt{onDataRange}. Restrictions that are met by XML attributes must not have a keyword \texttt{onClass}. Instead, the optional keyword \texttt{onDataRange} can be used to specify which data type the value of the attribute must have (i.e., XML attribute restrictions). The keyword \texttt{onClass} is used to identify restrictions that are met by XML child elements. Linking to the class further describes the XML child element and identifies its restrictions and inheritance. Since OWL properties are required for restrictions in an ontology, the keyword \texttt{onProperty} in the verification for these restrictions can be ignored (i.e., XML-child-item-restrictions). Furthermore, it must be ensured that restrictions are not used for direct inheritance, even if they are in the context \texttt{subClassOf}. Attribute restrictions can not include inheritance (\texttt{onClass}), and XML child restrictions only affect inheritance for the XML child element.

\begin{itemize}
\item Each XML element from the XML of the process model must be mapped to the ontology as a class. An XML element is an XML node that can contain additional XML nodes (XML attributes or further XML child elements).

\item Each attribute used in the XML of the process model must be stored in the ontology as data or object property. A property in an ontology is associated with and describes an ontology class. Properties are represented in XML as attributes of the XML elements of the process model. If the property is to be represented via an XML attribute, it must be displayed as a data property. If the property is to be represented as an XML child element, this property should be defined as an object property in the ontology.

\item Restrictions define a property of the class more closely and restrict it in terms of allowed occurring frequency. Restrictions that are met by XML attributes must not have the keyword \texttt{onClass}. Optionally, the expected data type can be specified with the keyword \texttt{onDataRange}. Restrictions that are met by XML child elements must have the keyword \texttt{onClass}. The link describes the XML child element in more detail and must fulfill all defined restrictions from the inherited links of the linked class.

\end{itemize}

\begin{figure}[htbp]
\centering
\includegraphics[keepaspectratio,width=250pt,height=0.75\textheight]{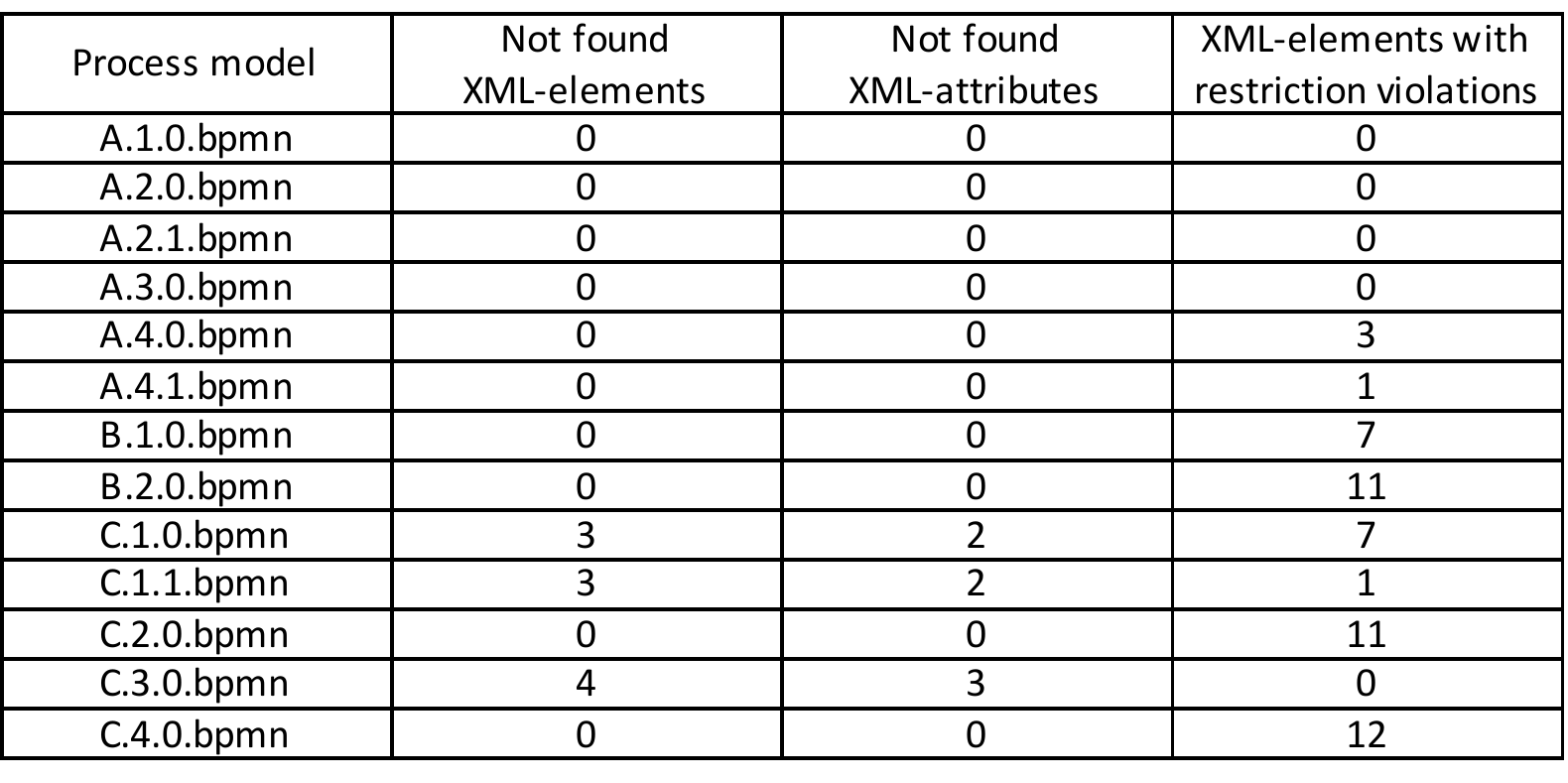}
\caption{Final test results of comparing the set of the BPMN-MIWG test models with the developed ontology of the BPMN standard.}
\label{testresults}
\end{figure}

The remaining restriction violations can be explained by the deviant implementation of the standard in the tested process models; that means how the process models are serialized. For example, the classes \texttt{Category}, \texttt{Collaboration}, \texttt{DataStore}, \texttt{Message}, and \texttt{Message Flow} show restriction violations because the attribute \texttt{name} is in the standard document as mandatory defined. However, the attribute is not included in the tested process models. The missing assignment of the XML elements and XML attributes in the process models C.1.0.bpmn, C.1.1.bpmn, and C.3.0.bpmn. Can be explained by the missing manufacturer-specific namespace for XML child elements of the XML element \texttt{extensionElements}.

Even after the ontology verification, however, an incorrect configuration of the ontology cannot be ruled out 100\%, as only the presence of errors is shown by tests. In the course of analyzing the standard, no structure for uniquely identifying the mapping of the values from the attribute tables to XML attributes or XML elements could be identified. It can be concluded that the standard does not clearly state whether a described attribute should be implemented as an XML attribute or an XML element. This can lead to misinterpretations of the standard. For example, the attributes \texttt{sourceRef} and \texttt{targetRef} of the class \texttt{SequenceFlow} are interpreted as XML attributes, whereas for the class \texttt{DataInputAssosiation} it is implemented as an XML element.

Finally we can conclude that the provided ontology matches the standard document beside the fact that is some critique about the standard text related to some inconsistencies in presentation; this can also be concluded from the fact that after the release of the standard document it was nearly impossible to exchange models between applications from different vendors and the BPMN Model Interchange Working Group would not be needed if anything is clear. Nevertheless, the work of all tool providers leads to clarification and a new version of the standard amalgams from this effort. The analysis of the provided BPMN ontology accurately shows what to improve in the definition of the standard to give an exact description of the semantics and syntax of BPMN.

\subsection{Transforming BPMN to S-BPM Models (OWL to OWL)}
\label{transformingbpmntos-bpmmodelsowltoowl}

\begin{figure}[htbp]
\centering
\includegraphics[keepaspectratio,width=250pt,height=0.75\textheight]{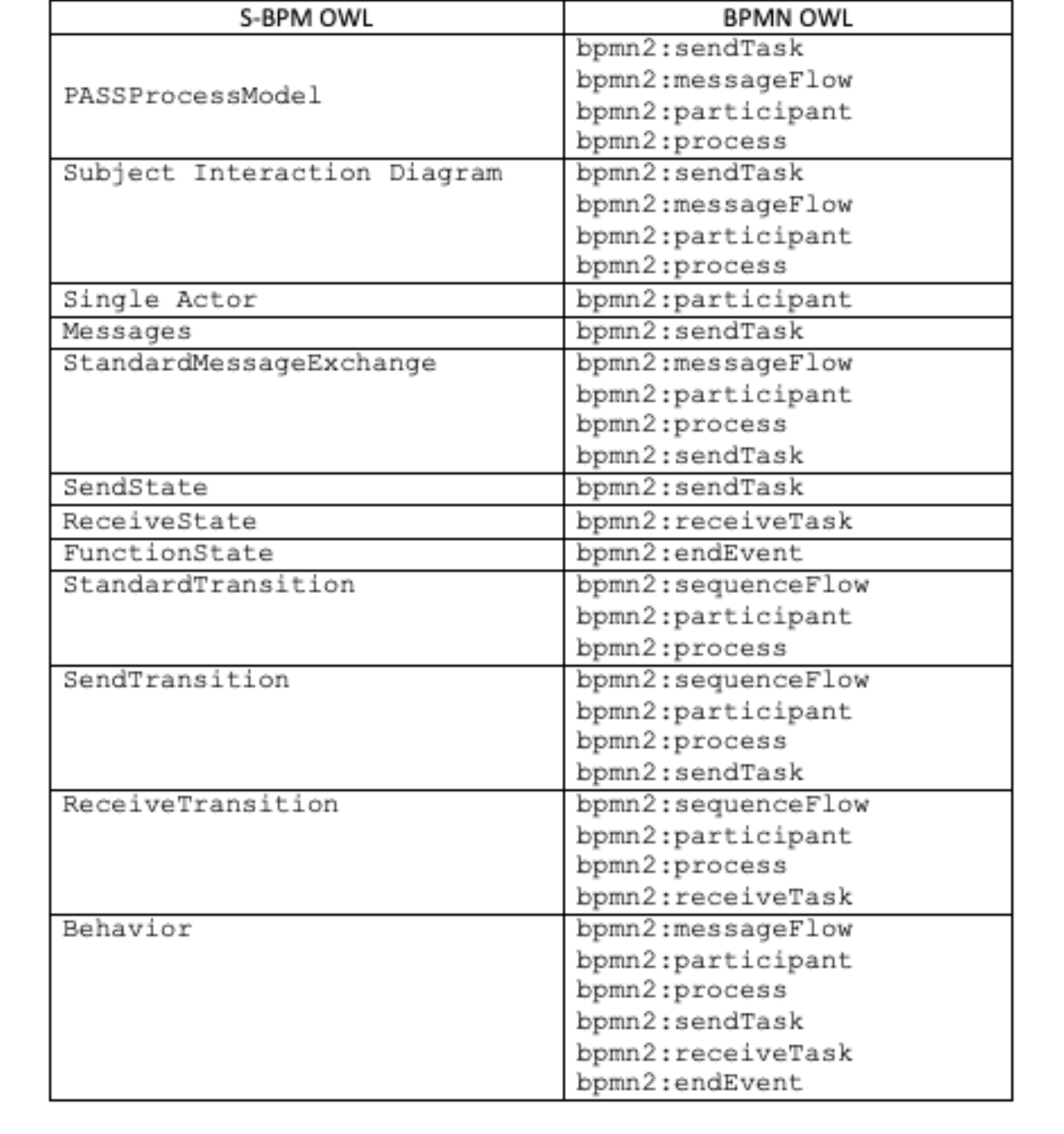}
\caption{Comparison of the matching concepts between BPMN and S-BPM based on OWL elements.}
\label{compareconcepts}
\end{figure}

A comparison (ontological analysis) of the underlying concepts of the modeling languages BPMN and S-BPM shows that a comparison seems to be possible. Nevertheless, a 1:1 mapping is not possible, but a pattern matching. Only a restricted set of BPMN elements is needed to model a process based on subject-oriented concepts which is not a problem because it has been proved, as discussed, that S-BPM can describe all workflow and interaction patterns; furthermore, all patterns can be executed on a developed workflow engine which demonstrates the formal and expressive underlying concepts of subject-orientation. Table \autoref{compareconcepts} compares the matching elements of the two ontologies (or concepts).

Thus, only a subset of BPMN elements is needed (it is helpful if the chosen modeling toll can be configured accordingly). The limited subset is limited to the BPMN elements pool, message, start event, activity, data store, exclusive or, throwing and catching intermediate message event, event-based gateway, and end event. A serialized BPMN model need to be transformed into an OWL file which conforms to the previously discussed BPMN ontology. Afterward, the OWL-file can be used to convert the model into an S-BPM representation (and back again). For the transformation, an application had to be developed~\citep{reitergit2}.

\subsubsection{Model Transformation}
\label{modeltransformation}

\begin{figure}[htbp]
\centering
\includegraphics[keepaspectratio,width=150pt,height=0.75\textheight]{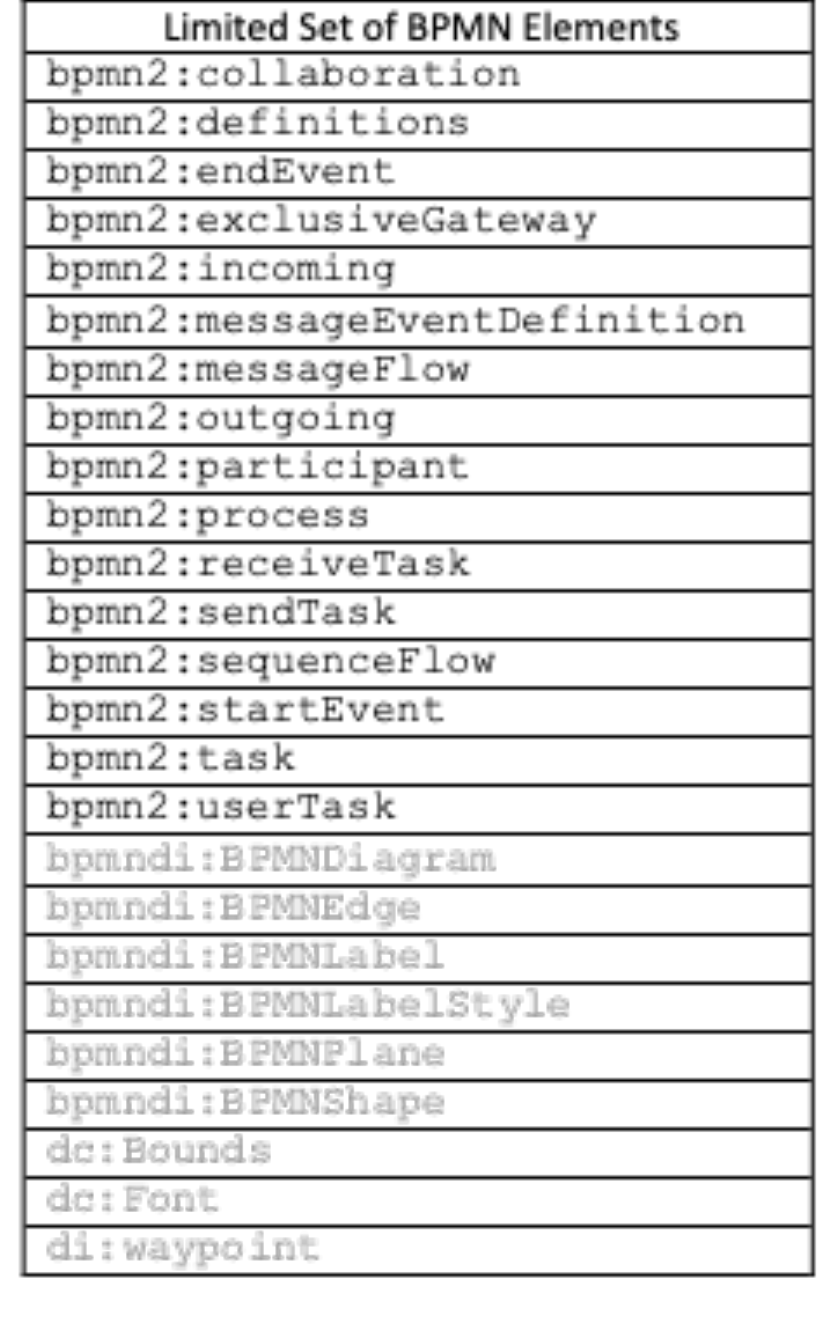}
\caption{A collection of usable and tested BPMN elements which can be used to transform a model to S-BPM.}
\label{limitedsetofbpmnelements}
\end{figure}

Transforming a BPMN file means to parse the corresponding XML file to identify patterns for which a corresponding S-BPM pattern can be identified. For example, in order to create a Subject Interaction Diagram, the BPMN elements bpmn2:sendTask, bpmn2:messageFlow, bpmn2:participant and bpmn2:process are needed. An overview of the elements which have been considered in our transformation process is summarized in table \autoref{limitedsetofbpmnelements}. This selection can be enhanced to consider more elements in the transformation process. In this work we have restricted ourself to consider only those elements which represent the core concepts of subject-orientation (i.e., the core concepts of an actor based view).

\subsubsection{Proof of Concept}
\label{proofofconcept}

\begin{figure}[htbp]
\centering
\includegraphics[keepaspectratio,width=250pt,height=0.75\textheight]{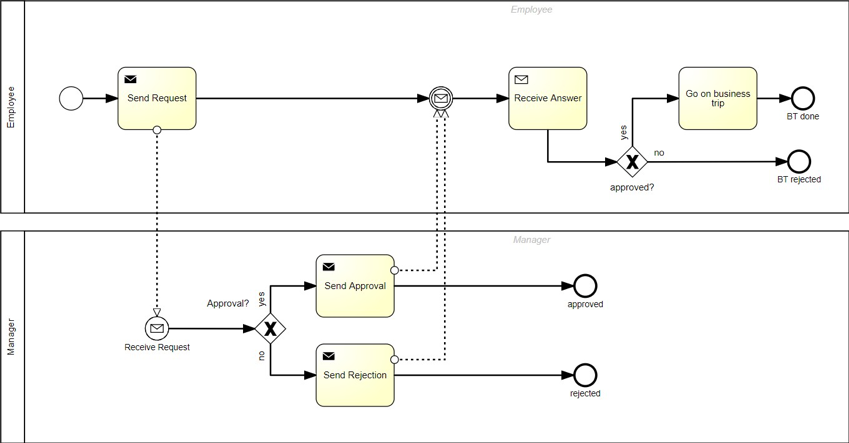}
\caption{A simple BPMN model used as input for a proof of concept.}
\label{bpmnconversionmodel}
\end{figure}

For the transformation process from BPMN to S-BPM an application has been developed~\citep{reiter2018}; the transformation process is done via two ontologies, as discussed previously: a BPMN model is transformed and stored as OWL ontology, and afterwards the resulting BPMN OWL file is tranformed into an S-BPM OWL file. Some explorative testings have been conducted and model \autoref{bpmnconversionmodel} is used to demonstrate the results. It can be concluded that based on a conceptual mapping via ontologies it is possible to transform a business process model from one modeling notation into another one.

It is not possible to compare the models, which are represented as OWL-files, based on visual inspection as there is no grafical representation of the generates S-BPM ontology available up to now. A proof of the translation can only be done based on inspection of the generated OWL files. And it can be stated that the conversioin works for the tested model as depicted in (\#bpmnconversionmodel). Nevertheless, it can be summarized that a conversion is principally possible, based on the developed concepts. But, at the time of writing still some work has to be done developing a reliable tool for the translation process.

\begin{figure}[htbp]
\centering
\includegraphics[keepaspectratio,width=250pt,height=0.75\textheight]{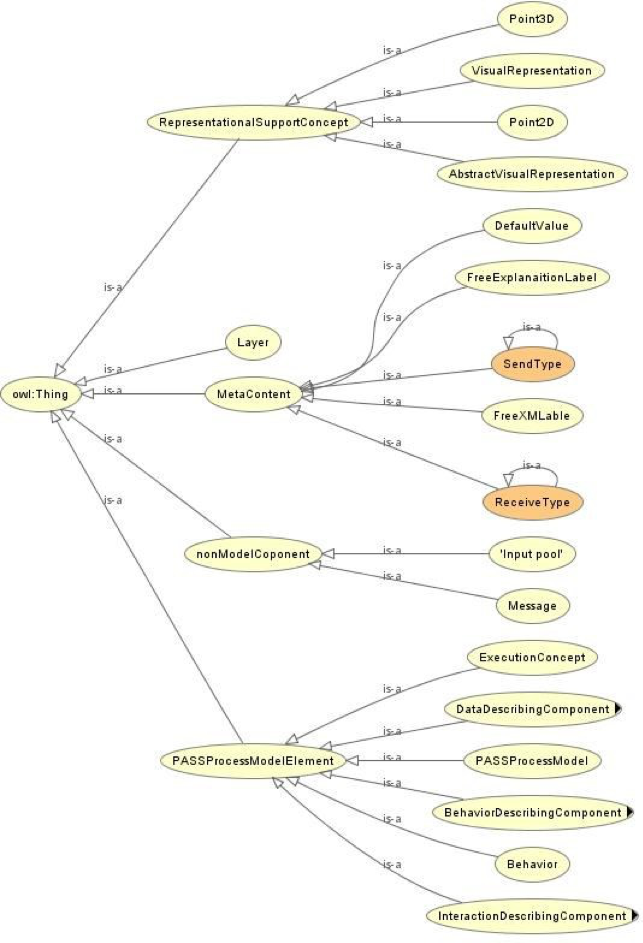}
\caption{The transformed S-BPM OWL.}
\label{sbpmconversionontviz}
\end{figure}

\section{Conclusion and Discussion}
\label{conclusionanddiscussion}

In this work, we have discussed several topics related to modeling and executing business processes. First, we have summarized some of the core concepts which easily can be identified as states and state transitions. Nevertheless, there are some more sophisticated concepts needed to create useful and executable business process models. That is the domain of business process modeling languages such as BPMN as an industry standard and S-BPM as a proprietary alternative, for example.

In this work, we have demonstrated that based on a conceptual comparison of modeling notations it is possible to translate a process model from one notation to another one. This is possible as there is some ``reality'' which needs to be modeled by all useful notations and therefore all modeling languages share some common concepts. It could also be demonstrated that a formal ontology is an appropriate way to define the concepts of a modeling notation in such a way that an ontological analysis can be conducted and a transformation process can be determined.

Even that it is possible to transform models from BPMN to S-BPM, they are built on some rather different paradigms, and it is fundamental to understand the advantages and disadvantages of each modeling notation. We think that business processes should be more comfortable to be modeled and should be implicit formally defined and ready for execution which seems to be not a strength of BPMN.

We also present some approaches to designing a typical architecture for workflow management systems as there seems to be some lack of research. The WfMC proposed a reference architecture more than 20 years ago, and we think that some of the proposed capabilities of WfMS have been forgotten, such as multiple workflow engines, for example. We propose to rethink actual design principles and to strive towards multi-enterprise and cloud-based architectures. In principle, a typical architecture despite the used modeling notation can be defined, as discussed in this work. Nevertheless, the enacted modeling notation pretends the design and capabilities of the workflow engine itself. In our proposed prototypical WfMS based on microservices, it would be possible to design an architecture which could execute models based on several business process notations as the workflow engine is encapsulated as exchangeable microservice in the overall architecture.

In summary, we can say that all research questions stated in the beginning (H1-H5) can be positively confirmed. We have provided links to all prototypical tools developed to answer the questions in this work and invite all interested readers to adapt and improve them as needed.

\bibliographystyle{IEEEtran}
\bibliography{IEEEabrv,GP-Modell}

\end{document}